\documentclass[twocolumn]{aastex6}
\usepackage{graphicx}
\usepackage{amssymb,amsfonts,amsmath,amstext,amsgen,amsopn,amsxtra,indentfirst,times,longtable}

\begin{document}

\title{Atmospheric Chemistry of Secondary and Hybrid Atmospheres \\ of Super Earths and Sub-Neptunes}

\author{Meng Tian\altaffilmark{1,2,5,6}}
\author{Kevin Heng\altaffilmark{2,3,4,6}}
\altaffiltext{1}{University of Bern, Center for Space and Habitability, Gesellschaftsstrasse 6, CH-3012, Bern, Switzerland.}
\altaffiltext{2}{Ludwig Maximilian University, Faculty of Physics, University Observatory Munich, Scheinerstrasse 1, Munich D-81679, Germany.}
\altaffiltext{3}{University of Warwick, Department of Physics, Astronomy \& Astrophysics Group, Coventry CV4 7AL, United Kingdom.}
\altaffiltext{4}{University of Bern, ARTORG Center for Biomedical Engineering Research, Murtenstrasse 50, CH-3008, Bern, Switzerland.}
\altaffiltext{5}{CSH Fellow in Geochemistry}
\altaffiltext{6}{Email: Meng.Tian@physik.lmu.de, Kevin.Heng@physik.lmu.de}

\begin{abstract}
The atmospheres of small exoplanets likely derive from a combination of geochemical outgassing and primordial gases left over from formation.  Secondary atmospheres, such as those of Earth, Mars and Venus, are sourced by outgassing.  Persistent outgassing into long-lived, primordial, hydrogen-helium envelopes produces hybrid atmospheres of which there are no examples in the Solar System.  We construct a unified theoretical framework for calculating the outgassing chemistry of both secondary and hybrid atmospheres, where the input parameters are the surface pressure, oxidation and sulfidation states of the mantle, as well as the primordial atmospheric hydrogen, helium and nitrogen content.  Non-ideal gases (quantified by the fugacity coefficient) and non-ideal mixing of gaseous components (quantified by the activity coefficient) are considered.  Both secondary and hybrid atmospheres exhibit a rich diversity of chemistries, including hydrogen-dominated atmospheres.  The abundance ratio of carbon dioxide to carbon monoxide serves as a powerful diagnostic for the oxygen fugacity of the mantle, which may conceivably be constrained by James Webb Space Telescope spectra in the near future.  Methane-dominated atmospheres are difficult to produce and require specific conditions: atmospheric surface pressures exceeding $\sim 10$ bar, a reduced (poorly oxidised) mantle and diminished magma temperatures (compared to modern Earth).  Future work should include photochemistry in these calculations and clarify the general role of atmospheric escape.  Exoplanet science should quantify the relationship between the mass and oxygen fugacity for a sample of super Earths and sub-Neptunes; such an empirical relationship already exists for Solar System bodies.
\end{abstract}


\section{Introduction}
\label{sect:intro}

\subsection{Motivation I: unified framework for outgassing theory}

Does the Solar System mislead us? The planets of our Solar System come in two flavors: gas/ice giants and rocky bodies.  The gas and ice giants have primary atmospheres with compositions that largely reflect the chemistry of the primordial nebula out of which they formed.  The terrestrial planets have secondary atmospheres with compositions that are largely determined by outgassing from their mantles and geochemical cycles (which cycle volatiles between the atmosphere, surface and interior).  Hybrid atmospheres, where the influences of both channels are comparable, do not exist in our Solar System.  Since we expect to hunt for biosignature gases in secondary atmospheres (e.g., \citealt{seager13}) that considerably deviate from Earth-like conditions, e.g., surface pressure and mantle oxygen fugacity, it becomes imperative to develop a theory of outgassing chemistry that is capable of exploring the diverse set of physical and chemical conditions anticipated in exoplanets, since the outgassed species may be false positives for biosignatures.  In other words, if biosignature hunting on exoplanets proceeds via astronomical remote sensing then biosignatures are spectral anomalies relative to a geochemical background.

Calculations of outgassing chemistry is a mature topic in the geochemical literature (e.g., \citealt{french66,ok77,holloway81,holloway87,cc93,huizenga11}).  However, studies tend to focus on Earth-centric conditions (e.g., \citealt{Iacono12,gs14,gaillard22}).  Furthermore, the calculations are typically performed using computer codes that include dozens of species and hundreds, if not thousands, of chemical reactions \citep[e.g.,][]{schaefer12, fegley16, schaefer17}, which hinders understanding from first principles.  In the astrophysical and climate literatures \citep{held05}, it is common practice to construct a hierarchy of theoretical models such that the relative transparency of simpler models (e.g., \citealt{french66}) may be used to sweep parameter space and inform the detailed explorations of more complex models.

To the best of our knowledge, a simple, unified theoretical framework for computing the outgassing chemistry of both secondary and hybrid atmospheres does not exist in the exoplanet literature.  Part of the challenge lies in curating a consistent set of thermodynamic definitions, notations and experimental data in calculations of mixed-phase equilibrium chemistry.  It is thus the goal of the present study to construct such a framework, which will inform future explorations of outgassing calculations for predicting the chemistry of secondary and hybrid atmospheres. 

\subsection{Motivation II: puzzle of super Earths and sub-Neptunes}

The seemingly clean dichotomy of planet types in the Solar System has been broken by the discovery of super Earths and sub-Neptunes, which are exoplanets with a continuum of radii between 1 and 4 times the radius of Earth \citep{howard12,fressin13,petigura13}.  Such exoplanets are common (see recent review by \citealt{bean21}).  Exoplanets with radii above 1.5--1.6 $R_\oplus$ \citep{rogers15,fulton17} appear to have puffy atmospheres dominated by hydrogen and/or helium \citep{owen19}---these are the sub-Neptunes.  Below this critical radius, the bulk densities of these exoplanets are consistent with a predominantly rocky body---these are the super Earths.  There is an ongoing debate on whether this population of exoplanets is sculpted by photo-evaporation or core-powered mass loss \citep{rogers21}. Furthermore, \cite{kite19} proposed that considerable H$_2$ dissolution into magma oceans stalls sub-Neptune growth via H$_2$ accretion, leading to a ``radius cliff" (Figure 1 in their study) and thus an over-abundance of sub-Neptunes in the population statistics.

\begin{figure}[!h]
\vspace{-0.1in}
\begin{center}
\includegraphics[width=\columnwidth]{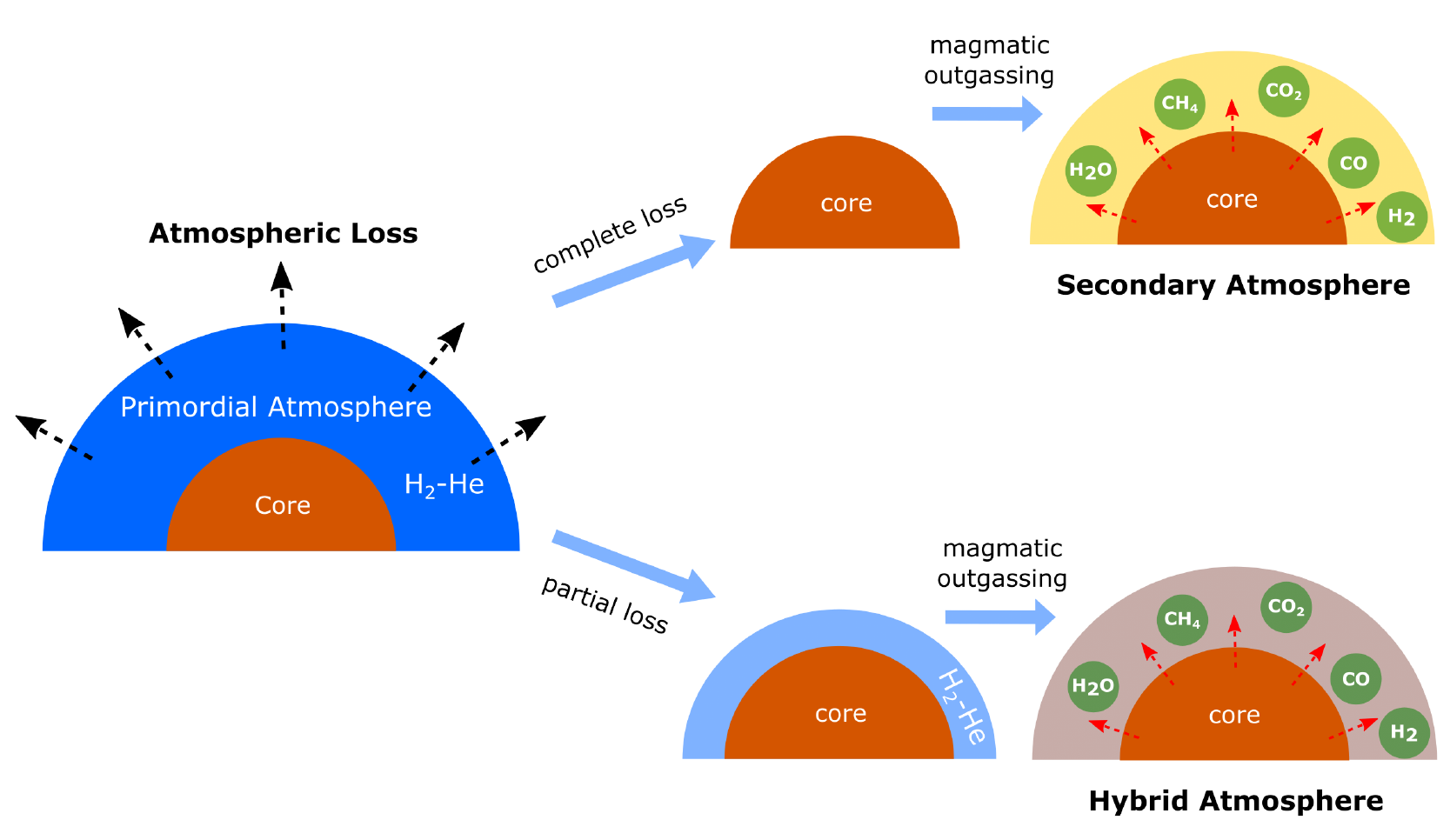}
\end{center}
\vspace{-0.15in}
\caption{Schematic depicting secondary versus hybrid atmospheres.  Secondary atmospheres are fully sourced by geochemical outgassing, while hybrid atmospheres derive from outgassing into a primordial hydrogen-helium envelope left over from the process of formation and evolution.}
\label{fig:schematic}
\end{figure}

The atmospheric chemistry of super Earths and sub-Neptunes remain largely unknown as current observations with the Hubble Space Telescope provide only loose constraints (e.g., \citealt{fh18,benneke19,mevans21}, and see \citealt{bean21} for a summary).  These exoplanets will be the subject of intense scrutiny by the James Webb Space Telescope (JWST) in the near future.  Do super Earths have secondary atmospheres and are some of them hydrogen- and/or helium-dominated \citep{hu15}?  Are some of these atmospheres hybrid between primary and secondary atmospheres (Figure \ref{fig:schematic})?  Models for predicting the chemistry of hybrid atmospheres are still in their infancy (e.g., \citealt{kite20}).  As the anticipated flood of JWST data on these objects arrive, it is imperative that a hierarchy of theoretical models for predicting outgassing chemistry is available.  

\subsection{Motivation III: non-ideal effects}

\begin{figure}%[!t]
\begin{center}
\vspace{-0.2in} 
\includegraphics[width=\columnwidth]{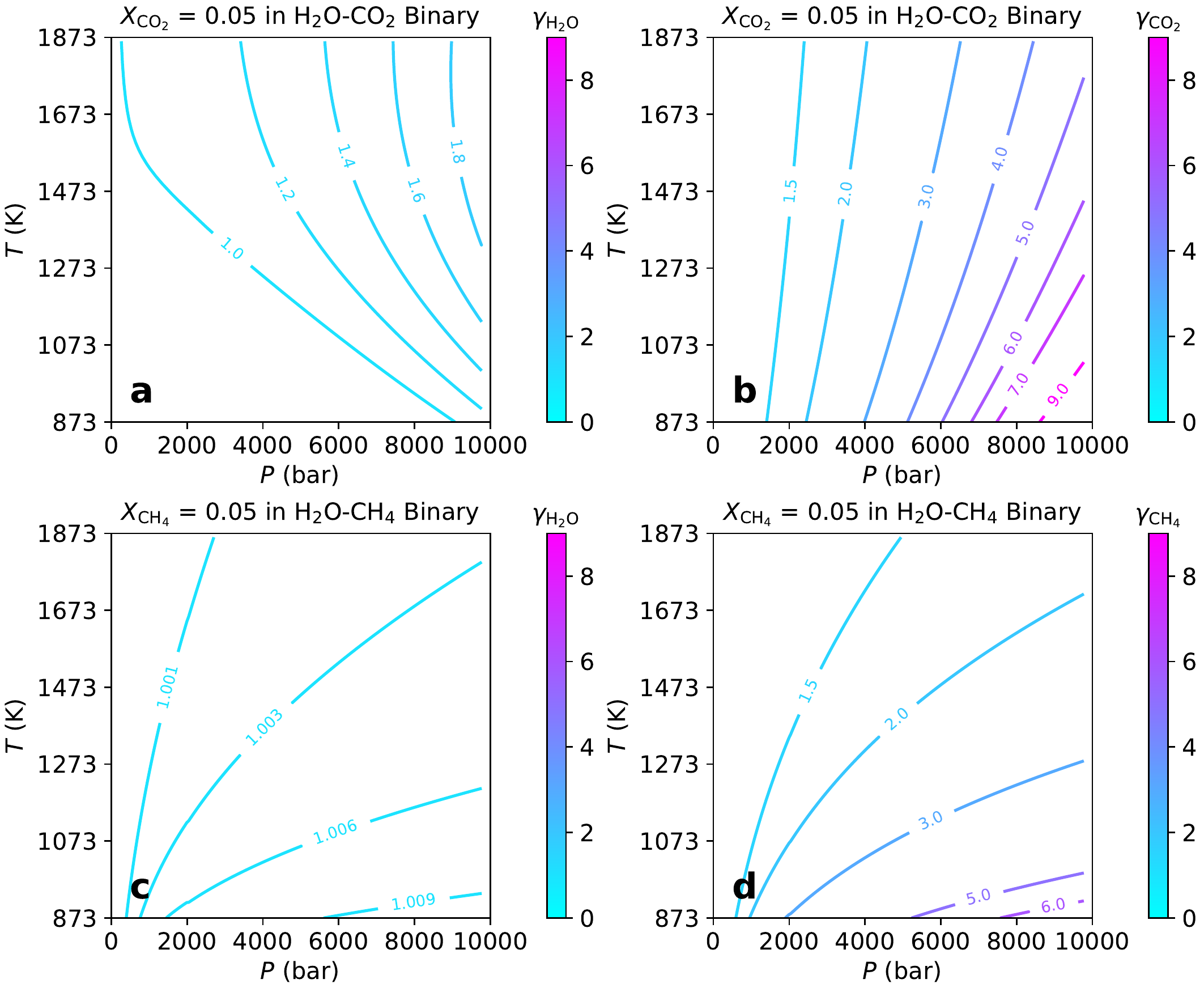}
\end{center}
\vspace{-0.2in}
\caption{Examples of activity coefficients (which we denote by $\gamma_i$ in the present study), which quantify departures from ideal mixing of gaseous components.  In addition to their dependence on temperature and pressure, activity coefficients depend on the specific mixture of molecules being considered.  Here, we show $\gamma_i$ contours from H$_2$O-CO$_2$ (top row) and H$_2$O-CH$_4$ (bottom row) binary mixtures.  The top and bottom rows show activity coefficients for fixed $X_{\rm CO_2} = 0.05$ and $X_{\rm CH_4} = 0.05$, respectively, in order to better visualise the multi-dimensional space of possibilities.  Note that a linear scale in pressure has been chosen for better visualisation of the contours.  Ideal mixing occurs when $\gamma_i=1$.}
%\vspace{-0.15in}
\label{fig:nonideal-a}
\end{figure}

\begin{figure}%[!t]
\begin{center}
%\vspace{-0.1in} 
\includegraphics[width=\columnwidth]{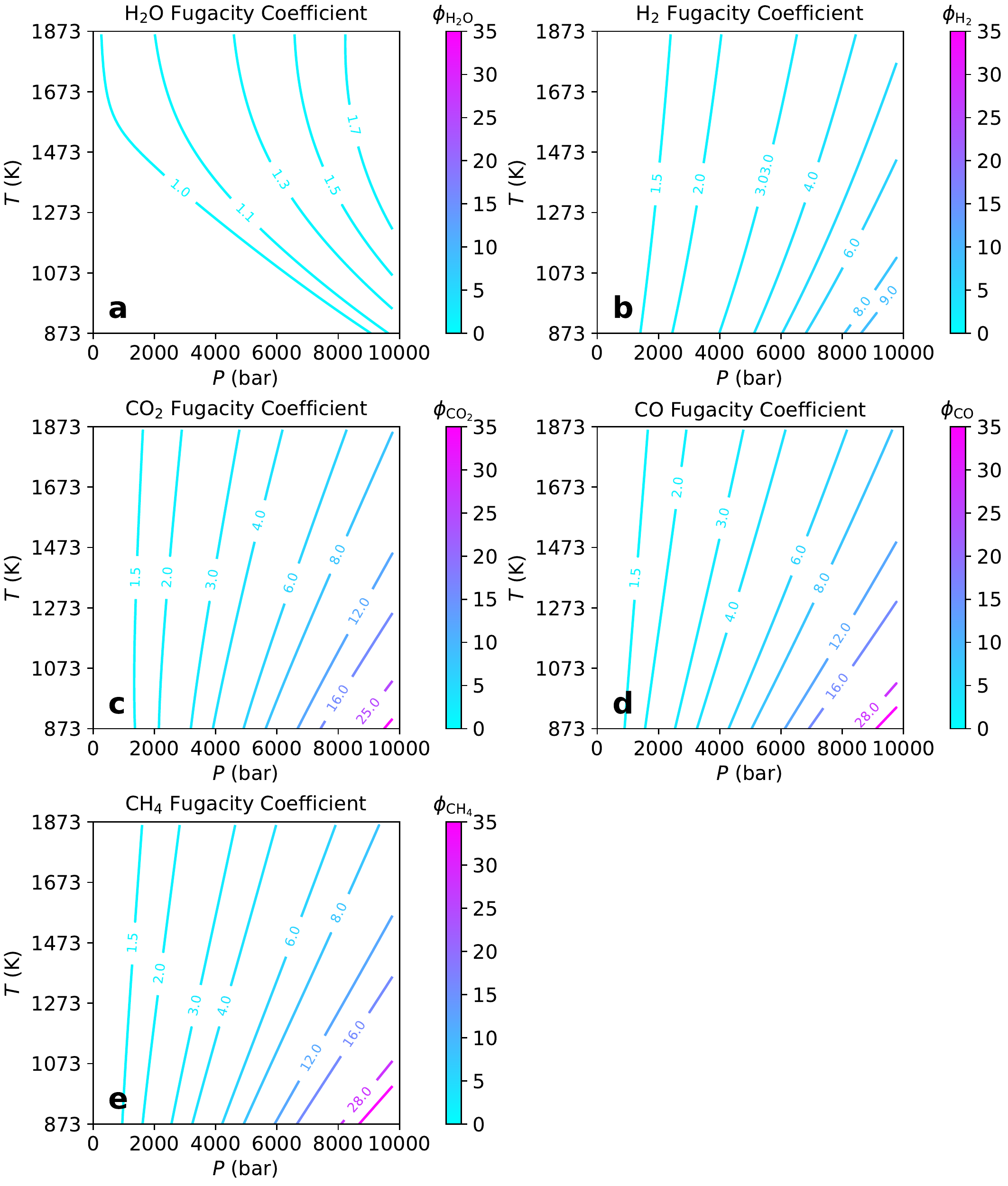}
\end{center}
\vspace{-0.2in}
\caption{Examples of fugacity coefficients (which we denote by $\phi_i$ in the present study), which quantify departures from the ideal-gas equation of state.  Fugacity coefficients depend only on temperature and pressure and may be specified for each species in isolation: (a) water, (b) molecular hydrogen, (c) carbon dioxide, (d) carbon monoxide, (e) methane.  Note that a linear scale in pressure has been chosen for better visualisation of the contours.  Each species behaves like an ideal gas when $\phi_i=1$.}
%\vspace{-0.15in}
\label{fig:nonideal-f}
\end{figure}

Almost without exception in the current literature, theoretical models of exoplanetary atmospheres assume ideal gases with constituent atoms and molecules that are ideally mixed.  Departures from such ideal behaviours are quantified by the fugacity and activity coefficients, respectively.  While ideal mixing may be a reasonable approximation at surface pressures $\lesssim 1000$ bar (Figure \ref{fig:nonideal-a}), the assumption of an ideal gas may be inaccurate for simple molecules (Figure \ref{fig:nonideal-f}).  \cite{kite19} has previously elucidated the relevance of the non-ideal-gas behavior of molecular hydrogen for sub-Neptunes.  To the best of our current ability, we include these non-ideal effects via the fugacity coefficient (for departures from an ideal gas) and activity coefficient (for departures from ideal mixing of gaseous components); our efforts are limited by the currently available experimental data on these coefficients.  In particular, activity coefficients are specific to the mixture of molecules being considered and are difficult to obtain or non-existent for arbitrary mixtures.

\subsection{Structure of paper}

Section \ref{sect:formalism} describes our formalism, which places secondary and hybrid atmospheres on an equal theoretical footing.  It elucidates the governing equations, clarifies the thermodynamic definitions (including some confusion over the activity coefficient), states the numerical solution method and discusses the relevant range of values for the atmospheric surface pressure and melt temperature.  We examine both carbon-hydrogen-oxygen (C-H-O) and carbon-hydrogen-oxygen-nitrogen-sulfur (C-H-O-N-S) systems in turn.  In Section \ref{sect:data}, we review and curate the thermodynamic data that are critical for performing the outgassing calculations.  Section \ref{sect:results} is devoted to a comprehensive exploration of parameter space for both secondary and hybrid atmospheres.  In doing so, we discover the difficulty of producing methane-dominated atmospheres, which motivates a more in-depth investigation.  In Section \ref{sect:discussion}, we compare our current study to previous work, explore its observational implications and describe opportunities for future work.

\section{Formalism}
\label{sect:formalism}

\subsection{Review: generalised thermodynamic quantities}

While none of the thermodynamics described here is novel, there are multiple definitions of the fugacity and activity present in the literature that are sometimes difficult to reconcile.  Conceptually, the fugacity is the generalisation of the pressure accounting for non-ideal-gas effects occurring at high pressures (e.g., Figure \ref{fig:nonideal-f}).  The activity describes non-ideal mixing in a gaseous system with multiple species (e.g., Figure \ref{fig:nonideal-a}).  The assumptions of an ideal gas and ideal mixing is based on the simplification that the constituent gas molecules possess only kinetic, and not potential, energy (e.g., page 78 of \citealt{devoe20}).  This assumes that inter-molecular forces and their contribution to the Gibbs free energy are negligible \citep{holloway87}.  By definition, an ideal gas assumes ideal mixing of its components, implying that the fugacity and activity coefficients, which we will define shortly, are unity (e.g., page 114 of \citealt{denbigh81}).

%If a multi-component gas is ideal, it also implies ideal mixing; however, ideal mixing does not imply an ideal-gas equation of state (see Appendix B).

From the first law of thermodynamics, one may derive for an ideal gas with a single species (e.g., \citealt{denbigh81,heng16}),
\begin{equation}
G = G_0 + {\cal R} T \ln\Psi,
\label{eq:gibbs}
\end{equation}
where $G$ is the specific Gibbs free energy (Gibbs free energy per unit mass), $G_0 \equiv G(P_0)$, $P_0$ is the reference pressure (usually set to 1 atm or 1 bar), ${\cal R}$ is the specific gas constant and $T$ is the temperature.  The task is to generalise $\Psi = P/P_0$ for gas mixtures with non-ideal-gas behavior.

As far as possible, we respect the established notation in the geochemical literature.  The fugacity is commonly denoted as $f$.  \citet{holloway77} and \citet{devoe20} use $\phi$ and $\gamma$ for the fugacity coefficient and activity coefficient, respectively.  In the current study, we follow this convention.

\subsubsection{Fugacity}

For a pure gas (one species),  
\begin{equation}
G = G_0 + \int^P_{P_0} \frac{1}{\rho} ~dP,
\label{eq:gibbs_pure}
\end{equation}
where $\rho$ is the mass density and $1/\rho$ is the volume per unit mass.  We call the second term after the equality the ``volume integral term".  For an ideal gas, $\rho = P/{\cal R}T$ and one obtains equation (\ref{eq:gibbs}) and $\Psi = P/P_0$.  For non-ideal equations of state, it is not possible to write down a general closed-form expression for $\Psi$.  Instead, one defines a quantity known as the fugacity \citep{lewis1901},
\begin{equation}
f = \phi_i P,
\label{eq:fugacityCoef}
\end{equation}
where $\phi_i$ is the fugacity coefficient of the pure gas composed of species $i$.  Fugacity coefficients are specific to the chemical species being considered.  For an ideal gas, $\phi_i=1$.  

The generalisation of equation (\ref{eq:gibbs}) is done by \textit{analogy} with no underlying mathematical rigor.  Essentially, one does the ad hoc substitution (equation [29] of \citealt{lewis1901}),
\begin{equation}
\frac{P}{P_0} \rightarrow \frac{f}{f_0},
\end{equation}
where $f_0 = \phi_{i,0} P_0$ and $\phi_{i,0} \equiv \phi_i(P_0)$.

\subsubsection{Activity}

At a pressure $P$ and temperature $T$, the Gibbs free energy of a gas component $i$ in a gaseous mixture is (page 115 of \citealt{denbigh81}),
\begin{equation}
G_{\rm mix} = G + {\cal R} T \ln{X_i},
\end{equation}
where $G$ is given by equation (\ref{eq:gibbs_pure}) and $X_i$ is the volume mixing ratio (relative abundance by number) of the $i$-th gaseous species.  One now defines the activity as the generalisation of $X_i$,
\begin{equation}
a_i = \gamma_i X_i,
\label{eq:activityCoef}
\end{equation}
where $\gamma_i$ is the activity coefficient of the $i$-th gaseous species.  Activity coefficients are specific to the chemical species \textit{and} mixture of species being considered. For ideal mixing, $\gamma_i=1$.  

Again, by \textit{analogy}, the specific Gibbs free energy of the $i$-th gaseous species in a gas mixture with non-ideal mixing of its constituents is (page 287 of \citealt{denbigh81})
\begin{equation}
G_{\rm mix} = G + {\cal R} T \ln{a_i}.
\label{eq:gibbs_activity}
\end{equation}

\subsubsection{Equilibrium constant}

As we will see later in Sections \ref{subsect:pure_gas_system} and \ref{subsect:french1966}, the equilibrium constant $K_{\rm eq}$ is constructed from the $\Psi$ of the reactants and products in a chemical reaction.  It is related to the Gibbs free energy of the reaction $\Delta G_r$ by (page 352 of \citealt{devoe20})
\begin{equation}
K_{\rm eq} = e^{-\frac{\Delta G_r}{{\cal R} T}}.
\label{eq:keq}
\end{equation}

For gases, one may write
\begin{equation}
G_{\rm mix} = G_0 + {\cal R} T \ln{\left( \frac{a_i f}{f_0} \right)},
\end{equation}
such that
\begin{equation}
\Psi = \frac{a_i f}{f_0} = \frac{f_i}{P_0}.
\label{eq:psai-f}
\end{equation}
We have written the fugacity of the $i$-th gaseous species as $f_i = a_i f$. Since $P_0$ is typically chosen to be 1 bar, it is reasonable to set $\phi_{i,0}=1$ (a gas at 1 bar behaves like an ideal gas) and thus $f_0 = \phi_{i,0} P_0 = P_0$. Given that pure-gas fugacity $f$ is a generalisation of total pressure $P$ and that activity $a_i$ is a generalisation of volume mixing ratio $X_i$, the fugacity of the $i$-th gaseous species $f_i = a_i f$ can be deemed a generalisation of partial pressure. It is worth emphasizing that $\Psi$ is \textit{not} the activity.  Some confusion stems from the fact that the reduced expression of $\Psi = f_i/P_0$ is sometimes\footnote{\texttt{https://en.wikipedia.org/wiki/Thermodynamic\_activity}} referred to as the activity (see also page 287 of \citealt{denbigh81}).  This apparent ambiguity originates from the mathematical freedom to combine different terms into the argument of the natural logarithm.

%\textbf{[MT: In geochem literature, $f_i$ for a gas component in mixture typically includes both activity coefficient and fugacity coefficient, i.e., $f_i = a_i f = \gamma_i X_i \phi P = \gamma_i \phi P_i$. Thus $\Psi = f_i/P_0$]}

For a species in its solid phase, the fugacity is not explicitly stated and the Gibbs free energy is instead,
\begin{equation}
G = G^\prime + {\cal R} T \ln{a_s},
\label{eq:solidG}
\end{equation}
where the quantity $G^\prime$ depends on the pressure $P$ (and is not referenced to $P_0$, unlike for gases) and the activity associated with the solid is $a_s$.  We simply \textit{choose} $\Psi = a_s$ while paying attention to the definition of $G^\prime$.  The exact expression for $G^\prime$ is provided in Section \ref{sect:data}.  In the current study, we consider only the activity associated with graphite.

\subsection{Constructing the equilibrium constant from $\Psi$ in gaseous C-H-O system}
\label{subsect:pure_gas_system}

Consider the standard net chemical reaction for converting methane to carbon monoxide (e.g., \citealt{bs99,lodders02,henglyons}),
\begin{equation}
\mbox{CH}_4 + \mbox{H}_2\mbox{O} \leftrightarrows \mbox{CO} + 3 \mbox{H}_2.
\end{equation}
In the limit of an ideal gas ($\phi_i=1$) with ideal mixing ($\gamma_i=1$), we have $\Psi = P_i/P_0$.

We may write down the Gibbs free energies associated with the products and reactants,
\begin{equation}
\begin{split}
G_{\rm CH_4} &= G_{\rm CH_4,0} + {\cal R} T \ln{\Psi_{\rm CH_4}}, \\
G_{\rm H_2O} &= G_{\rm H_2O,0} + {\cal R} T \ln{\Psi_{\rm H_2O}}, \\
G_{\rm CO} &= G_{\rm CO,0} + {\cal R} T \ln{\Psi_{\rm CO}}, \\
G_{\rm H_2} &= G_{\rm H_2,0} + {\cal R} T \ln{\Psi_{\rm H_2}}. \\
\end{split}
\end{equation}
The differences in Gibbs free energies are
\begin{equation}
\begin{split}
\Delta G &= G_{\rm CO} + 3G_{\rm H_2} - G_{\rm CH_4} - G_{\rm H_2O}, \\
\Delta G_0 &= G_{\rm CO,0} + 3G_{\rm H_2,0} - G_{\rm CH_4,0} - G_{\rm H_2O,0}.
\end{split}
\end{equation}

It follows that
\begin{equation}
\Delta G - \Delta G_0 = {\cal R} T \ln{\left( \frac{\Psi_{\rm CO}\Psi_{\rm H_2}^3}{\Psi_{\rm CH_4}\Psi_{\rm H_2O}} \right)}.
\end{equation}
Following the reasoning by \cite{heng16}, the system attains chemical equilibrium by adjusting to $\Delta G=0$.  $\Delta G_0$ is interpreted as the change of Gibbs free energy of formation at the reference pressure $P_0$.  By comparing the preceding equation with equation (\ref{eq:keq}), the equilibrium constant is
\begin{equation}
K_{\rm eq} = \frac{\Psi_{\rm CO}\Psi_{\rm H_2}^3}{\Psi_{\rm CH_4}\Psi_{\rm H_2O}},
\end{equation}
in agreement with equations (6) and (26) of \cite{henglyons}.  The Gibbs free energy of the reaction is $\Delta G_r = \Delta G_0$.  Generally, the approach of constructing $K_{\rm eq}$ from $\Psi$ is consistent with equation (13) of \cite{heng16}.

\subsection{Constructing equilibrium constants in mixed-phase C-H-O system of \cite{french66}}
\label{subsect:french1966}

\cite{french66} considered the system of net chemical reactions,
\begin{equation}
\begin{split}
\mbox{C} + \mbox{O}_2 &\leftrightarrows \mbox{CO}_2, \\
\mbox{C} + 0.5 \mbox{O}_2 &\leftrightarrows \mbox{CO}, \\
\mbox{C} + 2 \mbox{H}_2 &\leftrightarrows \mbox{CH}_4, \\
\mbox{H}_2 + 0.5 \mbox{O}_2 &\leftrightarrows \mbox{H}_2\mbox{O}. \\
\end{split}
\label{eq:french1966}
\end{equation}
In this system, carbon is present in its solid phase as graphite.  \cite{french66} assumed ideal mixing ($\gamma_i=1$).

If we focus on the first net reaction, the Gibbs free energies are
\begin{equation}
\begin{split}
G_{\rm C} &= G^\prime_{\rm C} + {\cal R} T \ln{a_{\rm C}}, \\
G_{\rm O_2} &= G_{\rm O_2,0} + {\cal R} T \ln{\left(\frac{f_{\rm O_2}}{P_0}\right)}, \\
G_{\rm CO_2} &= G_{\rm CO_2,0} + {\cal R} T \ln{\left(\frac{f_{\rm CO_2}}{P_0}\right)}. \\
\end{split}
\end{equation}
The critical detail is that $G^\prime_{\rm C}(P,T)$ depends on the pressure $P$, whereas $G_{\rm O_2,0}(P_0)$ and $G_{\rm CO_2,0}(P_0)$ depend only on the reference pressure $P_0$.

The differences in Gibbs free energies is
\begin{equation}
\Delta G = G_{\rm CO_2} - G_{\rm O_2} - G_{\rm C}.
\end{equation}
If we again argue that $\Delta G=0$ when chemical equilibrium is attained \citep{heng16}, then the expression relating the equilibrium constant to the Gibbs free energy of the reaction follows,
\begin{equation}
K_{\rm eq,1} = \frac{f_{\rm CO_2}}{f_{\rm O_2} a_{\rm C}} = e^{-\frac{\Delta G_r}{{\cal R} T}},
\end{equation}
where the Gibbs free energy of the reaction is
\begin{equation}
\label{eq:examplerxnG}
\Delta G_r = G_{\rm CO_2,0} - G_{\rm O_2,0} - G^\prime_{\rm C}.
\end{equation}
In addition to the usual Gibbs free energies of formation associated with the gaseous species, which are defined at the reference pressure $P_0$, one considers the Gibbs free energies of formation associated with graphite, which is defined at the pressure of interest $P$.  If we further set $a_{\rm C}=1$, equation (1) of \cite{french66} is recovered.

For the other three net chemical reactions in equation (\ref{eq:french1966}), the equilibrium constants are
\begin{equation}
\label{eq:Gibbs-fugacity}
\begin{split}
K_{\rm eq,2} &= \frac{f_{\rm CO}}{\left( f_{\rm O_2} P_0 \right)^{1/2}a_{\rm C}}, \\
K_{\rm eq,3} &= \frac{f_{\rm CH_4} P_0}{f_{\rm H_2}^2a_{\rm C}}, \\
K_{\rm eq,4} &= \frac{f_{\rm H_2O} P_0^{1/2}}{f_{\rm H_2} f_{\rm O_2}^{1/2}}. \\
\end{split}
\end{equation}
If one again sets $a_{\rm C}=1$ and omits writing the factors of $P_0$ explicitly (by setting them to unity), then one recovers equations (2) to (4) of \cite{french66}.

\subsection{Review: oxygen and sulfur fugacities}
\label{subsect:fugacities}

The oxygen fugacity ($f_{\rm O_2}$) quantifies how reduced or oxidised the mantle of an exoplanet is.  It may be interpreted as the equivalent partial pressure of oxygen even if the oxygen is locked up in solids in the form of minerals.   A representative reaction for iron in the core of the Earth and iron oxide (w\"{u}stite) in its mantle is \citep{ww05}
\begin{equation} \label{eq:IW-rxn}
\mbox{Fe} + 0.5 \mbox{O}_2 \leftrightarrows \mbox{FeO}.
\end{equation}
If factors of $P_0$ are ignored, then the oxygen fugacity buffered by this reaction has the following equilibrium constant,
\begin{equation}
K_{\rm eq} = \frac{a_{\rm FeO}}{a_{\rm Fe} f^{1/2}_{\rm O_2}}.
\end{equation}
By using equation (\ref{eq:keq}) and performing a change of base of the logarithm (from base 2 to 10), we obtain
\begin{equation}
\log f_{\rm O_2} = \frac{2}{\ln{10}} \frac{\Delta G_r}{{\cal R} T} + 2 \log{\left( \frac{a_{\rm FeO}}{a_{\rm Fe}} \right)}.
\label{eq:fO2_log10}
\end{equation}

An idealisation known as the iron-w\"{u}stite (IW) buffer is defined when iron and w\"{u}stite occur in their pure forms, such that their activities may be set to unity \citep{ww05},
\begin{equation}
\mbox{IW} \equiv \log{f^{\rm IW}_{\rm O_2}} = \frac{2}{\ln{10}} \frac{\Delta G_r}{{\cal R} T},
\label{eq:IW}
\end{equation}
where we follow the convention in the geochemical literature of abbreviating $\log{f^{\rm IW}_{\rm O_2}}$ as ``IW".  The oxygen fugacity may span 50 orders of magnitude in value \citep{frost91} and it is both inconvenient and non-intuitive to quote its absolute value (in bar).  Rather, geochemists favor quoting the oxygen fugacity relative to simple\footnote{Buffers are simple in the sense that the activities of the solid phases involved are always unity.} buffers such as IW.  From a computational standpoint, IW is a convenient reference point that allows $\log{f_{\rm O_2}}$ to be reported in terms of intuitive, order-of-unity numbers.

It is possible to use equation (\ref{eq:fO2_log10}) to estimate the oxygen fugacity associated with the core-mantle boundary or core formation of the Earth.  By interpreting the activities as relative abundances by number, we may approximate $a_{\rm FeO}\approx 0.08$ (mantle) and $a_{\rm Fe} \approx 0.8$ (core) \citep{ww05} such that $\log{(a_{\rm FeO}/a_{\rm Fe})=-1}$.  Equation (\ref{eq:fO2_log10}) then yields $\log f_{\rm O_2} = 2 \Delta G_r / {\cal R} T \ln{10} - 2$.  Using equation (\ref{eq:IW}), we obtain $\log f_{\rm O_2} = \mbox{IW}-2$ for the core-mantle boundary of the Earth or the mantle during Earth's core formation \citep{ww05}.  Physically, the iron oxide in the mantle and the iron in the core jointly \textit{buffer} the oxygen fugacity.  It is worth noting that the oxygen fugacity of the Earth varies both temporally and spatially.  For modern Earth, the upper mantle has an estimated oxygen fugacity of $\mbox{IW}+3$ to $\mbox{IW}+7$ \citep{fm08}. Another buffer that is commonly used is the FMQ buffer, where F, M and Q represent fayalite (Fe$_2$SiO$_4$), magnetite (Fe$_3$O$_4$) and quartz (SiO$_2$), respectively \citep{frost91}.  $\mbox{IW}+5$ converts to about FMQ \citep{frost91, fm08}.  Thus, even for modern Earth the oxygen fugacity varies by about 9 orders of magnitude from the core-mantle boundary to the upper mantle.  Temporally, it is believed that the oxygen fugacity of the upper mantle increased to its current value by at least 3.6 billion years ago \citep{delano01}.

Since $\Delta G_r$ generally depends on temperature and pressure, it is unsurprising that IW has the same dependences.  To obtain an absolute value for IW, \cite{ballhaus91} provides a convenient, empirical fitting function for IW based on the experimental data from \citet{oneill87},
\begin{equation}
\begin{split}
\mbox{IW} =& 14.07 - 28784 \left(\frac{T}{1 \mbox{ K}}\right)^{-1} - 2.04\log{\left(\frac{T}{1 \mbox{ K}}\right)} \\
& + 0.053 \left(\frac{P}{1 \mbox{ bar}}\right) \left(\frac{T}{1 \mbox{ K}}\right)^{-1} + 3\times 10^{-6} \left(\frac{P}{1 \mbox{ bar}}\right),
\end{split}
\label{eq:IW-fo2}
\end{equation}
which is valid for $900 \le T/\mbox{K} \le 1420$. The pressure dependence is derived by volume integration.  It is understood that IW, as represented above, has physical units of the logarithm (base 10) of pressure in bar.

Analogously, a sulfur fugacity buffer may be written down based on pyrrhotite (FeS) and pyrite (FeS$_2$) \citep{fb87}, 
\begin{equation} \label{eq:PP-rxn}
\mbox{FeS} + 0.5 \mbox{S}_2 \leftrightarrows \mbox{FeS}_2,
\end{equation}
which we abbreviate as ``PP" for convenience.  \citet{fr76} provides an empirical fitting function based on the experimental data from \citet{tb64},
\begin{equation}
\begin{split}
{\mbox{PP}} \equiv \log f_{\rm S_2}^{\rm PP} =& - 16073 \left(\frac{T}{1 \mbox{ K}}\right)^{-1} + 15.74 \\
&+ 0.06 \left[ \left(\frac{P}{1 \mbox{ bar}}\right) - 1 \right] \left(\frac{T}{1 \mbox{ K}}\right)^{-1},
\end{split}
\label{eq:PP-fs2}
\end{equation}
which is valid for $598 \le T/\mbox{K} \le 1016$.  The pressure dependence is derived by volume integration.  It is again understood that PP, as represented above, has physical units of the logarithm (base 10) of pressure in bar.

Compared to the oxygen fugacity, it is worth noting that data on the sulfur fugacity in different geological settings are sparse, although estimates span an impressive range of values from $\log{f_{\rm S_2}} \approx -20$ ($\mbox{PP}-8$) in submarine hydrothermal vents \citep{keith14}, $\log{f_{\rm S_2}} \approx -5$ to $-2$ (${\rm PP}-3$ to ${\rm PP}$) associated with metamorphic degassing \citep{po89} to $\log{f_{\rm S_2}} \approx 4$ ($\mbox{PP}+1$) at high pressures ($\sim 10^4$ bar) associated with the sulfide-saturation conditions of basaltic melts \citep[][sample G280]{sd14}.

The sulfur and oxygen fugacities may be related if the FeO content of the melt is known \citep{bockrath04}, a caveat that is also noted in Section 3.1 of \cite{liggins20}.  Since we do not explicitly model the melt composition, this additional layer of complexity is left to future work and we regard $f_{\rm S_2}$ and $f_{\rm O_2}$ as independent input parameters. 

\subsection{Generalised model for mixed-phase C-H-O-N-S system}

\subsubsection{Net chemical reactions}

Using the \cite{french66} model as a baseline, we consider the following, expanded set of net chemical reactions \citep{holloway81,bs99,lodders02,moses11,gs14},
\begin{equation}
\begin{split}
\mbox{C} + \mbox{O}_2 &\leftrightarrows \mbox{CO}_2, \\
\mbox{C} + 0.5 \mbox{O}_2 &\leftrightarrows \mbox{CO}, \\
\mbox{C} + 2 \mbox{H}_2 &\leftrightarrows \mbox{CH}_4, \\
\mbox{H}_2 + 0.5 \mbox{O}_2 &\leftrightarrows \mbox{H}_2\mbox{O}, \\
\mbox{N}_2 + 3 \mbox{H}_2 &\leftrightarrows 2 \mbox{NH}_3, \\
\mbox{NH}_3 + \mbox{CH}_4 &\leftrightarrows \mbox{HCN} + 3\mbox{H}_2, \\
0.5 \mbox{S}_2 + \mbox{O}_2 &\leftrightarrows \mbox{SO}_2, \\
0.5 \mbox{S}_2 + \mbox{H}_2\mbox{O} &\leftrightarrows \mbox{H}_2\mbox{S} + 0.5 \mbox{O}_2. \\
\end{split}
\label{eq:reactions}
\end{equation}
Similar to the oxygen fugacity, we treat the sulfur fugacity $f_{\rm S_2}$ as an input parameter \citep[e.g.,][]{holloway81}.  The nitrogen content of the atmosphere is parametrised via the partial pressure of nitrogen $P_{\rm N_2}$.  Assuming these two additional input parameters allows the number of unknowns and equations to be equal.

\subsubsection{Equilibrium constants}

Modern thermodynamic data allow us to consider non-ideal mixing ($\gamma_i \ne 1$) and non-ideal gas behavior ($\phi_i \ne 1$), unlike for \cite{french66}.  For compactness of notation, we define
\begin{equation}
\alpha_i \equiv \gamma_i \phi_i.
\end{equation}
This motivates us to write down more generalised expressions for the equilibrium constants, 
\begin{equation}
\begin{split}
K_{\rm eq,1} &= \frac{\alpha_{\rm CO_2} P_{\rm CO_2}}{\gamma_{\rm O_2} f_{\rm O_2} a_{\rm C}}, \\
K_{\rm eq,2} &= \frac{\alpha_{\rm CO} P_{\rm CO}}{\left( \gamma_{\rm O_2} f_{\rm O_2} P_0 \right)^{1/2}a_{\rm C}}, \\
K_{\rm eq,3} &= \frac{\alpha_{\rm CH_4} P_{\rm CH_4} P_0}{\left( \alpha_{\rm H_2} P_{\rm H_2}\right)^2 a_{\rm C}}, \\
K_{\rm eq,4} &= \frac{\alpha_{\rm H_2O} P_{\rm H_2O} P_0^{1/2}}{\alpha_{\rm H_2} P_{\rm H_2} \left( \gamma_{\rm O_2}f_{\rm O_2}\right)^{1/2}}. \\
K_{\rm eq,5} &= \frac{\left( \alpha_{\rm NH_3} P_{\rm NH_3} P_0 \right)^2}{\alpha_{\rm N_2} P_{\rm N_2} \left( \alpha_{\rm H_2} P_{\rm H_2} \right)^3 }, \\
K_{\rm eq,6} &= \frac{\alpha_{\rm HCN} P_{\rm HCN} \left( \alpha_{\rm H_2} P_{\rm H_2} \right)^3 }{\alpha_{\rm NH_3} \alpha_{\rm CH_4} P_{\rm NH_3} P_{\rm CH_4} P^2_0}, \\
K_{\rm eq,7} &= \frac{\alpha_{\rm SO_2} P_{\rm SO_2} P^{1/2}_0}{\gamma_{\rm O_2} f_{\rm O_2} \left( \gamma_{\rm S_2} f_{\rm S_2} \right)^{1/2} }, \\
K_{\rm eq,8} &= \frac{\alpha_{\rm H_2S} P_{\rm H_2S} \left( \gamma_{\rm O_2} f_{\rm O_2} \right)^{1/2}}{\alpha_{\rm H_2O} P_{\rm H_2O} \left( \gamma_{\rm S_2} f_{\rm S_2} \right)^{1/2}}. \\
\end{split}
\label{eq:all-K}
\end{equation}
We note the convention in the geochemical literature (the use of equation [\ref{eq:psai-f}]; e.g., \citealt{fb87}) is to absorb the activity coefficient ($\gamma_i$) into the fugacity ($f_i$). By departing from this practice, we separate the effects of non-ideal mixing (expressed through $\gamma_i$) from non-ideal-gas equations of state (expressed through $\phi_i$) by explicitly writing $f_i = \phi_i P_i$.
%\begin{equation}
%\begin{split}
%K_{\rm eq} &= \frac{\alpha_{\rm CO_2} P_{\rm CO_2}}{f_{\rm O_2} a_{\rm C}}, \\
%K_{\rm eq,2} &= \frac{\alpha_{\rm CO} P_{\rm CO}}{\left( f_{\rm O_2} P_0 \right)^{1/2}a_{\rm C}}, \\
%K_{\rm eq,3} &= \frac{\alpha_{\rm CH_4} P_{\rm CH_4} P_0}{\left( \alpha_{\rm H_2} P_{\rm H_2}\right)^2 a_{\rm C}}, \\
%K_{\rm eq,4} &= \frac{\alpha_{\rm H_2O} P_{\rm H_2O} P_0^{1/2}}{\alpha_{\rm H_2} P_{\rm H_2} \left( f_{\rm O_2}\right)^{1/2}}. \\
%K_{\rm eq,5} &= \frac{\left( \alpha_{\rm NH_3} P_{\rm NH_3} P_0 \right)^2}{\alpha_{\rm N_2} P_{\rm N_2} \left( \alpha_{\rm H_2} P_{\rm H_2} \right)^3 }, \\
%K_{\rm eq,6} &= \frac{\alpha_{\rm HCN} P_{\rm HCN} \left( \alpha_{\rm H_2} P_{\rm H_2} \right)^3 }{\alpha_{\rm NH_3} \alpha_{\rm CH_4} P_{\rm NH_3} P_{\rm CH_4} P^2_0}, \\
%K_{\rm eq,7} &= \frac{\alpha_{\rm SO_2} P_{\rm SO_2} P^{1/2}_0}{f_{\rm O_2} \left( f_{\rm S_2} \right)^{1/2} }, \\
%K_{\rm eq,8} &= \frac{\alpha_{\rm H_2S} P_{\rm H_2S} \left( f_{\rm O_2} \right)^{1/2}}{\alpha_{\rm H_2O} P_{\rm H_2O} \left( f_{\rm S_2} \right)^{1/2}}. \\
%\end{split}
%\end{equation}

\subsubsection{Partial and total pressures}

By rewriting the expressions for the equilibrium constants, one may write down expressions for the partial pressures of the various species,

%\textbf{\color{red} [MT: $P_0$ in the expression for $P_{\rm SO_2}$ should be replaced with $P_0^{\frac{1}{2}}$.]}
\begin{equation}
\begin{split}
P_{\rm CO_2} &= \frac{a_{\rm C} K_{\rm eq,1} \gamma_{\rm O_2} f_{\rm O_2} }{\alpha_{\rm CO_2}}, \\
P_{\rm CO} &= \frac{a_{\rm C} K_{\rm eq,2} \left( \gamma_{\rm O_2} f_{\rm O_2} P_0 \right)^{1/2} }{\alpha_{\rm CO}}, \\
P_{\rm CH_4} &= \frac{a_{\rm C} K_{\rm eq,3} \left( \alpha_{\rm H_2} P_{\rm H_2} \right)^2}{\alpha_{\rm CH_4} P_0}, \\
P_{\rm H_2O} &= \frac{K_{\rm eq,4} \alpha_{\rm H_2} P_{\rm H_2} \left( \gamma_{\rm O_2} f_{\rm O_2} \right)^{1/2}}{\alpha_{\rm H_2O} P^{1/2}_0}, \\
P_{\rm NH_3} &= \frac{K^{1/2}_{\rm eq,5} \left( \alpha_{\rm N_2} P_{\rm N_2} \right)^{1/2} \left( \alpha_{\rm H_2} P_{\rm H_2} \right)^{3/2}}{\alpha_{\rm NH_3} P_0}, \\
P_{\rm HCN} &= \frac{K_{\rm eq,6} \alpha_{\rm NH_3} \alpha_{\rm CH_4} P_{\rm NH_3} P_{\rm CH_4} P^2_0}{\alpha_{\rm HCN} \left( \alpha_{\rm H_2} P_{\rm H_2} \right)^3}, \\
P_{\rm SO_2} &= \frac{K_{\rm eq,7} \gamma_{\rm O_2} f_{\rm O_2} \left( \gamma_{\rm S_2} f_{\rm S_2} \right)^{1/2}}{\alpha_{\rm SO_2} P_0^{1/2}}, \\
P_{\rm H_2S} &= \frac{K_{\rm eq,8} \alpha_{\rm H_2O} P_{\rm H_2O} \left( \gamma_{\rm S_2} f_{\rm S_2} \right)^{1/2}}{\alpha_{\rm H_2S} \left( \gamma_{\rm O_2} f_{\rm O_2} \right)^{1/2} }, \\
\end{split}
\label{eq:partial_pressures}
\end{equation}
%\begin{equation}
%\begin{split}
%P_{\rm CO_2} &= \frac{a_{\rm C} K_1 f_{\rm O_2} }{\alpha_{\rm CO_2}}, \\
%P_{\rm CO} &= \frac{a_{\rm C} K_2 \left( f_{\rm O_2} P_0 \right)^{1/2} }{\alpha_{\rm CO}}, \\
%P_{\rm CH_4} &= \frac{a_{\rm C} K_3 \left( \alpha_{\rm H_2} P_{\rm H_2} \right)^2}{\alpha_{\rm CH_4} P_0}, \\
%P_{\rm H_2O} &= \frac{K_4 \alpha_{\rm H_2} P_{\rm H_2} \left( f_{\rm O_2} \right)^{1/2}}{\alpha_{\rm H_2O} P^{1/2}_0}, \\
%P_{\rm NH_3} &= \frac{K^{1/2}_5 \left( \alpha_{\rm N_2} P_{\rm N_2} \right)^{1/2} \left( \alpha_{\rm H_2} P_{\rm H_2} \right)^{3/2}}{\alpha_{\rm NH_3} P_0}, \\
%P_{\rm HCN} &= \frac{K_6 \alpha_{\rm NH_3} \alpha_{\rm CH_4} P_{\rm NH_3} P_{\rm CH_4} P^2_0}{\alpha_{\rm HCN} \left( \alpha_{\rm H_2} P_{\rm H_2} \right)^3}, \\
%P_{\rm SO_2} &= \frac{K_7 f_{\rm O_2} \left( f_{\rm S_2} \right)^{1/2}}{\alpha_{\rm SO_2} P_0^{1/2}}, \\
%P_{\rm H_2S} &= \frac{K_8 \alpha_{\rm H_2O} P_{\rm H_2O} \left( f_{\rm S_2} \right)^{1/2}}{\alpha_{\rm H_2S} \left( f_{\rm O_2} \right)^{1/2} }, \\
%\end{split}
%\label{eq:partial_pressures}
%\end{equation}
which depend on the parameters $f_{\rm O_2}$, $f_{\rm S_2}$ and $P_{\rm N_2}$.  We note that $P_{\rm HCN} \propto P_{\rm H_2}^{1/2}$.

The total pressure is given by
\begin{equation}
\begin{split}
P =& P_{\rm H_2} + P_{\rm He} + P_{\rm O_2} + P_{\rm S_2} + P_{\rm N_2} + P_{\rm CO_2} + P_{\rm CO} \\
&+ P_{\rm CH_4} + P_{\rm H_2O} + P_{\rm NH_3} + P_{\rm HCN} + P_{\rm SO_2} + P_{\rm H_2S}. \\
\end{split}
\label{eq:total_pressure}
\end{equation}

\subsubsection{Solution method}
\label{sec:solmethod}

The solution method depends on whether one is solving for a secondary or hybrid atmosphere.  Following \cite{french66}, if one considers a secondary atmosphere then the partial pressure of molecular hydrogen ($P_{\rm H_2}$) is a quantity that one solves for.  In the original C-H-O system considered by \cite{french66}, one solves a quadratic equation for $P_{\rm H_2}$.  When we include the additional molecular species beyond what \cite{french66} considered, we need to numerically solve equation (\ref{eq:total_pressure}), which provides a quartic equation in $x$ where $x \equiv P_{\rm H_2}^{1/2}$.  The input parameters are $T$, $P$, $f_{\rm O_2}$, $f_{\rm S_2}$ and $P_{\rm N_2}$.

For a hybrid atmosphere, one imagines a primordial envelope of molecular hydrogen that is the consequence of the formation and evolutionary history of the exoplanet.  In the absence of such a robust theory of formation and evolution, we \textit{prescribe} the value of $P_{\rm H_2}$ but solve for $P$.  The input parameters are $T$, $P_{\rm H_2}$, $f_{\rm O_2}$, $f_{\rm S_2}$ and $P_{\rm N_2}$.  This is done by iteration, where one makes a first guess for $P$ and updates the contributions to it and the equilibrium constants (which depend on $P$) until convergence is attained.  Physically, one is solving for a system in which one has a mantle with fixed oxygen and sulfur fugacities interacting with a primordial envelope of hydrogen and helium. %{\color{blue} [On a related technical note: if we separate out $\gamma_{\rm O_2}$ from $f_{\rm O_2}$ and $\gamma_{\rm S_2}$ from $f_{\rm S_2}$, as in equations (32) and (33), then our calculations are equivalent to prescribing the products $\gamma_{\rm O_2} f_{\rm O_2}$ and $\gamma_{\rm S_2} f_{\rm S_2}$ as input parameters, or equivalent to prescribing $f_{\rm O_2}$ and $f_{\rm S_2}$ but setting $\gamma_{\rm O_2}=\gamma_{\rm S_2}=1$.]} 

Our implementation is as follows:
\begin{itemize}

\item Choose values for $T$ and $P$ (secondary atmospheres) or $P_{\rm H_2}$ (hybrid atmospheres).

\item Using equation (\ref{eq:fO2_log10}), calculate $f_{\rm O_2}$. By definition, $\gamma_{\rm O_2}=1$ in this buffer reaction.

\item Similarly, calculate $f_{\rm S_2}$ using equation (\ref{eq:PP-fs2}).  Again, $\gamma_{\rm S_2}=1$ by definition for this buffer reaction.

\item For all species other than O$_2$ and S$_2$, calculate $\gamma_i$ based on guessed or iterated atmospheric compositions.  Since data are unavailable for N- and S-bearing molecules, we set $\phi_i=\gamma_i=1$ for these species (see discussion in Section \ref{sect:gamma}). Iterate between $\gamma_i$ and atmosphere composition (partial pressures $P_i$ ) until convergence attains.

\end{itemize}

The partial pressure $P_{\rm He}$ is an additional input parameter that considers the presence of helium as an inert gas contributing to the atmospheric pressure.  When solving for a secondary atmosphere, $P$ is replaced by $P-P_{\rm He}$, which alters the solution for $P_{\rm H_2}$.  When solving for a hybrid atmosphere, the assumed relative content of hydrogen versus helium affects the partial pressures of various gases directly.  The inclusion of helium is motivated by the work of \cite{hu15}, who postulated that atmospheric escape over $\sim 0.1$ Gyr timescales may evolve primordial H$_2$-He atmospheres to He-dominated ones.  However, as the parameter space is already very broad, its influence and implications are deferred to a future study and we set $P_{\rm He}=0$ throughout this study.  

The partial pressure of molecular nitrogen $P_{\rm N_2}$ plays a similar role with the key difference that it participates in the net chemical reactions, while helium does not.

It is clear from the above solution procedure that our outgassing model is zero-dimensional, meaning that it does not resolve the respective temperature and pressure gradients in the atmosphere and the melt-producing interior, and hence is a geochemical ``box" model. Moreover, it's noteworthy that for the ``box" of melt-bearing interior, gas dissolution into the melts is neglected, which differs from previous studies \citep{gs14, ortenzi20, bower22, gaillard22}. We intentionally avoid considering gas solubilities in this study for two reasons. First, consideration of gas solubility requires prescription of the bulk volatile budget of an exoplanet (in absolute mass terms), which is subject to large uncertainties. This is also the reason why \cite{ortenzi20}, \cite{bower22} and \cite{gaillard22} explored a range of initial volatile inventories. In other words, including gas solubility into the calculations in turn introduces more free parameters that are difficult to assign values to. Second, the method ignoring gas solubility and thus without volatile budget constraints corresponds to global Gibbs energy minimization, whereas the method that includes volatile budgets corresponds to ``constrained" Gibbs energy minimization; the difference is that between phase diagram and pseudosection as elaborated in \cite{powell98}. Within the hierarchical modeling approach \citep{held05}, our model is one step simpler than models that include gas dissolution into melts. Nonetheless, in Appendix \ref{append:solubility}, we provide an example using the hydrogen-oxygen (H-O) subsystem to illustrate how gas dissolution into melt can be incorporated into the current modeling framework. 

\subsection{Relevant range of temperatures and pressures for outgassing}
\label{sec:ptrange}

The input pressure $P$ is interpreted as the surface pressure of the atmosphere \citep{gs14}.  We consider $P=1$ mbar to $10^4$ bar.  The lower limit is arbitrary and empirically inspired by the Martian atmosphere.  The upper limit is motivated by the estimate that a hydrogen-dominated atmosphere with $\sim 1\%$ of the mass of the exoplanet is massive enough to double the radius of its core \citep{owen19}.  Denoting its radius by $R$ and its surface gravity by $g$, the mass of the atmosphere of an exoplanet is $M_{\rm atm}=4 \pi R^2 P/g$.  It follows that the ratio of the atmospheric mass to the total mass of the exoplanet is
\begin{equation}
\frac{M_{\rm atm}}{M} = 0.026 ~\left( \frac{P}{10^4 \mbox{ bar}} \right) \left( \frac{R}{1.6 ~R_\oplus} \frac{\rho}{2.0 \mbox{ g cm}^{-3}} \right)^{-2}.
\end{equation}
It is thus not unreasonable to regard sub-Neptunes as exoplanets with surface pressures $\sim 10^4$ bar.

The input temperature $T$ is interpreted as that of the melt from which the secondary atmospheres are outgassed.  We use the term ``melt" to refer to liquids at both low and high temperatures; the term ``magma" refers to a high-temperature melt.  Its range of values is bracketed by two plausible melting scenarios associated with rocky exoplanets.  The first scenario is a fully molten magma ocean \citep{hirschmann12, kg19, sossi20, gaillard22, bower22}.  The second scenario considers volcanic outgassing from a largely solidified body \citep{sr93, gs14, liggins20, honing21}.  The lower temperature bound that we adopt is $T=873$ K, which corresponds roughly to the minimum partial melting temperature (solidus) of wet granite under the pressure range of 1 mbar to $10^4$ bar \citep{hw73}.  To fully melt a ``dry" peridotite, the temperature (liquidus) is about $1973$ K at a pressure of 1 bar.  Using Figure 7a of \cite{takahashi93}, we determine that this temperature value will not change much at 1 mbar and scale its value to about $2073$ K at $10^4$ bar.  It is known that volatiles in a melt can reduce the liquidus by about $180$--$300$ K \citep{hort98}, which motivates us to set the upper temperature bound for melts to $1873$ K to accommodate the scenario of a planetary-scale magma ocean.  Temperatures higher than 1873 K will likely involve non-negligible silicate vaporization (e.g., \citealt{sf03, schaefer12, herbort20}), which is beyond the scope of this study.

Motivated by the possibility that melt compositions may be different from peridotitic (e.g., basalt), \cite{gaillard22} chose $T=1773$ K.  The study of \cite{gs14} chose $T=1573$ K, which we adopt and approximate as 1600 K to emphasise the somewhat ad hoc choice of melt temperature in the absence of a full interior-mantle model.  In plots where we have to fix the temperature and vary other parameters, we adopt $T=1600$ K, an intermediate value between the lower (873 K) and upper (1873 K) bounds of melt temperatures.

\section{Thermodynamic Data}
\label{sect:data}

\begin{table*}
\begin{center}
\caption{Fitting Coefficients for Computing Heat Capacity and Gibbs Free Energy}
\label{tab:gibbs}
%\resizebox{\textwidth}{!}{
\begin{tabular}{lcccccccccccc}
\hline
\hline
%$T$ (K) & CO$_2$ & CO & CH$_4$ & H$_2$O & NH$_3$ & HCN & SO$_2$ & H$_2$S & H$_2$ & He & N$_2$ & C$^\dagger$ \\
Quantity & $H_{i,0}$ & $S_{i,0}$ & $A_{1,i}$ & $A_{2,i}$ & $A_{3,i}$ & $A_{4,i}$ \\ 
Physical Units & J mol$^{-1}$ & J K$^{-1}$ mol$^{-1}$ & J K$^{-1}$ mol$^{-1}$ & J K$^{-2}$ mol$^{-1}$ & J K mol$^{-1}$ & J K$^{1/2}$ mol$^{-1}$ \\
\hline
 CO$_2$ & -393510 & 213.7 & 87.8 & -0.002644 & 706400 & -998.9 \\
 CO & -110530 & 197.67 & 45.7 & -0.000097 & 662700 & -414.7\\
 CH$_4$ & -74810 & 186.26 & 150.1 & 0.002062 & 3427700 & -2650.4 \\
 H$_2$O & -241810 & 188.8 & 40.1 & 0.008656 & 487500 & -251.2 \\
 NH$_3$ & -45898 & 192.774 & 101.6 & -0.000281 & 656995 & -1316.9 \\
 HCN & 135143 & 201.828 & 74.86 & -0.000203 & 295205 & -745.8 \\
 SO$_2$ & -296842 & 248.212 & 72.48 & -0.000642 & 191992 & -581.9 \\
 H$_2$S & -20300 & 205.77 & 47.4 & 0.01024 & 615900 & -397.8 \\
 H$_2$ & 0 & 130.7 & 23.3 & 0.004627 & 0 & 76.3 & & \\
 S$_2$ & 128540 & 231 & 37.1 & 0.002398 & -161000 & -65 \\
 N$_2$ & 0 & 191.609 & 40.14 & 0.000217 & 119849 & -224.7 \\
 C (Graphite)  & 0 & 5.85 & 51.0 & -0.004428 & 488600 & -805.5 \\
\hline
\hline
\end{tabular}%}\\
%\vspace{0.1in}
\end{center}
Note: $H_{i,0}$ and $S_{i,0}$ are stated for $T_0 = 298.15$ K and $P_0=1$ bar. For SO$_2$, N$_2$, NH$_3$ and HCN, $H_{i,0}$ and $S_{i,0}$ data are directly taken from the JANAF database, and $A_{j,i}$ ($j=1,2,3,4$) are from regression of isobaric (at 1 bar) heat capacity data from the JANAF database. All other data from \cite{hp98}.
\vspace{0.1in}
\end{table*}

In the current section, we describe how the heat capacity at constant pressure, entropy, enthalpy of formation, equation of state, fugacity coefficient and activity coefficient are calculated using established databases.  However, this requires the elucidation of additional formalism concerning the computation of the Gibbs free energy in order to place all of these different ingredients in context.

\subsection{Preamble}

For both gaseous and solid phases, the Gibbs free energy is given by equation (A1) of \cite{powell78},
\begin{equation}
G_i =  G_{i,0} + \int_{P_0}^P \left.\frac{\partial G_i}{\partial P} \right\vert_T ~dP,
\end{equation}
where $G_i = G_i(T,P)$ and $G_{i,0} \equiv  G_i(T,P_0)$.  For convenience, we write the integrand as $V_i$.  The Gibbs free energy at the reference pressure $P_0$ may be further expanded as $G_{i,0} =  H_i(T,P_0) - T S_i(T,P_0)$, where $H_i(T,P_0)$ and $S_i (T,P_0)$ are the enthalpy and entropy at the reference pressure, respectively.  If we further expand $H_i(T,P_0)$ and $S_i (T,P_0)$ about a reference temperature $T_0$, then we obtain
\begin{equation}
\begin{split}
G_{i,0} &= H_{i,0} - T S_{i,0} + \int^T_{T_0} \frac{\partial H_i(T,P_0)}{\partial T} ~dT \\
&- T \int^T_{T_0} \frac{\partial S_i(T,P_0)}{\partial T} ~dT ,
\end{split}
\end{equation}
where we have defined $H_{i,0} \equiv H_i(T_0, P_0)$ and $S_{i,0} \equiv S_i(T_0, P_0)$ for convenience.  We note that
\begin{equation}
\begin{split}
\frac{\partial H_i(T,P_0)}{\partial T} &= C_{P,i}(T,P_0), \\
\frac{\partial S_i(T,P_0)}{\partial T} &= \frac{C_{P,i}(T,P_0)}{T},
\end{split}
\end{equation}
where $C_{P,i}(T,P_0)$ is the isobaric heat capacity of species $i$ at the reference pressure.  It follows that the full expression for the Gibbs free energy is \citep{powell78}
\begin{equation}
\begin{split}
G_i =& H_{i,0} - T S_{i,0} + \int_{T_0}^T C_{P, i}(T,P_0) ~dT \\
&- T \int_{T_0}^T \frac{C_{P, i}(T,P_0)}{T} ~dT+\int_{P_0}^P V_i ~dP.
\end{split}
\label{eq:Gibbs_full}
\end{equation}

In the evaluation of net chemical reactions, it is not the absolute energy levels that are relevant.  Rather, it is the difference in energies between the reactants and the products that inform the equilibrium constants.  It is analogous to how potential energies are always defined relative to a reference value.  In \cite{heng16} and \cite{henglyons}, the \textit{change} in $G_{i,0}$ is interpreted as the Gibbs free energy of formation\footnote{Publicly available from the JANAF database at \texttt{https://janaf.nist.gov/}.}.  In other words, the reference level of $G_{i,0}$ is irrelevant because we are interested only in its difference between the reactants and products of the net chemical reaction.  Similarly, since the reference level is irrelevant we are free to interpret $H_{i,0}$ as the enthalpy of formation (e.g., Appendix A of \citealt{powell78}), as long as we are cognizant of the fact that we ultimately wish to compute relative, rather than absolute, energies.  In the current study, our approach is to use $H_{i,0}$ and $S_{i,0}$ to compute $G_i$ using equation (\ref{eq:Gibbs_full}).  We consider this approach to be more general as it is applicable beyond gaseous phases and explicitly allows for non-ideal behavior to be computed.

\subsection{Heat capacity at constant pressure}

To cast equation (\ref{eq:Gibbs_full}) in a more useful form, we need an expression for the heat capacity at constant pressure.  Based on experimental data, the study of \cite{hp98} provides empirical fitting functions,
\begin{equation}
 C_{P, i} \left(T,P_0 \right) = A_{1,i} + A_{2,i} T + A_{3,i} T^{-2} + A_{4,i} T^{-1/2},
\label{eq:cp_fit}
\end{equation}
which are valid for $T \le 2273$ K.  Table \ref{tab:gibbs} summarises the fitting coefficients ($A_{j,i}$ where $j=1,2,3,4$) for the gases considered in the current study, as well as for graphite. For H$_2$, O$_2$, CO, CO$_2$, H$_2$O, CH$_4$, H$_2$S and S$_2$, all the thermodynamic data are from the extended dataset by \cite{hp98, hp11}. For other S- and N-bearing species, i.e., SO$_2$, N$_2$, NH$_3$ and HCN, we derive the fitting coefficients by performing regression to the isobaric heat capacity data at 1 bar from the JANAF database; the enthalpy and entropy data are directly from the JANAF database.

By substituting equation (\ref{eq:cp_fit}) into (\ref{eq:Gibbs_full}), we obtain
\begin{equation}
\begin{split}
G_i =& H_{i,0} - T S_{i,0} \\
&+ A_{1,i} (T - T_0) + \frac{A_{2,i}}{2} \left(T^2 - T^2_0\right) \\
&- A_{3,i} \left(\frac{1}{T} - \frac{1}{T_0} \right) + 2 A_{4,i} \left( \sqrt{T}-\sqrt{T_0}\right) \\
&- A_{1,i} T \ln{\left( \frac{T}{T_0} \right)} - A_{2,i} T \left( T - T_0 \right) \\
&+ \frac{A_{3,i} T}{2} \left(\frac{1}{T^2} - \frac{1}{T^2_0} \right) + 2 A_{4,i} T \left( \frac{1}{\sqrt{T}} - \frac{1}{\sqrt{T_0}} \right) \\
&+\int_{P_0}^P V_i ~dP.
\end{split}
\label{eq:Gibbs_full_2}
\end{equation}
The preceding equation elucidates the various ingredients needed to numerically evaluate $G_i$: the entropy and enthalpy of formation (first line) and equation of state (last line).

For a pure gas, the conventional practice is to separate out the volume integral term in equation (\ref{eq:Gibbs_full_2}), because this term depends on whether the gas behaves ideally or non-ideally.  Subsequently, in the computation of the equilibrium constant only $G_{i,0}$ enters $\Delta G_r$ whereas the volume integral becomes part of the equilibrium constant.  For solids, the $G^\prime$ term in equation \eqref{eq:solidG} is exactly given by $G_i$ in equation (\ref{eq:Gibbs_full_2}).  Thus, it is $G_i$, rather than $G_{i,0}$, that enters $\Delta G_r$.

\begin{figure*}[h!]
\begin{center}
\vspace{-0.2in} 
\includegraphics[width=1.85\columnwidth]{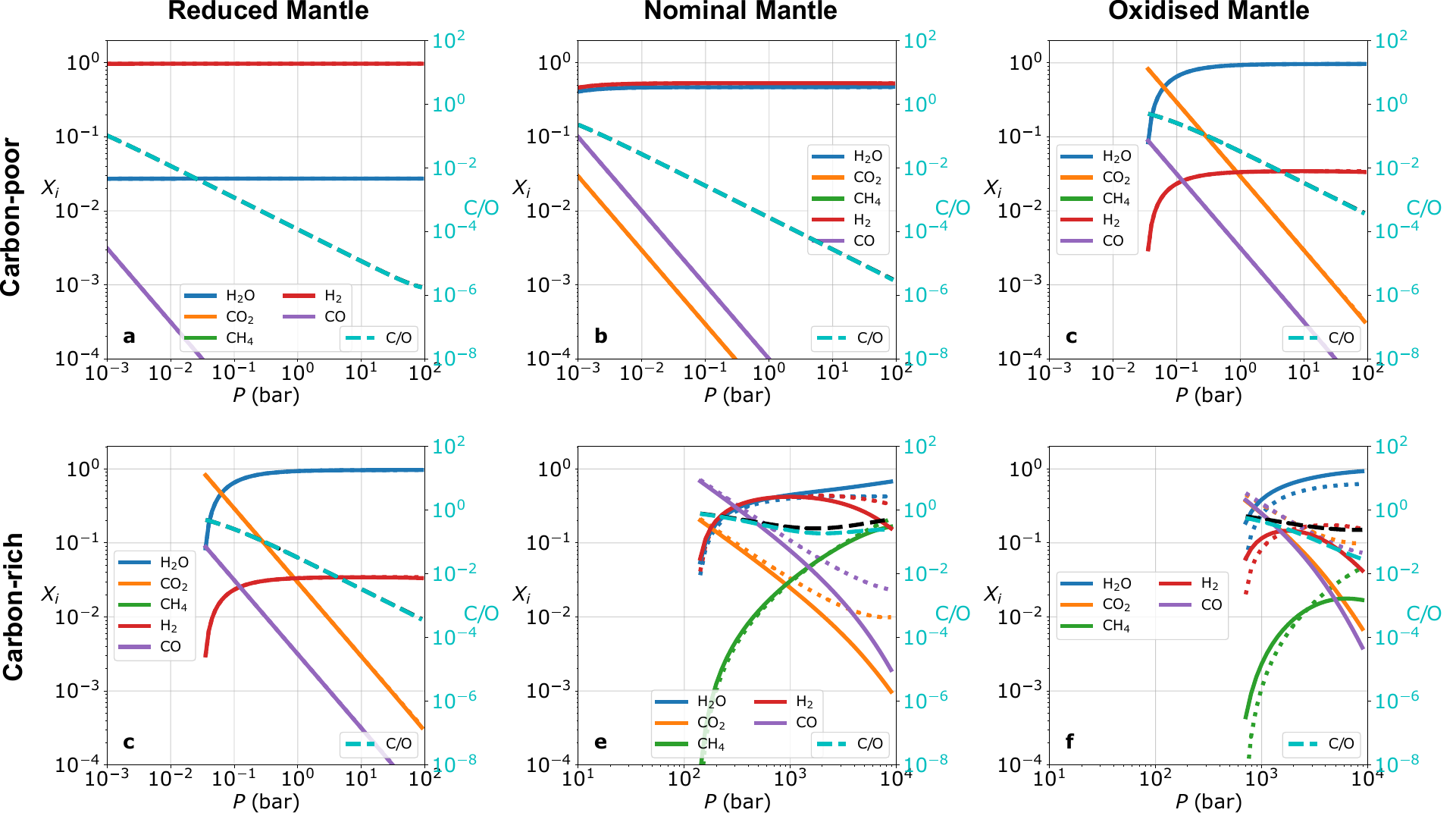}
\end{center}
%\vspace{-0.2in}
\caption{Examples of secondary atmospheres in the C-H-O chemical system, where the volume mixing ratios (relative abundances by number) of gases are shown as a function of the prescribed atmospheric surface pressure.  The top and bottom rows are for low- and high-carbon content in the mantle, respectively.  The first, second and third columns are for reduced, nominal and oxidised mantles, respectively.  See text for specific parameter values.  Regions of the plots where no curves exist are because the computed partial pressures of CO and CO$_2$ exceed the prescribed total pressure, implying that no mathematical solutions exist.  Solid and dotted curves correspond to calculations with fully non-ideal effects (see text for details) and the assumption of an ideal gas with ideally-mixed constituents, respectively.}
%\vspace{-0.15in}
\label{fig:CHO_secondary_vary_P}
\end{figure*}

\begin{figure*}[h!]
\begin{center}
% \vspace{-0.2in} 
% \includegraphics[width=0.65\columnwidth]{f5a-r1.pdf}
% \includegraphics[width=0.65\columnwidth]{f5b-r1.pdf}
% \includegraphics[width=0.65\columnwidth]{f5c-r1.pdf}
% \includegraphics[width=0.65\columnwidth]{f5d-r1.pdf}
% \includegraphics[width=0.65\columnwidth]{f5e-r1.pdf}
% \includegraphics[width=0.65\columnwidth]{f5f-r1.pdf}
\includegraphics[width=1.85\columnwidth]{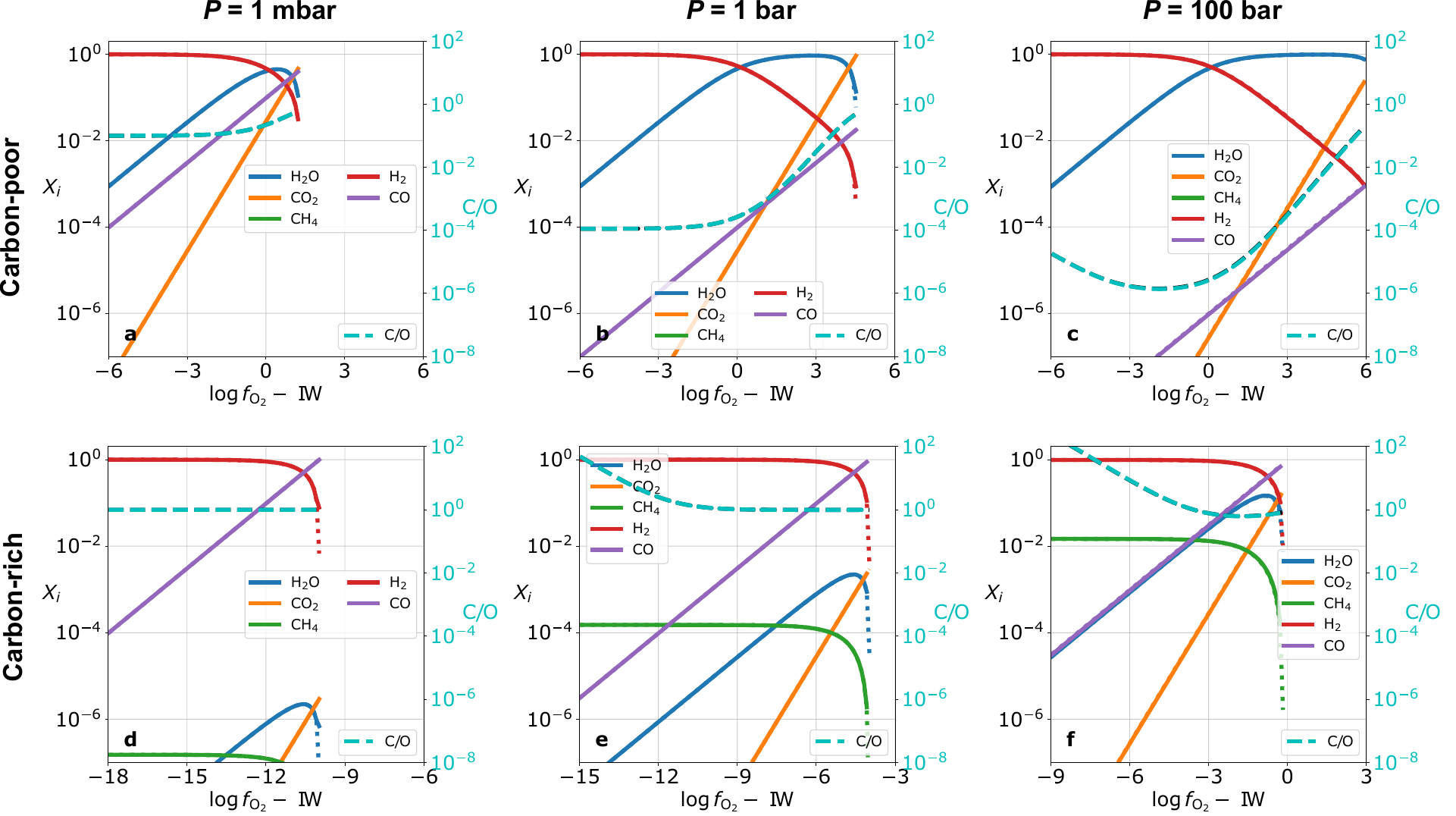}
\end{center}
% \vspace{-0.2in}
\caption{Same as Figure \ref{fig:CHO_secondary_vary_P} for secondary atmospheres in the C-H-O chemical system, but with volume mixing ratios as a function of the oxygen fugacity of the mantle.  The first, second and third columns are for atmospheric surface pressures of 1 mbar (Mars-like), 1 bar (Earth-like) and 100 bar (Venus-like), respectively. The heterogeneous ranges of values for the oxygen fugacity on the horizontal axes were chosen to minimise displaying regions of parameter space where no mathematical solutions exist.}
%\vspace{-0.15in}
\label{fig:CHO_secondary_vary_fO2}
\end{figure*}

\begin{figure*}%[h!]
\begin{center}
% \vspace{-0.2in} 
% \includegraphics[width=0.65\columnwidth]{f6a-r1.pdf}
% \includegraphics[width=0.65\columnwidth]{f6b-r1.pdf}
% \includegraphics[width=0.65\columnwidth]{f6c-r1.pdf}
% \includegraphics[width=0.65\columnwidth]{f6d-r1.pdf}
% \includegraphics[width=0.65\columnwidth]{f6e-r1.pdf}
% \includegraphics[width=0.65\columnwidth]{f6f-r1.pdf}
\includegraphics[width=1.85\columnwidth]{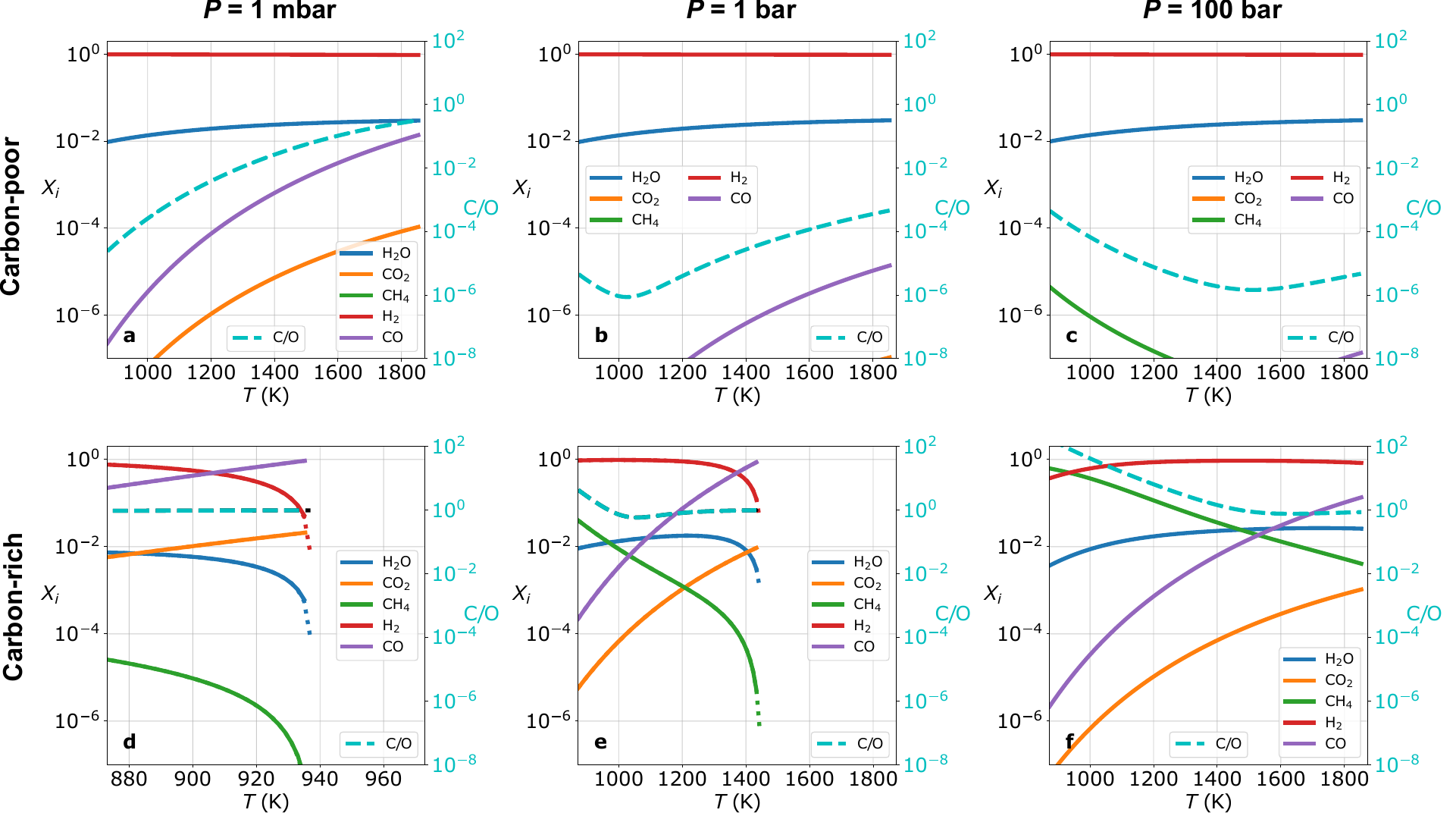}
\end{center}
% \vspace{-0.2in}
\caption{Same as Figure \ref{fig:CHO_secondary_vary_fO2} for secondary atmospheres in the C-H-O chemical system, but with volume mixing ratios as a function of the melt temperature.  For display purposes, a reduced mantle ($\log{f_{\rm O_2}} = \mbox{IW} - 3$) was arbitrarily chosen because it minimises the regions of parameter space with no mathematical solutions.}
%\vspace{-0.15in}
\label{fig:CHO_secondary_vary_T}
\end{figure*}

\subsection{Entropy and enthalpy of formation}

The entropy ($S_{i,0}$) and enthalpy of formation ($H_{i,0}$) are taken from \cite{hp98}, except for the nitrogen- and sulfur-bearing species for which we obtain them from the JANAF database.  These quantities are stated for the reference temperature of $T_0 = 298.15$ K and the reference pressure of $P_0=1$ bar; they are tabulated in Table \ref{tab:gibbs}.

\subsection{Equation of state}

For non-ideal gases, we use the Compensated-Redlich-Kwong (CORK) equation of state \citep{hp91} for H$_2$O, CO$_2$, CH$_4$, CO and H$_2$.  Specifically, this refers to $V_i (T,P)$.  For nitrogen- and sulfur-bearing species, we assume ideal gases due to the absence of data.  For solids, we use the equation of state for graphite provided by \cite{hp98}.

In practice, we note that solids have much smaller molar volume and are less compressible than gases, implying that the volume integral is sometimes neglected under lower pressures (e.g., page 815 of \citealt{sr93}) and $G_i \approx G_{i,0}$ is obtained.

\subsection{Fugacity coefficient}

With the equations of state of non-ideal gases in hand, one may obtain the fugacity coefficient of a pure gas by numerically evaluating the volume integral (page 186 of \citealt{devoe20}),
\begin{equation}
f = P_0 e^{\frac{1}{{\cal R} T} \int^P_{P_0} V ~dP},
\end{equation}
where we have set $f_0 = P_0 = 1$ bar.  Since $f = \phi_i P$, one obtains $\phi_i$ of species $i$ numerically.

\subsection{Activity coefficient}
\label{sect:gamma}

\cite{hp03} provide fitting functions of activity coefficients ($\gamma_i$) for CO-CO$_2$-CH$_4$-H$_2$O-H$_2$ mixtures, which are valid for $723 \le T/\mbox{ K} \le 2073$ and $500 \le P/\mbox{ bar} \le 4 \times 10^4$.  In practice, data on activity coefficients are sparser than for fugacity coefficients. Activity data for sulfur-bearing species in the C-H-O-S system are partially available \citep{evans10}, but they are unknown when nitrogen-bearing species (N$_2$, HCN and NH$_3$) are considered. Since a full treatment of non-ideality requires a complete knowledge of every pair-wise interaction among C-H-O-N-S-bearing species, like in the C-H-O system, existing studies choose to either ignore non-ideal mixing \citep{sl22} or only partially consider ideal mixing. For example, \cite{ague22} simulated the C-O-H-S subsystem where full non-ideality is considered for all the species except for SO$_2$. We follow this approach in the full C-O-H-S-N system to account for the full non-ideality of the C-O-H subsystem \citep{hp03}, whereas for S- and N-bearing species we assume full ideality ($\gamma_i= \phi_i = 1$) due to lack of non-ideal thermodynamic parameters. Such an approximation remains to be validated by and thus calls for future experimental measurements of activity coefficient data, especially for N- and S-bearing species in the full C-H-O-N-S system (e.g., \citealt{kite20}). 

\section{Results}
\label{sect:results}

Now that we have fully elucidated our theoretical framework for computing the atmospheric chemistry of secondary and hybrid atmospheres, we next explore their anticipated chemical diversity.  We first examine carbon-hydrogen-oxygen (C-H-O) systems followed by carbon-hydrogen-oxygen-nitrogen-sulfur (C-H-O-N-S) chemical systems.  The goal is to elucidate what is chemically possible when only considering an atmosphere-melt system in chemical equilibrium.  It is understood that, for realistic comparisons to observed systems, one needs to account for how photochemistry may alter the various chemical abundances of some irradiated atmospheres, which is out of scope of the present study.  We find that methane-dominated atmospheres are rare, which motivates a follow-up investigation into the properties required for their existence.

We present a suite of figures each consisting of a montage of 6 plots with some combination of the following parameter values: low ($a_{\rm C}=10^{-7}$) versus high ($a_{\rm C}=0.1$) carbon content of the mantle melt; reduced ($\log f_{\rm O_2}=\mbox{IW}-3$), nominal ($\log f_{\rm O_2}=\mbox{IW}$) and oxidised ($\log f_{\rm O_2}=\mbox{IW}+3$) mantle melts; $P=1$ mbar, 1 bar and 100 bar (inspired by Mars, Earth and Venus); $P_{\rm H_2}=1$ mbar, 1 bar and 100 bar (in the absence of a general theory of atmospheric escape). In plots where we have to pick a single melt temperature, we choose $T=1600$ K as explained previously in Section \ref{sec:ptrange}.  In plots where we have to pick a single oxygen fugacity of the mantle, we fix $\log f_{\rm O_2}=\mbox{IW}$ for illustration unless otherwise stated in the caption. Once a combination of parameter values are selected (e.g., $a_{\rm C} = 10^{-7}$, $\log f_{\rm O_2} = {\rm IW} - 3$, and $T = 1600$ K in Figure~\ref{fig:CHO_secondary_vary_P}a), we explore the chemical trend with an extra parameter (e.g., $P$ in Figure~\ref{fig:CHO_secondary_vary_P}a); the various parameter ranges used for examining trends are listed in Table \ref{tab:cho-results}. Overall, these choices of parameter values are inspired by and extended beyond the range of Solar System rocky planets, and the overarching intention here is to illustrate qualitative trends rather than model any specific exoplanet.

\begin{table*}
\begin{center}
\caption{Explored Parameter Ranges$^\dagger$ and Key Findings for the C-H-O System}
\label{tab:cho-results}
%\resizebox{\textwidth}{!}{
\begin{tabular}{ m{0.1\textwidth}  >{\centering\arraybackslash}m{0.4\textwidth}  >{\centering\arraybackslash}m{0.4\textwidth}}
\hline
\hline
Parameter & Range for Secondary Atmosphere & Range for Hybrid Atmosphere \\
\hline
$a_{\rm C}$ & carbon-rich: 0.1; carbon-poor:10$^{-7}$ & carbon-rich: 0.1; carbon-poor:10$^{-7}$ \\
$\log f_{\rm O_2}$ & between IW$-6$ and IW$+6$ & between IW$-6$ and IW$+6$ \\
$T$ & 873--1873 K & 873--1873 K\\
$P$ & 10$^{-3}$--10$^{4}$ bar & not applicable \\
$P_{\rm H_2}$ & not applicable & 10$^{-3}$--10$^{4}$ bar \\
\hline
Key Findings \newline &  H$_2$- and/or H$_2$O-rich atmospheres at low carbon content; \newline H$_2$ transits to H$_2$O and CO to CO$_2$ as $f_{\rm O_2}$ rises; \newline atmosphere chemistry sensitive to melt temperature & qualitative trends same as secondary atmospheres; \newline $P_{\rm H_2}$ rise at fixed $f_{\rm O_2}$ favors H$_2$O and suppresses CO and CO$_2$ (Figure \ref{fig:CHO_hybrid_vary_fO2}b \& c) \\
\hline
\hline
\end{tabular}%}\\
%\vspace{0.1in}
\end{center}
\vspace{-0.1in}
$^\dagger$Some ranges are changed to allow finding physically realistic solutions that are trend-revealing, e.g., the $\log f_{\rm O_2}$ range in Figure~\ref{fig:CHO_secondary_vary_fO2}d--f. 
\vspace{0.1in}
\end{table*}

The choice of these values for the carbon activity are difficult to reconcile with the absolute carbon content in mass, as we do not model mass conservation explicitly.  However, some physical interpretation may be provided.  A melt saturated with pure graphite corresponds to $a_{\rm C}=1$.  If $a_{\rm C}<1$, then the graphite is undersaturated and thus dissolves into co-existing melts as some form of carbon.  Since the activity is the generalisation of the relative abundance by number (volume mixing ratio for gases), one may interpret the carbon activity as the relative abundance of carbon in the melt.

The total atmospheric surface pressure (for secondary atmospheres only; $P$), hydrogen partial pressure (for hybrid atmospheres only; $P_{\rm H_2}$), helium partial pressure ($P_{\rm He}$) and nitrogen partial pressure ($P_{\rm N_2}$) encode our ignorance about a set of complex processes (radiative transfer, atmospheric mixing, atmospheric escape, etc), which are out of scope of the present study.

In the numerical solutions for secondary atmospheres, some combinations of parameter values for total pressure $P$, melt temperature $T$, carbon activity $a_{\rm C}$, and oxygen fugacity $f_{\rm O_2}$ do not admit physically realistic solutions, as exemplified by subsequent figures. This is explicable. With the prescribed $T$, $a_{\rm C}$ and $f_{\rm O_2}$, the partial pressures of CO and CO$_2$ can be straightforwardly determined through the first two net reactions in \eqref{eq:french1966}, and if the sum of these two partial pressures $P_{\rm CO_2} + P_{\rm CO} = a_{\rm C} K_{\rm eq,1}  f_{\rm O_2} + a_{\rm C} K_{\rm eq,2} \left( f_{\rm O_2} P_0 \right)^{1/2} $ is larger than the prescribed total pressure $P$, it becomes impossible for the solved partial pressures to be all positive. Physically, if $a_{\rm C}$ is prescribed to be 1, then the nonexistence of a realistic solution signifies graphite destablisation. This reasoning is also elucidated by equation (8) of \cite{french66} and is applicable to solutions for the full C-H-O-N-S systems. When numerically solving for hybrid atmospheres, the absence of a physically realistic solution is reflected by the ever-increasing total pressure from iteration to iteration (Section \ref{sec:solmethod}). 

\subsection{C-H-O Chemical Systems}

To build initial intuition for the chemistry of secondary and hybrid atmospheres, we consider the simplest non-trivial chemical system that consists of carbon (C), hydrogen (H) and oxygen (O).  We will see that key trends emerge that carry over to systems that include nitrogen and sulfur.

\subsubsection{Secondary atmospheres}

Figures \ref{fig:CHO_secondary_vary_P}, \ref{fig:CHO_secondary_vary_fO2} and \ref{fig:CHO_secondary_vary_T} examine the trends in the relative abundances of H$_2$, H$_2$O, CO, CO$_2$ and CH$_4$ as functions of the atmospheric surface pressure ($P$), oxygen fugacity of the mantle ($f_{\rm O_2}$) and melt temperature ($T$), respectively.  An atmosphere deriving from a reduced mantle with low carbon content is H$_2$- and H$_2$O-dominated, not unlike a gas-giant exoplanet (top-left panel of Figure \ref{fig:CHO_secondary_vary_P}).  As the mantle becomes increasingly oxidised (moving from the first to the third column of Figure \ref{fig:CHO_secondary_vary_P}), two important trends appear:
\begin{itemize}

\item The atmosphere transitions from being dominated by H$_2$ to being dominated by H$_2$O;

\item The major carbon carrier transitions from being CO to CO$_2$.

\end{itemize}
These trends persist even when the carbon content of the mantle becomes high (bottom row of Figure \ref{fig:CHO_secondary_vary_P}).  Methane appears when the carbon content is high, but its abundance is never comparable to those of the other species unless the surface pressure is high (Figure~\ref{fig:CHO_secondary_vary_P}d).  Oxidised mantles further suppress the appearance of methane (Figure~\ref{fig:CHO_secondary_vary_P}d--f).  Figure \ref{fig:CHO_secondary_vary_fO2} confirms the preceding trends.  The competition between CO and CO$_2$ motivates the use of their relative abundances as a diagnostic for the oxygen fugacity.

Figure \ref{fig:CHO_secondary_vary_T} demonstrates that the exact melt temperature assumed needs to be considered carefully, because the abundances of CO, CO$_2$ and CH$_4$ vary by orders of magnitude across a relatively modest range of $T$ values.  An important feature of Figures \ref{fig:CHO_secondary_vary_P}, \ref{fig:CHO_secondary_vary_fO2} and \ref{fig:CHO_secondary_vary_T} is that the carbon-to-oxygen (C/O) ratio of the gaseous phase varies by orders of magnitude.  Non-ideal effects appear to be minimal, unless large pressures are considered.

\subsubsection{Hybrid atmospheres}

\begin{figure*}[h!]
\begin{center}
\vspace{-0.1in} 
\includegraphics[width=1.85\columnwidth]{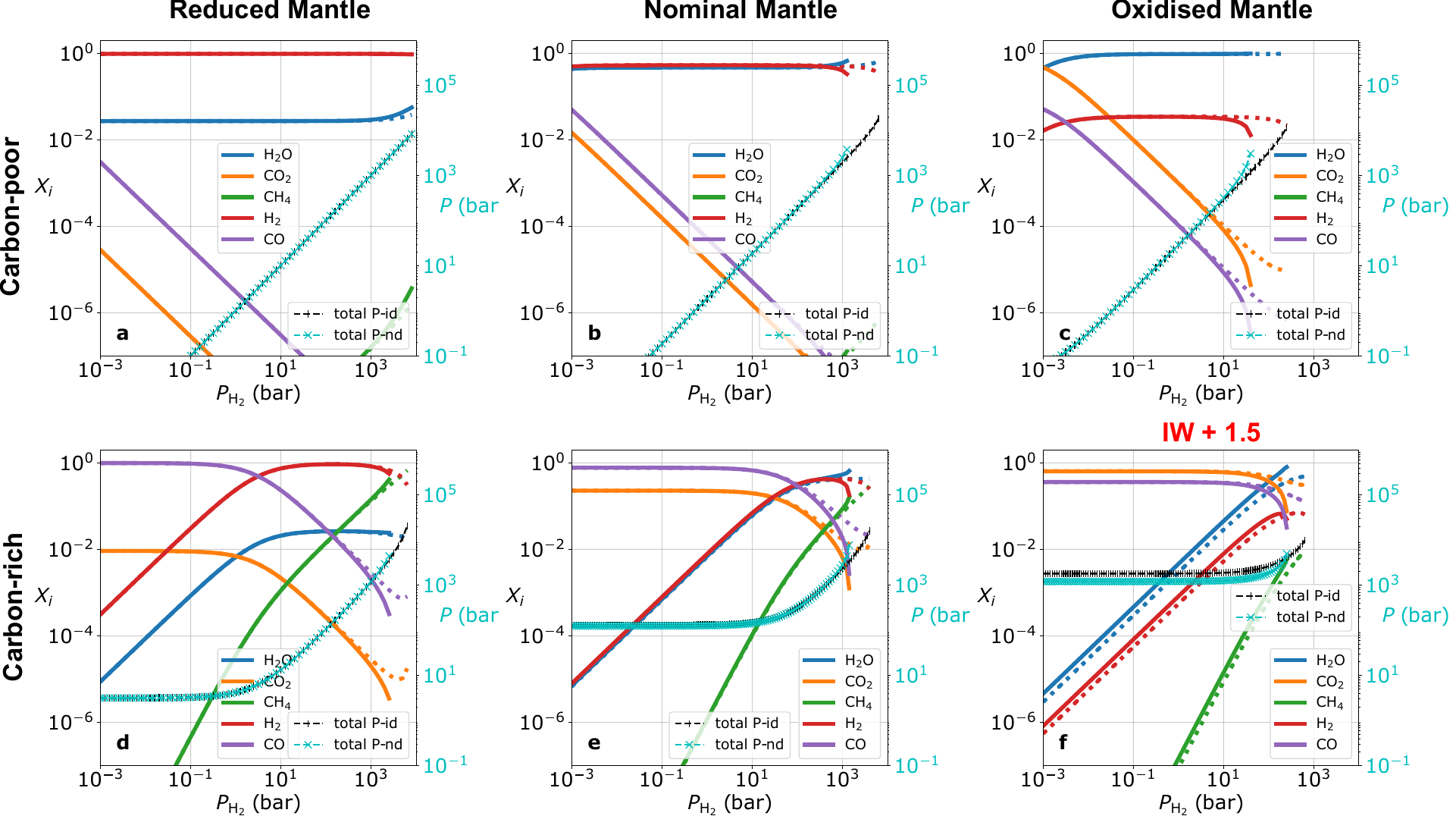}
\end{center}
% \vspace{-0.2in}
\caption{Examples of hybrid atmospheres in the C-H-O chemical system, where the volume mixing ratios (relative abundances by number) of gases are shown as a function of the prescribed atmospheric partial pressure of molecular hydrogen.  The top and bottom rows are for low- and high-carbon content in the mantle, respectively.  The first, second and third columns are for reduced, nominal and oxidised mantles, respectively.  See text for specific parameter values.  Regions of the plots where no curves exist are because the computed partial pressures of CH$_4$ and H$_2$O exceed the computed total pressure, implying that no mathematical solutions exist. The oxygen fugacity for panel f is prescribed at IW$+1.5$ instead of that of oxidised mantle (IW$+3$) to minimise no-solution parameter space. Solid and dotted curves correspond to calculations with fully non-ideal effects (see text for details) and the assumption of an ideal gas with ideally-mixed constituents, respectively. The solved total pressures for ideal (P-id) and non-ideal (P-nd) cases overlap with each other until $P_{\rm H_2} \gtrsim 10^3$ bar. }
%\vspace{-0.15in}
\label{fig:CHO_hybrid_vary_PH2}
\end{figure*}

\begin{figure*}[h!]
\begin{center}
% \vspace{-0.2in} 
% \includegraphics[width=0.6\columnwidth]{f8a.pdf}
% \includegraphics[width=0.6\columnwidth]{f8b.pdf}
% \includegraphics[width=0.6\columnwidth]{f8c.pdf}
% \includegraphics[width=0.6\columnwidth]{f8d.pdf}
% \includegraphics[width=0.6\columnwidth]{f8e.pdf}
% \includegraphics[width=0.6\columnwidth]{f8f.pdf}
\includegraphics[width=1.85\columnwidth]{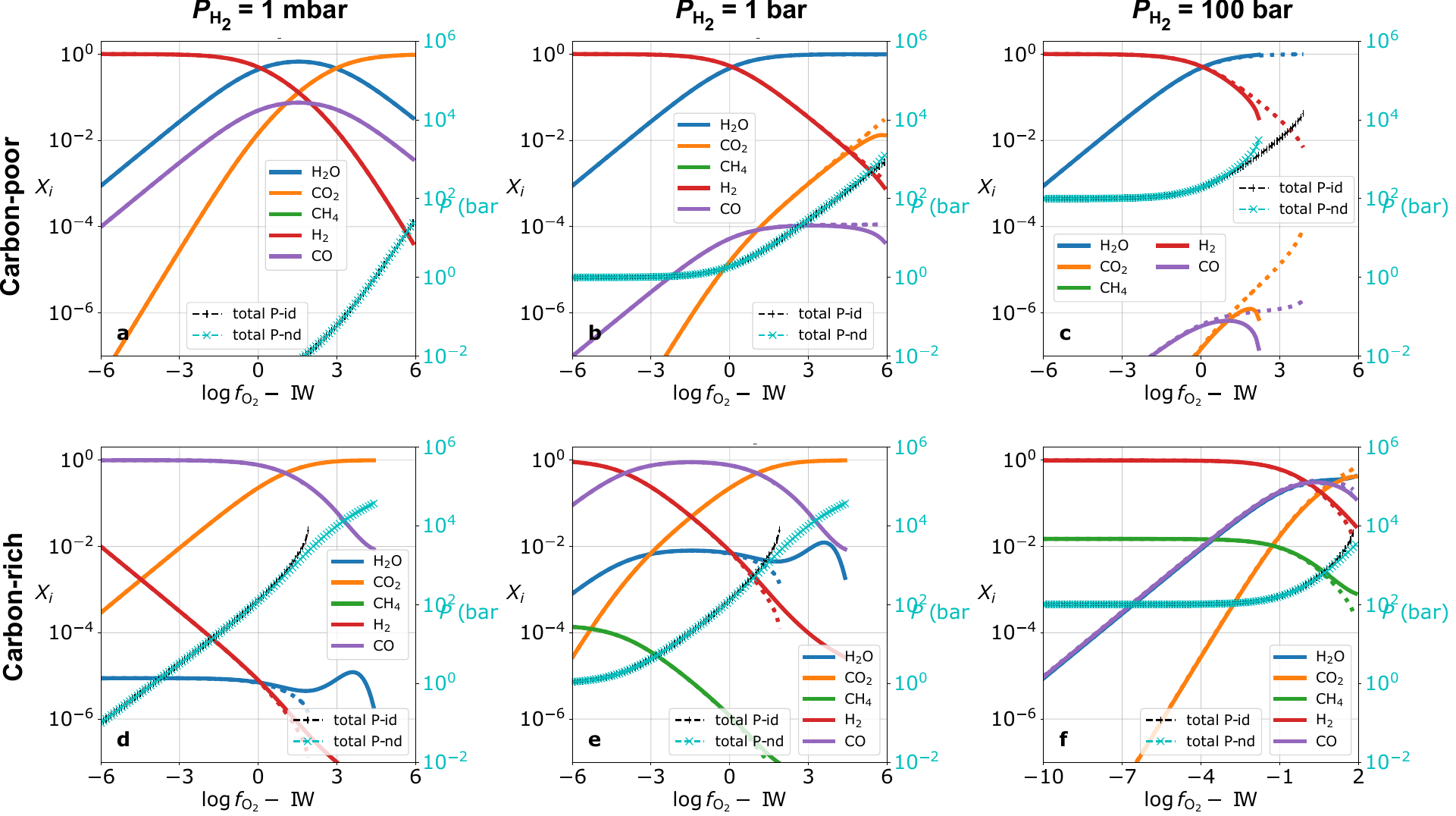}
\end{center}
% \vspace{-0.2in}
\caption{Same as Figure \ref{fig:CHO_hybrid_vary_PH2} for hybrid atmospheres in the C-H-O chemical system, but with volume mixing ratios as a function of the oxygen fugacity of the mantle.  The first, second and third columns are for atmospheric partial pressures of molecular hydrogen of 1 mbar, 1 bar and 100 bar, respectively. The solved total pressures for ideal (P-id) and non-ideal (P-nd) cases overlap with each other until $\log f_{\rm O_2} \gtrsim {\rm IW}+2$. }
%\vspace{-0.15in}
\label{fig:CHO_hybrid_vary_fO2}
\end{figure*}

\begin{figure*}%[h!]
\begin{center}
% \vspace{-0.2in} 
% \includegraphics[width=0.6\columnwidth]{f9a.pdf}
% \includegraphics[width=0.6\columnwidth]{f9b.pdf}
% \includegraphics[width=0.6\columnwidth]{f9c.pdf}
% \includegraphics[width=0.6\columnwidth]{f9d.pdf}
% \includegraphics[width=0.6\columnwidth]{f9e.pdf}
% \includegraphics[width=0.6\columnwidth]{f9f.pdf}
\includegraphics[width=1.85\columnwidth]{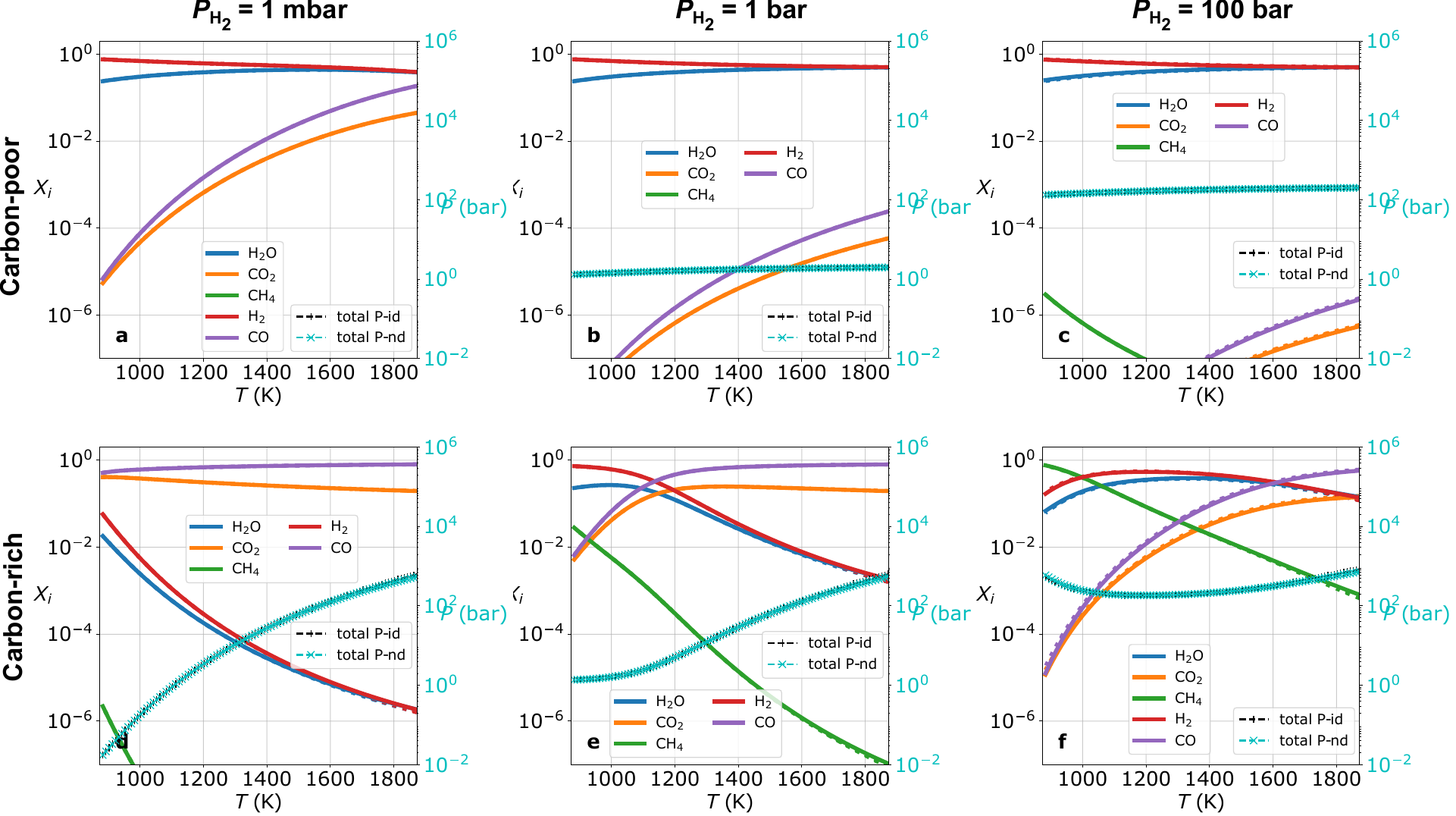}
\end{center}
% \vspace{-0.2in}
\caption{Same as Figure \ref{fig:CHO_hybrid_vary_fO2} for hybrid atmospheres in the C-H-O chemical system, but with a mantle of nominal oxidation state (see text for details) and volume mixing ratios as a function of the melt temperature. The solved total pressures for ideal (P-id) and non-ideal (P-nd) cases overlap for the entire $T$ range.}
%\vspace{-0.15in}
\label{fig:CHO_hybrid_vary_T}
\end{figure*}

The qualitative trends and lessons learned from our study of C-H-O secondary atmospheres carry over when we examine hybrid atmospheres in Figures \ref{fig:CHO_hybrid_vary_PH2}, \ref{fig:CHO_hybrid_vary_fO2} and \ref{fig:CHO_hybrid_vary_T}.  A key difference is that, when the partial pressure of atmospheric molecular hydrogen ($P_{\rm H_2}$) becomes large (e.g., Figure \ref{fig:CHO_hybrid_vary_fO2}b \& c), the production of CO and CO$_2$ is suppressed in favor of H$_2$O.  It is again emphasized that modest changes in the melt temperatures lead to order-of-magnitude changes in the relative abundances of the gaseous species (Figure \ref{fig:CHO_hybrid_vary_T}).

\subsection{C-H-O-N-S Chemical Systems}

The addition of nitrogen (N) and sulfur (S) to the chemical system results in 5 more gaseous species: hydrogen sulfide (H$_2$S), sulfur dioxide (SO$_2$), ammonia (NH$_3$), molecular nitrogen (N$_2$) and hydrogen cyanide (HCN).  This doubles the number of species compared to C-H-O systems.

\begin{table*}
\begin{center}
\caption{Explored Parameter Ranges$^\dagger$$^\ddagger$ and Key Findings for the C-H-O-N-S System}
\label{tab:chons-results}
%\resizebox{\textwidth}{!}{
\begin{tabular}{m{0.1\textwidth}  >{\centering\arraybackslash}m{0.50\textwidth}  >{\centering\arraybackslash}m{0.30\textwidth}}
\hline
\hline
Parameter & Range for Secondary Atmosphere & Range for Hybrid Atmosphere \\
\hline
$\log f_{\rm S_2}$ & between PP$-10$ and PP & between PP$-10$ and PP \\
$P_{\rm N_2}/P$ & 10$^{-4}$--1 & not applicable \\
$P_{\rm N_2}/P_{\rm H_2}$ & not applicable & 10$^{-4}$--1 \\
\hline
Key Findings\newline \newline & qualitative trends of the C-H-O system carries over; the H$_2$S/SO$_2$ ratio is sensitive to surface pressure and $f_{\rm O_2}$; the NH$_2$/HCN ratio sensitive to surface pressure but not to $f_{\rm O_2}$; the H$_2$S/SO$_2$ ratio is not sensitive to $f_{\rm S_2}$ but their abundances are; likewise, the NH$_2$/HCN ratio is not sensitive to $P_{\rm N_2}$ but their abundances are; & same as the secondary atmospheres \newline \newline \\
\hline
\hline
\end{tabular}%}\\
%\vspace{0.1in}
\end{center}
\vspace{-0.1in}
$^\dagger$ The range of $a_{\rm C}$, $\log f_{\rm O_2}$, $T$, and $P$ or $P_{\rm H_2}$ are the same as for the C-H-O system (Table \ref{tab:cho-results}).

$^\ddagger$Some ranges are changed to allow finding physically realistic solutions that are trend-revealing, e.g., the $\log f_{\rm O_2}$ range in Figure~\ref{fig:CHONS_secondary_vary_fO2}d. 
\vspace{0.1in}
\end{table*}

For illustration, we choose $P_{\rm N_2}/P=0.1$ (for secondary atmospheres) and $P_{\rm N_2}/P_{\rm H_2}=0.1$ (for hybrid atmospheres), although the chemical trends are also explored for $P_{\rm N_2}/P$ and $P_{\rm N_2}/P_{\rm H_2}$ in the range of 10$^{-4}$--1 (Table \ref{tab:chons-results}, and Figures~\ref{fig:CHONS_secondary_vary_PN2} \& \ref{fig:CHONS_hybrid_vary_PN2} ). While these choices are arbitrary, it is better than choosing an absolute value for $P_{\rm N_2}$ as we wish to avoid situations where the trends associated with nitrogen-bearing species appear artificially weak because an arbitrarily low value of the nitrogen partial pressure was chosen.

As already noted in Section \ref{subsect:fugacities}, the sulfur fugacity associated with the Earth spans an enormous range of values. Our choices of values for the sulfur fugacity are to allow us to visually display trends in an illustrative manner. Similar to the strategy of parameter value selection in the C-H-O system, we choose $\log f_{\rm S_2} =$ PP $- 10$ if a value of $\log f_{\rm S_2}$ needs to be fixed (e.g., Figure~\ref{fig:CHONS_secondary_vary_P}), and separately explore the chemical trends when $\log f_{\rm S_2}$ varies between PP $- 10$ and PP (Table \ref{tab:chons-results}, and Figures~\ref{fig:CHONS_secondary_vary_fS2} \& \ref{fig:CHONS_hybrid_vary_fS2}).

\subsubsection{Secondary atmospheres}

\begin{figure*}[h!]
\begin{center}
\vspace{-0.2in} 
\includegraphics[width=1.85\columnwidth]{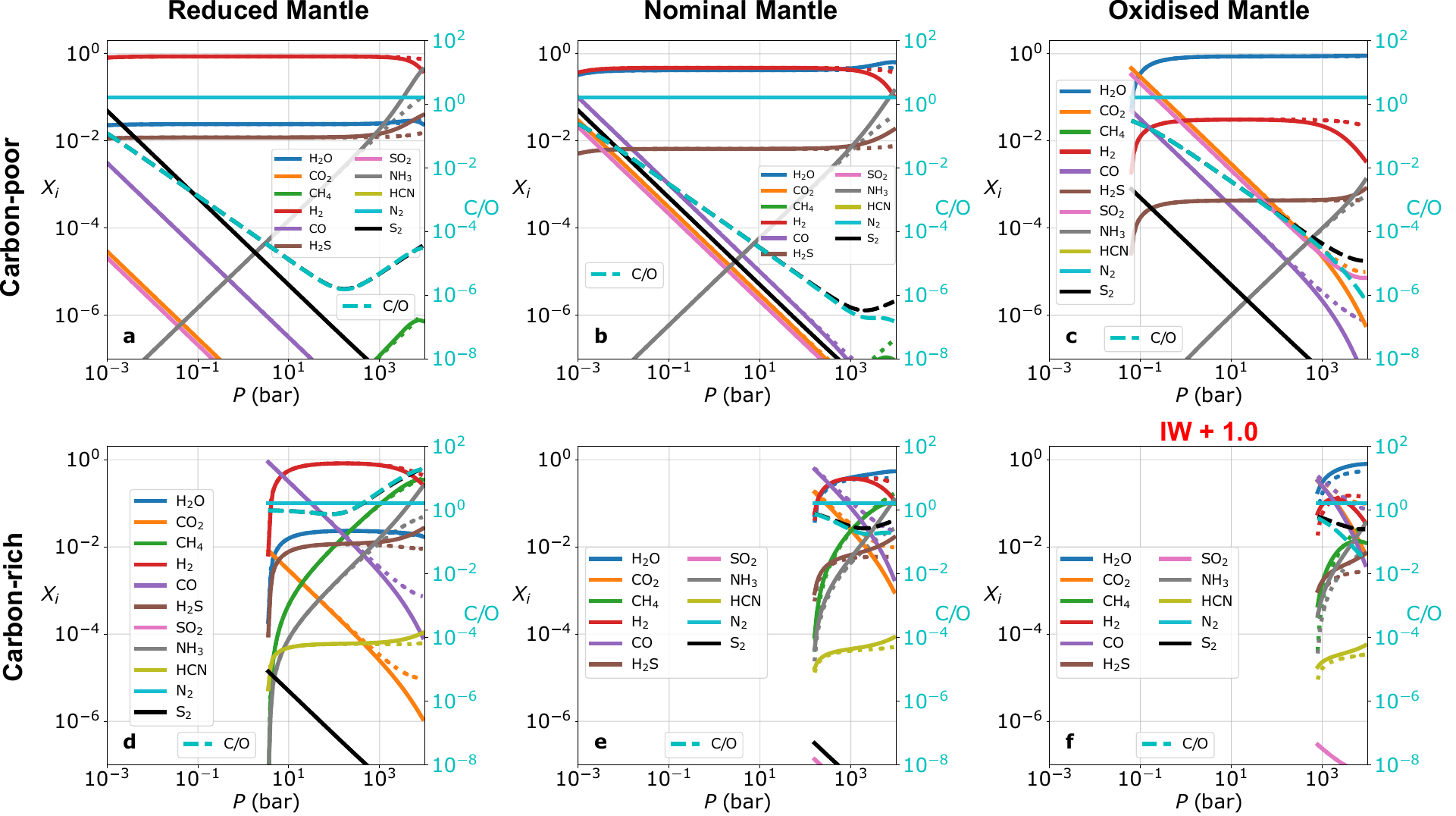}
\end{center}
 \vspace{-0.2in}
\caption{Examples of secondary atmospheres in the C-H-O-N-S chemical system, where the volume mixing ratios (relative abundances by number) of gases are shown as a function of the prescribed atmospheric surface pressure.  The sulfur fugacity is arbitrarily chosen to be $\log f_{\rm S_2} = {\rm PP} - 10$.  The top and bottom rows are for low- and high-carbon content in the mantle, respectively.  The first, second and third columns are for reduced, nominal and oxidised mantles, respectively.  See text for specific parameter values.  Regions of the plots where no curves exist are because the computed partial pressures of CO, CO$_2$ and H$_2$O exceed the prescribed total pressure, implying that no mathematical solutions exist.  The oxygen fugacity for panel f is prescribed at IW$+1$ instead of that of oxidised mantle (IW$+3$) to minimise no-solution parameter space. Solid and dotted curves correspond to calculations with partially non-ideal effects (see text for details) and the assumption of an ideal gas with ideally-mixed constituents, respectively.}
%\vspace{-0.15in}
\label{fig:CHONS_secondary_vary_P}
\end{figure*}

\begin{figure*}[h!]
\begin{center}
% \vspace{-0.2in} 
% \includegraphics[width=0.6\columnwidth]{f11a.pdf}
% \includegraphics[width=0.6\columnwidth]{f11b.pdf}
% \includegraphics[width=0.6\columnwidth]{f11c.pdf}
% \includegraphics[width=0.6\columnwidth]{f11d.pdf}
% \includegraphics[width=0.6\columnwidth]{f11e.pdf}
% \includegraphics[width=0.6\columnwidth]{f11f.pdf}
\includegraphics[width=1.85\columnwidth]{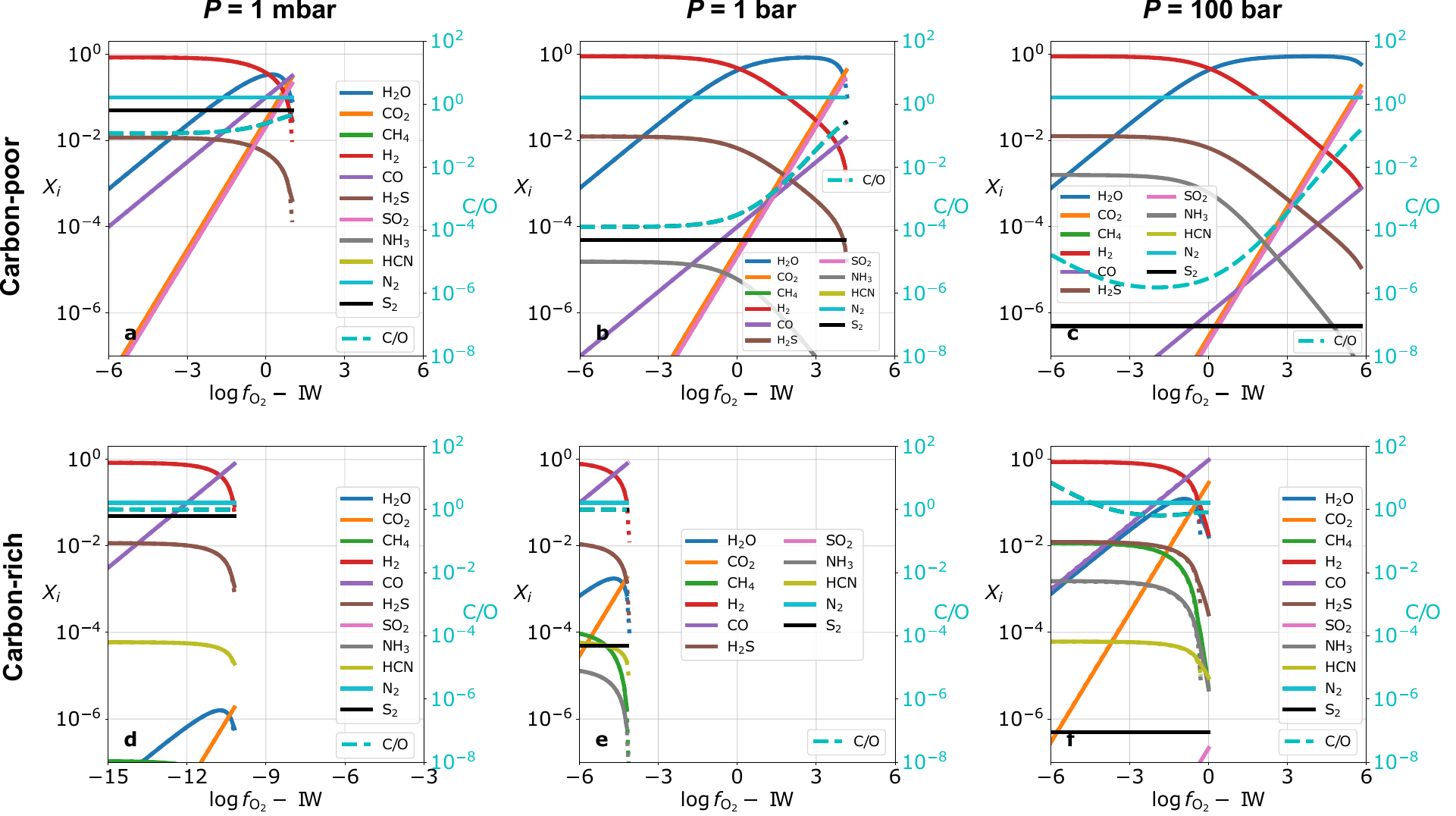}
\end{center}
% \vspace{-0.2in}
\caption{Examples of secondary atmospheres in the C-H-O-N-S chemical system, with volume mixing ratios as a function of the oxygen fugacity of the mantle.  The sulfur fugacity is arbitrarily chosen to be $\log f_{\rm S_2} = {\rm PP} - 10$.  The first, second and third columns are for atmospheric surface pressures of 1 mbar (Mars-like), 1 bar (Earth-like) and 100 bar (Venus-like), respectively. The heterogeneous ranges of values for the oxygen fugacity on the horizontal axes were chosen to minimise displaying regions of parameter space where no mathematical solutions exist.}
%\vspace{-0.15in}
\label{fig:CHONS_secondary_vary_fO2}
\end{figure*}

\begin{figure*}[h!]
\begin{center}
\vspace{-0.1in} 
\includegraphics[width=1.85\columnwidth]{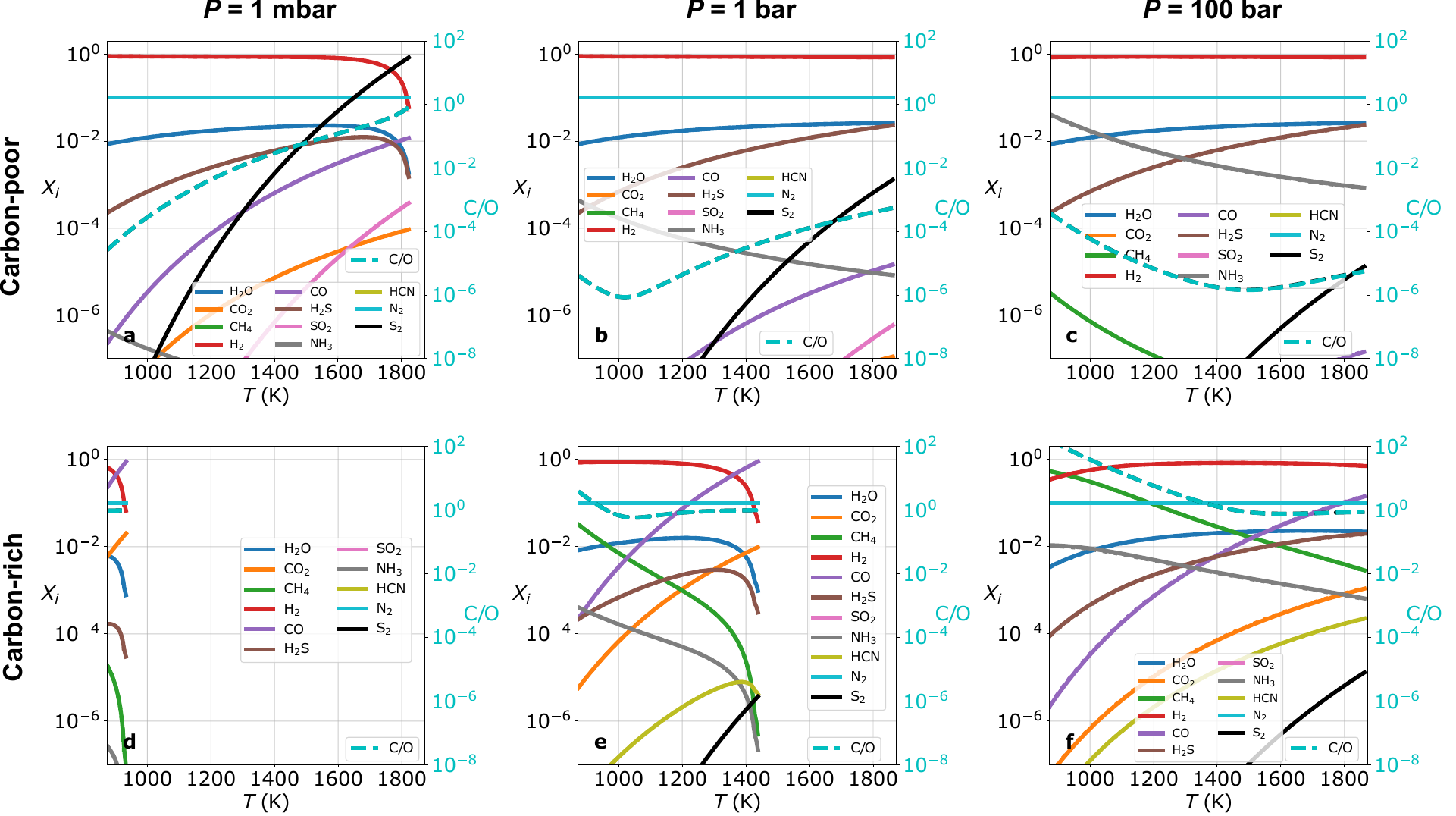}
\end{center}
% \vspace{-0.2in}
\caption{Same as Figure \ref{fig:CHONS_secondary_vary_fO2} for secondary atmospheres in the C-H-O-N-S chemical system, but with volume mixing ratios as a function of the melt temperature.   The sulfur fugacity is arbitrarily chosen to be $\log f_{\rm S_2} = {\rm PP} - 10$. For display purposes, a reduced mantle ($\log{f_{\rm O_2}} = \mbox{IW} - 3$) was arbitrarily chosen because it minimises the regions of parameter space with no mathematical solutions.}
%\vspace{-0.15in}
\label{fig:CHONS_secondary_vary_T}
\end{figure*}

\begin{figure*}[h!]
\begin{center}
% \vspace{-0.2in} 
% \includegraphics[width=0.6\columnwidth]{f13a.pdf}
% \includegraphics[width=0.6\columnwidth]{f13b.pdf}
% \includegraphics[width=0.6\columnwidth]{f13c.pdf}
% \includegraphics[width=0.6\columnwidth]{f13d.pdf}
% \includegraphics[width=0.6\columnwidth]{f13e.pdf}
% \includegraphics[width=0.6\columnwidth]{f13f.pdf}
\includegraphics[width=1.85\columnwidth]{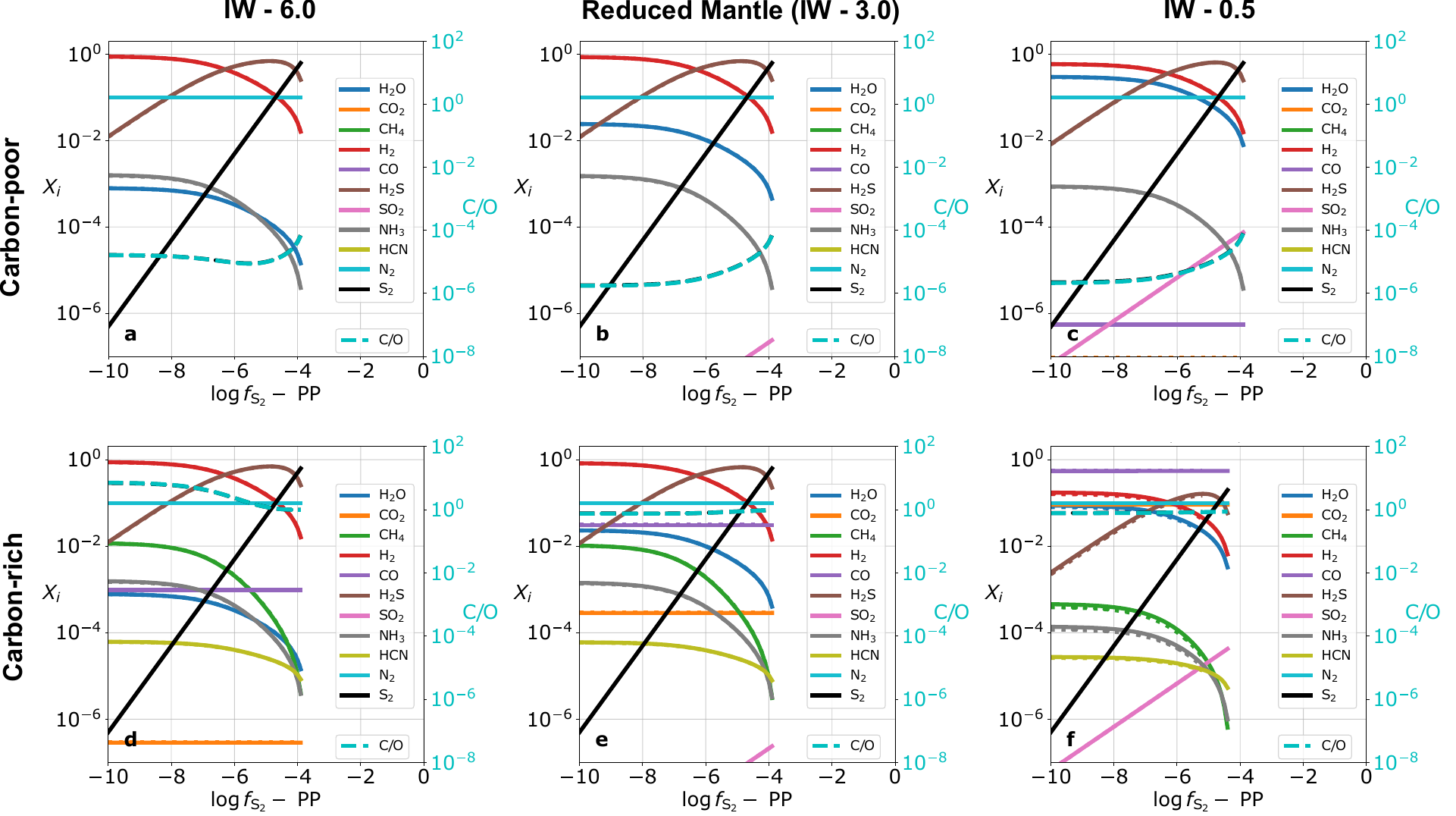}
\end{center}
% \vspace{-0.2in}
\caption{Same as Figure \ref{fig:CHONS_secondary_vary_P} for secondary atmospheres in the C-H-O-N-S chemical system, but with volume mixing ratios as a function of the sulfur fugacity of the mantle.  For display purposes, different values of the oxygen fugacity and a fixed total pressure $P=100$ bar were chosen because it minimises the regions of parameter space with no mathematical solutions. }
%\vspace{-0.15in}
\label{fig:CHONS_secondary_vary_fS2}
\end{figure*}

\begin{figure*}[h!]
\begin{center}
\vspace{-0.2in} 
\includegraphics[width=1.85\columnwidth]{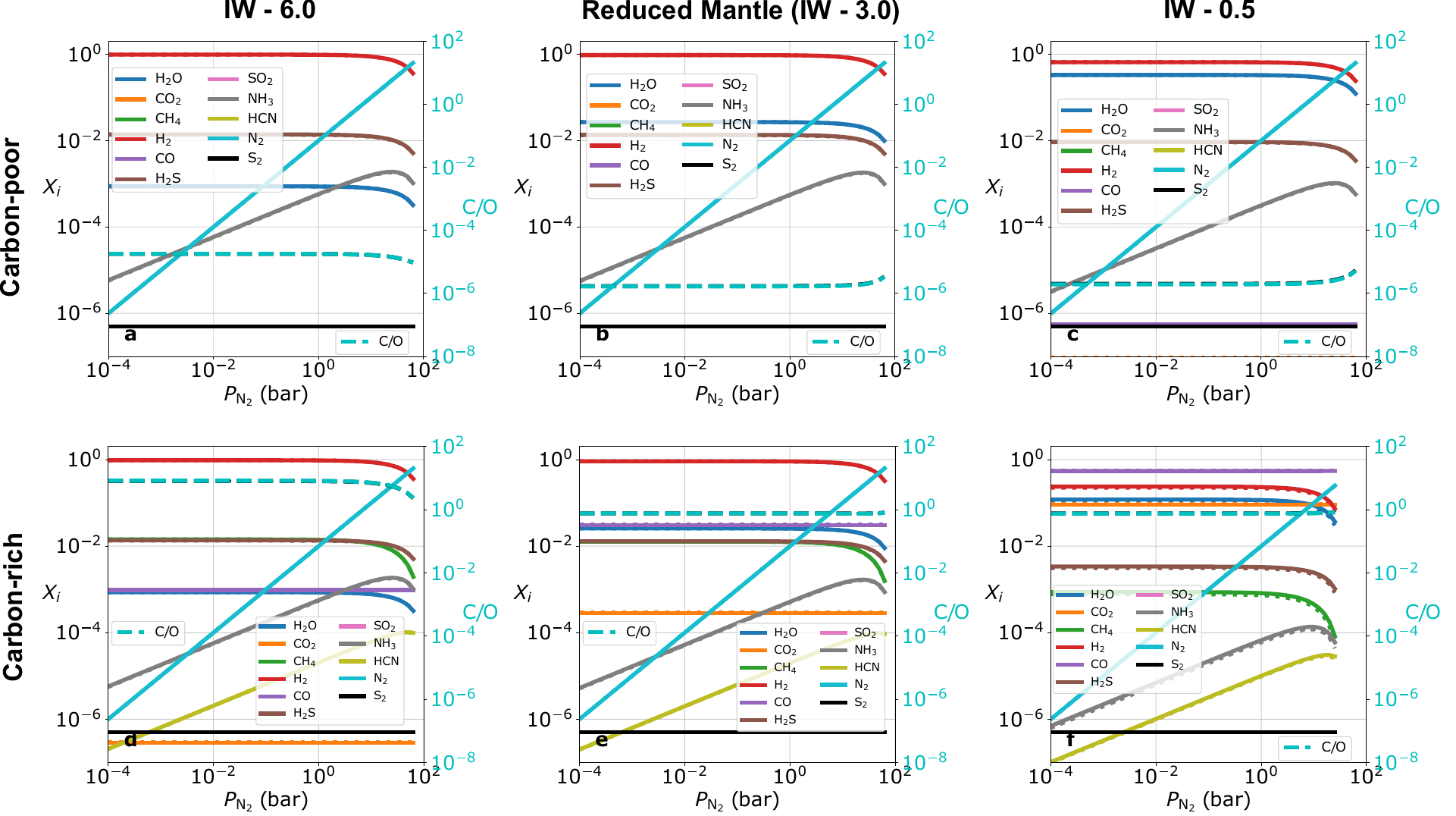}
\end{center}
\vspace{-0.2in}
\caption{Same as Figure \ref{fig:CHONS_secondary_vary_fS2} for secondary atmospheres in the C-H-O-N-S chemical system, but with volume mixing ratios as a function of the atmospheric partial pressure of molecular nitrogen.  The sulfur fugacity is arbitrarily chosen to be $\log f_{\rm S_2} = {\rm PP} - 10$.  For display purposes, different values of the oxygen fugacity and a fixed total pressure $P=100$ bar were chosen because it minimises the regions of parameter space with no mathematical solutions. }
%\vspace{-0.15in}
\label{fig:CHONS_secondary_vary_PN2}
\end{figure*}

Figures \ref{fig:CHONS_secondary_vary_P}, \ref{fig:CHONS_secondary_vary_fO2}, \ref{fig:CHONS_secondary_vary_T}, \ref{fig:CHONS_secondary_vary_fS2} and \ref{fig:CHONS_secondary_vary_PN2} examine the trends in the relative abundances of H$_2$, H$_2$O, CO, CO$_2$, CH$_4$, H$_2$S, SO$_2$, NH$_3$, N$_2$ and HCN as functions of the atmospheric surface pressure ($P$), oxygen fugacity of the mantle ($f_{\rm O_2}$), melt temperature ($T$), sulfur fugacity of the mantle ($f_{\rm S_2}$) and atmospheric partial pressure of molecular nitrogen ($P_{\rm N_2}$), respectively.  The qualitative trends and lessons learned from examining C-H-O chemical systems carry over.  Additionally, the following trends emerge:
\begin{itemize}

\item The competition between H$_2$S and SO$_2$ depends sensitively on the surface pressure (Figure \ref{fig:CHONS_secondary_vary_P}) and oxygen fugacity (Figure \ref{fig:CHONS_secondary_vary_fO2}).  

\item Likewise, the competition between NH$_3$ and HCN is sensitive to the surface pressure (Figure \ref{fig:CHONS_secondary_vary_P}).  As neither NH$_3$ nor HCN are oxygen carriers, their abundances are somewhat insensitive to the oxygen fugacity (Figure \ref{fig:CHONS_secondary_vary_fO2}).

\item The absolute abundances of H$_2$S and SO$_2$ are sensitive to the sulfur fugacity, but the ratio of their abundances is not (Figure \ref{fig:CHONS_secondary_vary_fS2}).  

\item Likewise, the absolute abundances of NH$_3$ and HCN are sensitive to the partial pressure of nitrogen, but the ratio of their abundances is not (Figure \ref{fig:CHONS_secondary_vary_PN2}).

\end{itemize}
As before, varying the melt temperature by a factor $\sim 2$ results in order-of-magnitude variations in the molecular abundances (Figure \ref{fig:CHONS_secondary_vary_T}).  The competition between H$_2$S and SO$_2$ motivates the use of their relative abundances as a supporting diagnostic for the oxygen fugacity.

\subsubsection{Hybrid atmospheres}

Figures \ref{fig:CHONS_hybrid_vary_PH2}, \ref{fig:CHONS_hybrid_vary_fO2}, \ref{fig:CHONS_hybrid_vary_T}, \ref{fig:CHONS_hybrid_vary_fS2} and \ref{fig:CHONS_hybrid_vary_PN2} examine the trends in the relative abundances of H$_2$, H$_2$O, CO, CO$_2$, CH$_4$, H$_2$S, SO$_2$, NH$_3$, N$_2$ and HCN as functions of the atmospheric partial pressure of molecular hydrogen ($P_{\rm H_2}$), oxygen fugacity of the mantle ($f_{\rm O_2}$), melt temperature ($T$), sulfur fugacity of the mantle ($f_{\rm S_2}$) and atmospheric partial pressure of molecular nitrogen ($P_{\rm N_2}$), respectively.  Since no new trends are uncovered beyond what we already learned from examining C-H-O systems and C-H-O-N-S secondary atmospheres, we present these calculations in Appendix \ref{append:fig} for completeness.

\subsection{Methane-rich atmospheres}

\begin{table*}
\begin{center}
\caption{Monte Carlo Sampled Parameter Space for the C-H-O-N-S System and Key Findings}
\label{tab:mc}
%\resizebox{\textwidth}{!}{
\begin{tabular}{ m{0.35\textwidth}  >{\centering\arraybackslash}m{0.50\textwidth}}
\hline
\hline
Parameter & Sample Distribution and Space \\
\hline
$\log a_{\rm C}$ & uniform$^\dagger$ between $-8$ and 0 \\
$\log f_{\rm O_2}$ & uniform between IW$-6$ and IW$+6$ \\
$\log f_{\rm S_2}$ & uniform between PP$-10$ and PP \\
$T$ & uniform between 873 and 1873 K\\
$\log (P/{\rm bar})$ for secondary atmospheres & uniform between $-3$ and 4 \\
$\log (P_{\rm N_2}/P)$ for secondary atmospheres & uniform between $-2$ and 0 \\
$\log (P_{\rm H_2}/{\rm bar})$ for hybrid atmospheres & uniform between $-3$ and 4 \\
$\log (P_{\rm N_2}/P_{\rm H_2})$ for hybrid atmospheres & uniform between $-2$ and 0 \\

\hline
Key Findings \newline \newline \newline \newline & joint $X_{\rm CO_2}/X_{\rm CO}$ and $X_{\rm H_2S}/X_{\rm SO_2}$ as a strong constraint on the $f_{\rm O_2}$ of a rocky mantle; complemented by $X_{\rm H_2O}/X_{\rm CH_4}$ and $X_{\rm NH_3}/X_{\rm HCN}$, melt carbon content and temperature may be further constrained; $f_{\rm S_2}$ and $P_{\rm N_2}$ are challenging to constrain from the four ratios. \newline \newline \newline methane-dominated atmospheres requires low melt temperatures ($T \lesssim 1000$ K), reduced mantles ($f_{\rm O_2} \lesssim \mbox{IW}$) and high atmospheric surface pressures exceeding $\sim 10$ bar. \\
\hline
\hline
\end{tabular}%}\\
%\vspace{-0.1in}
\end{center}
\vspace{-0.1in}
$^\dagger $ set $a_{\rm C} = 1$ when investigating methane-dominated atmospheres.
\end{table*}

\begin{figure*}[h!]
\begin{center}
\vspace{-0.2in} 
\includegraphics[width=0.72\columnwidth]{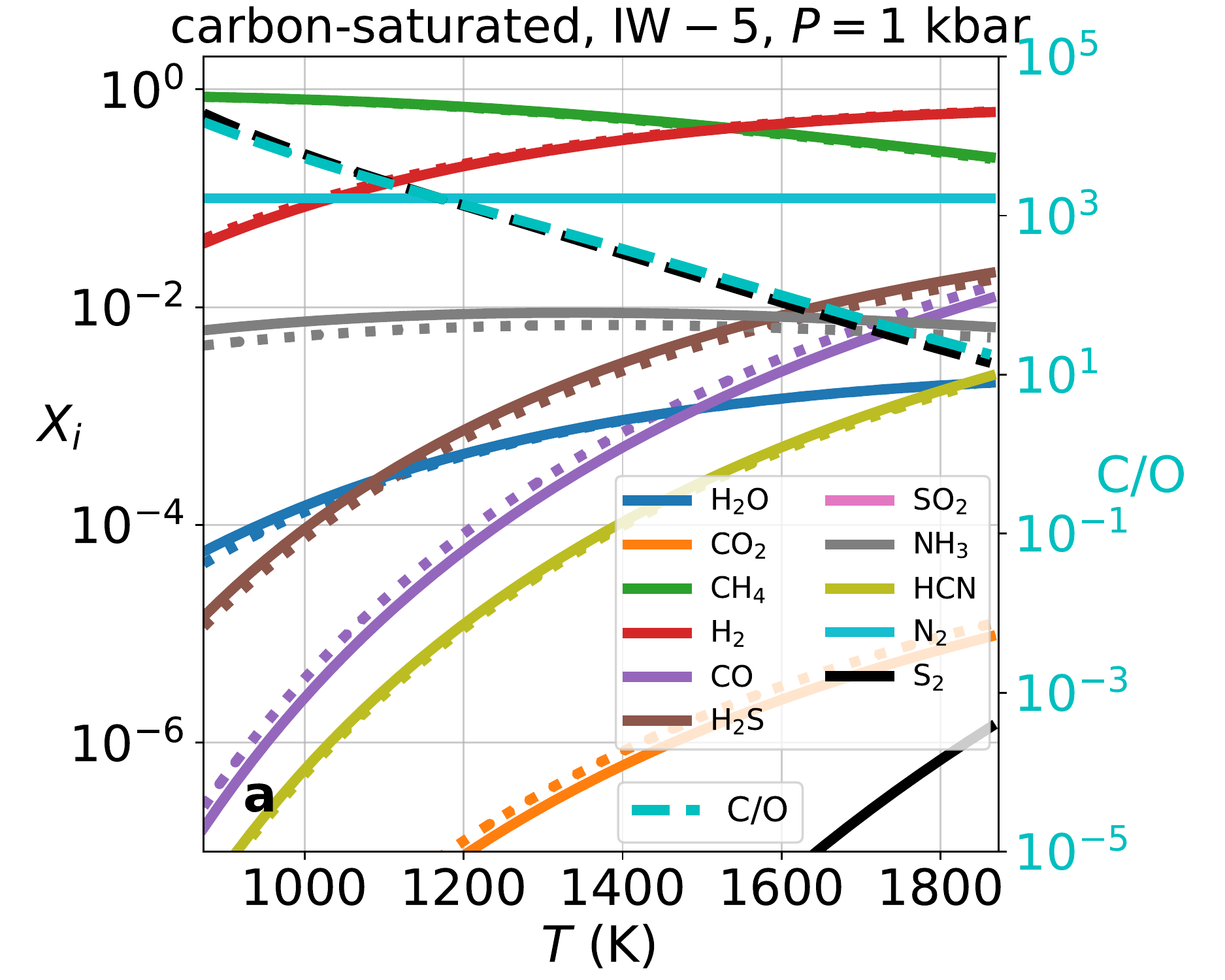}
\includegraphics[width=0.72\columnwidth]{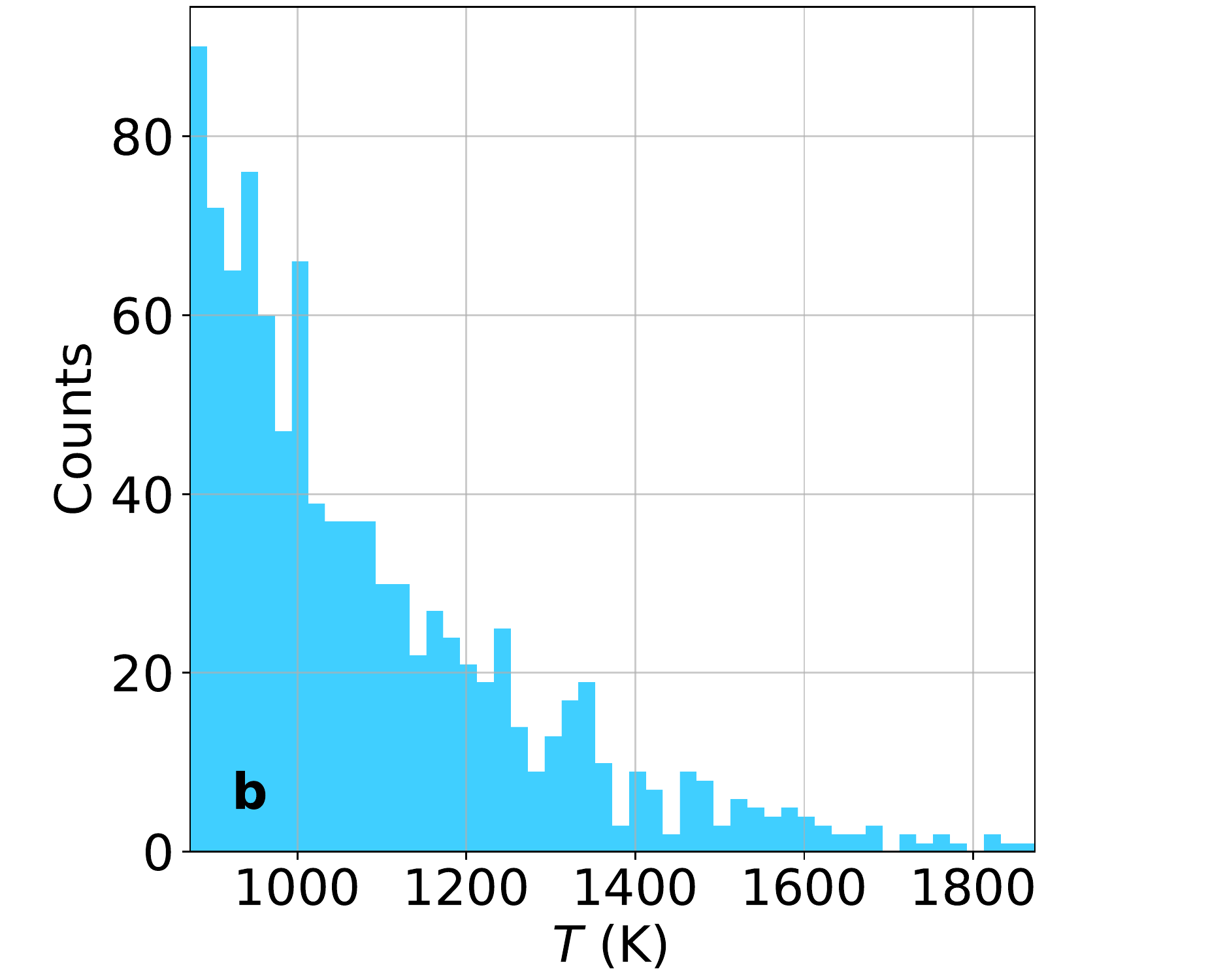}\\
\includegraphics[width=0.72\columnwidth]{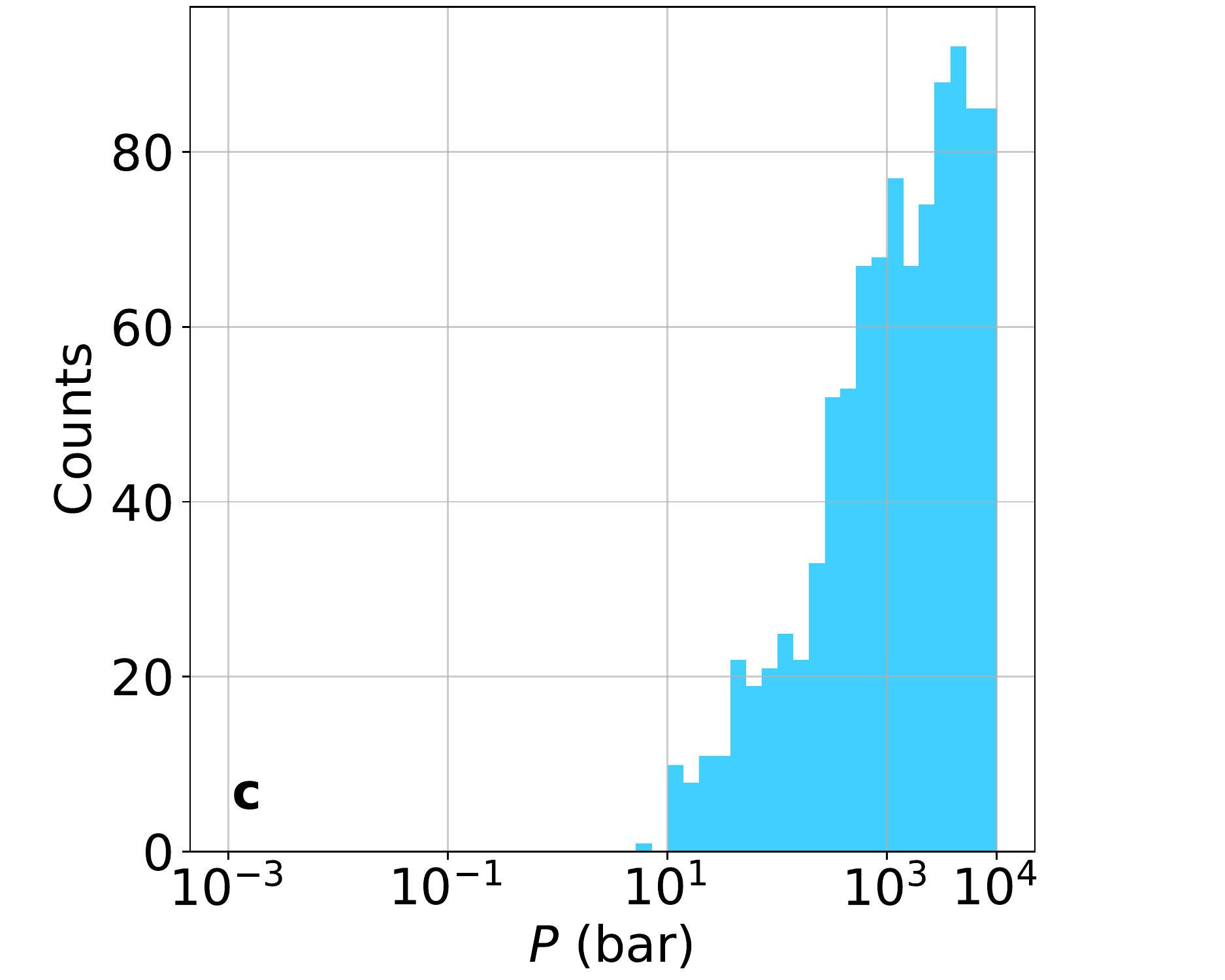}
\includegraphics[width=0.72\columnwidth]{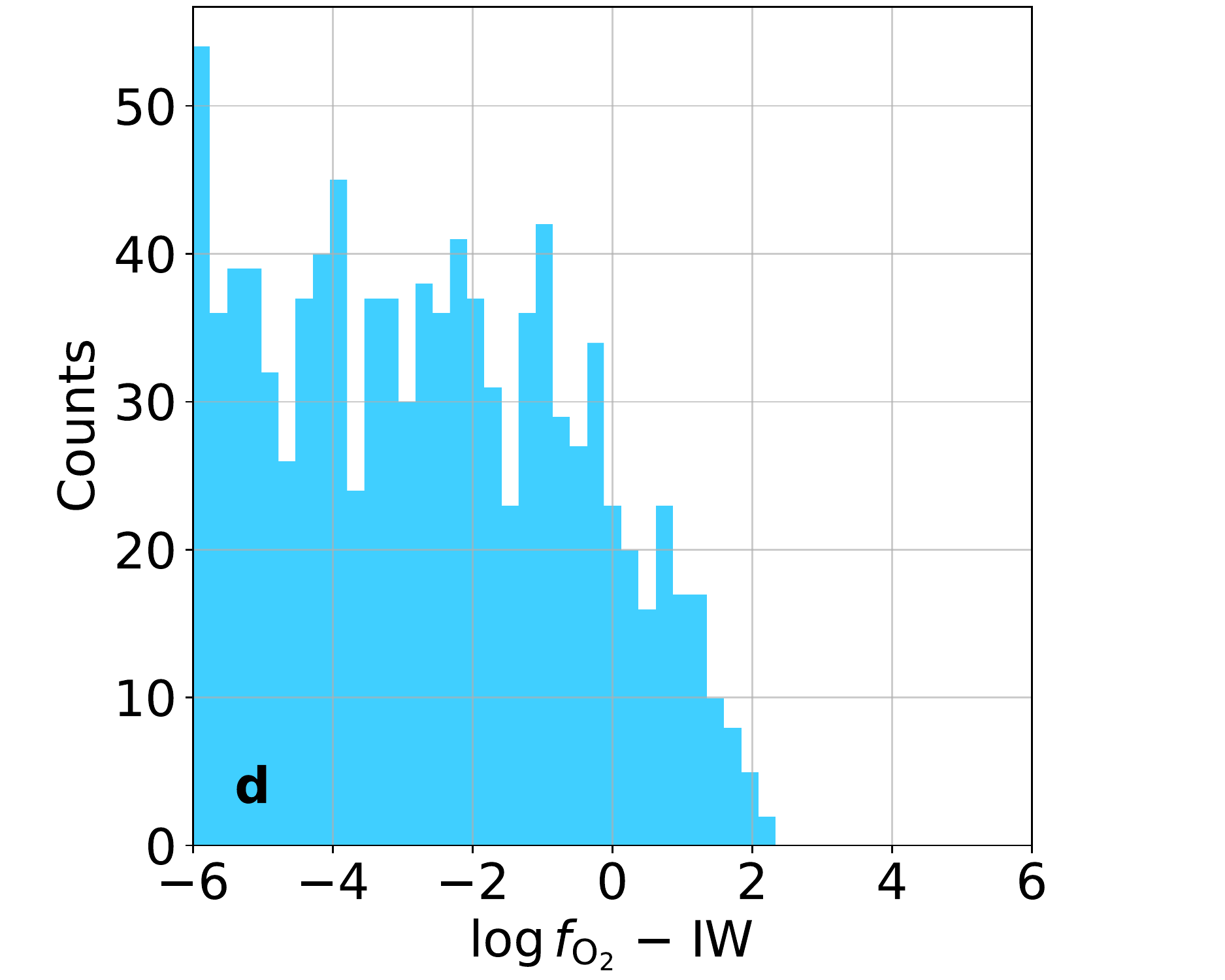}
\end{center}
% \vspace{-0.2in}
\caption{Investigating the occurrence of methane-dominated atmospheres for secondary atmospheres calculated using the C-H-O-N-S system.  The top-left panel shows examples of CH$_4$-dominated atmospheres ($P=1$ kbar, $\log f_{\rm O_2}=\mbox{IW}-5$, $\log f_{\rm S_2}=\mbox{PP}-10$, $a_{\rm C}=1$, $P_{\rm N_2} = 0.1 P$) for different melt temperatures, where the solid and dotted curves correspond to including non-ideal effects and assuming ideal gases, respectively.  The other panels show the distributions of Monte Carlo outcomes of the melt temperature (panel b), atmospheric surface pressure (panel c) and oxygen fugacity of the mantle (panel d) corresponding to methane-dominated atmospheres.}
%\vspace{-0.15in}
\label{fig:methane-dominated_secondary}
\end{figure*}

\begin{figure*}[h!]
\begin{center}
% \vspace{-0.2in} 
\includegraphics[width=0.72\columnwidth]{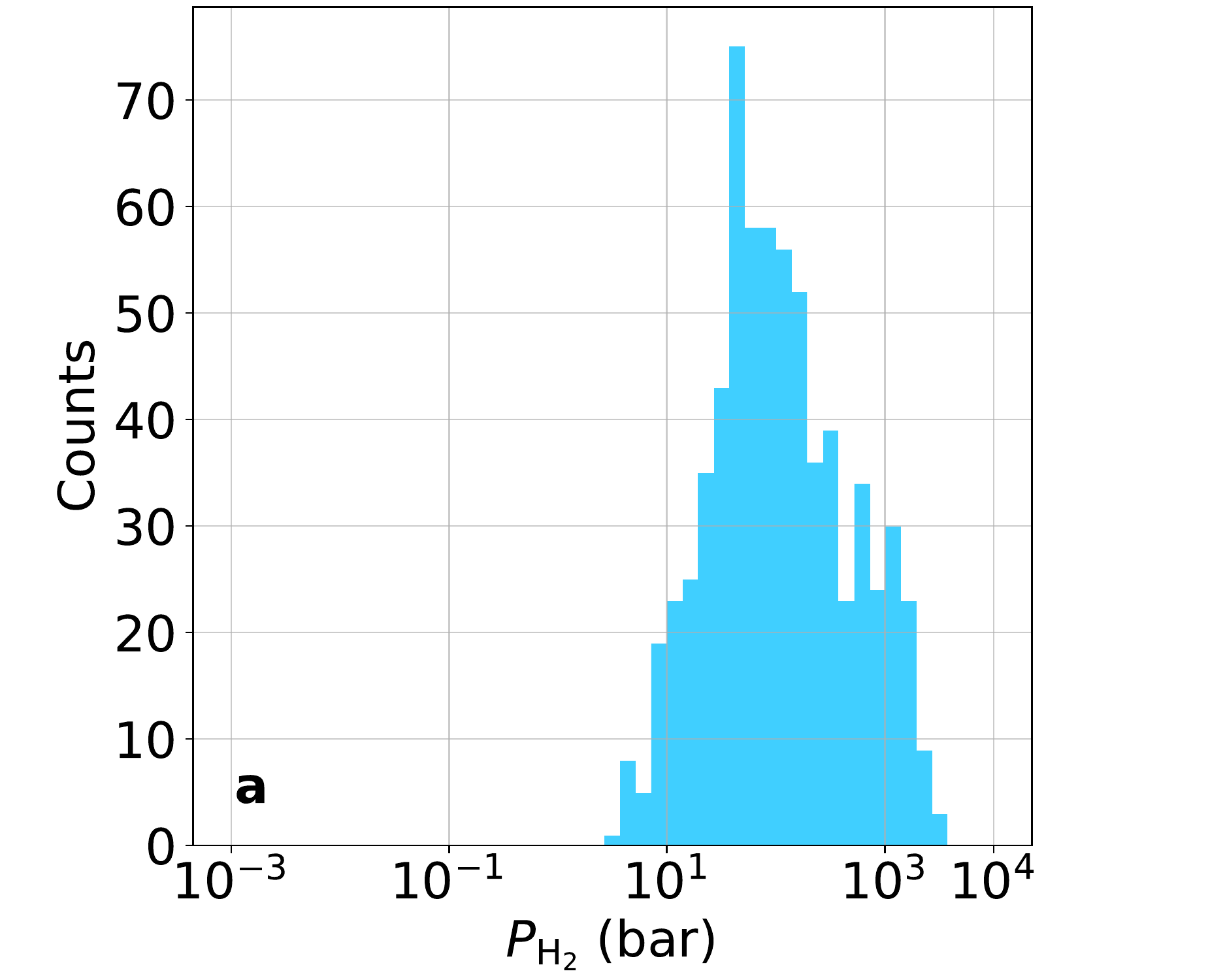}
\includegraphics[width=0.72\columnwidth]{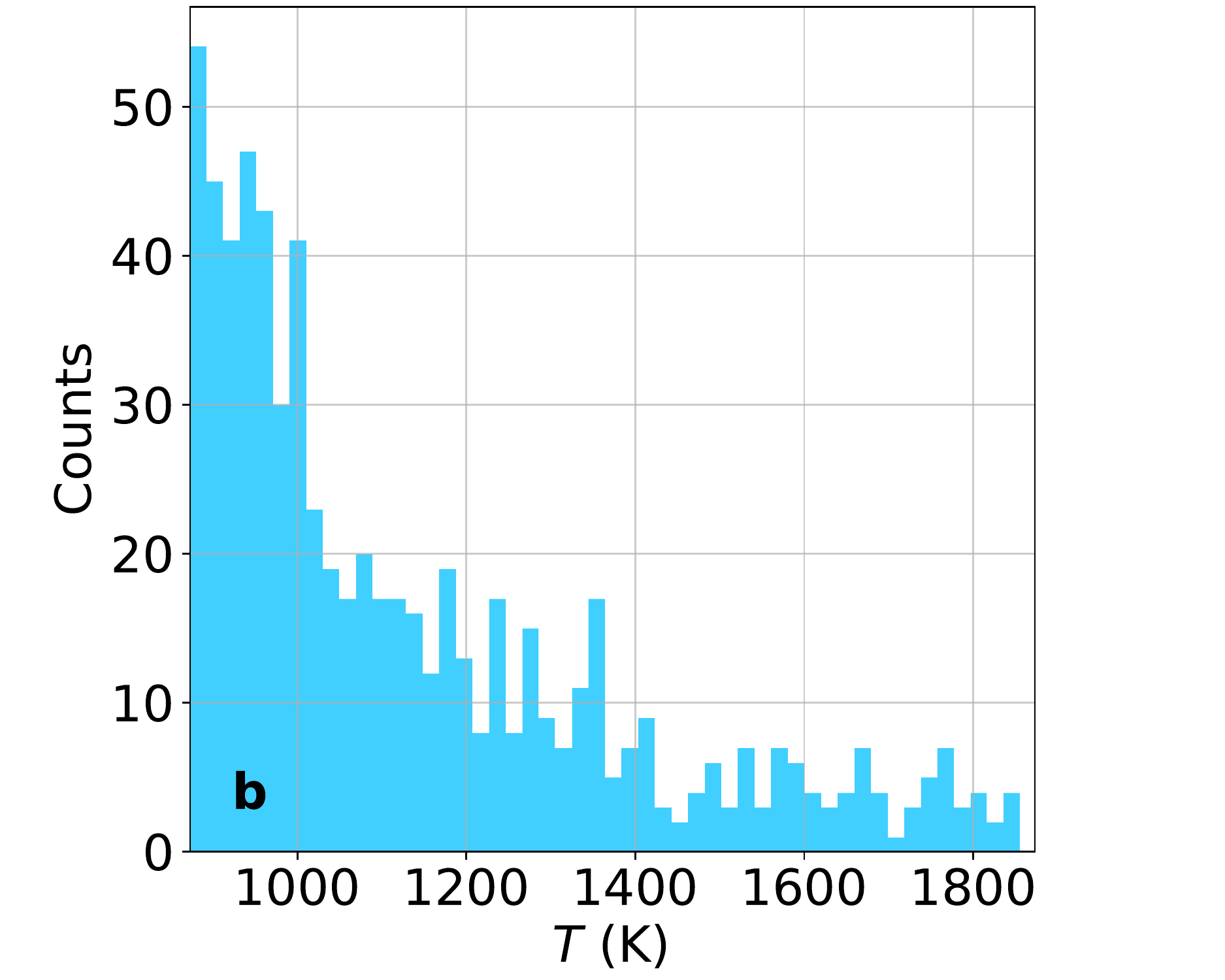}\\
\includegraphics[width=0.72\columnwidth]{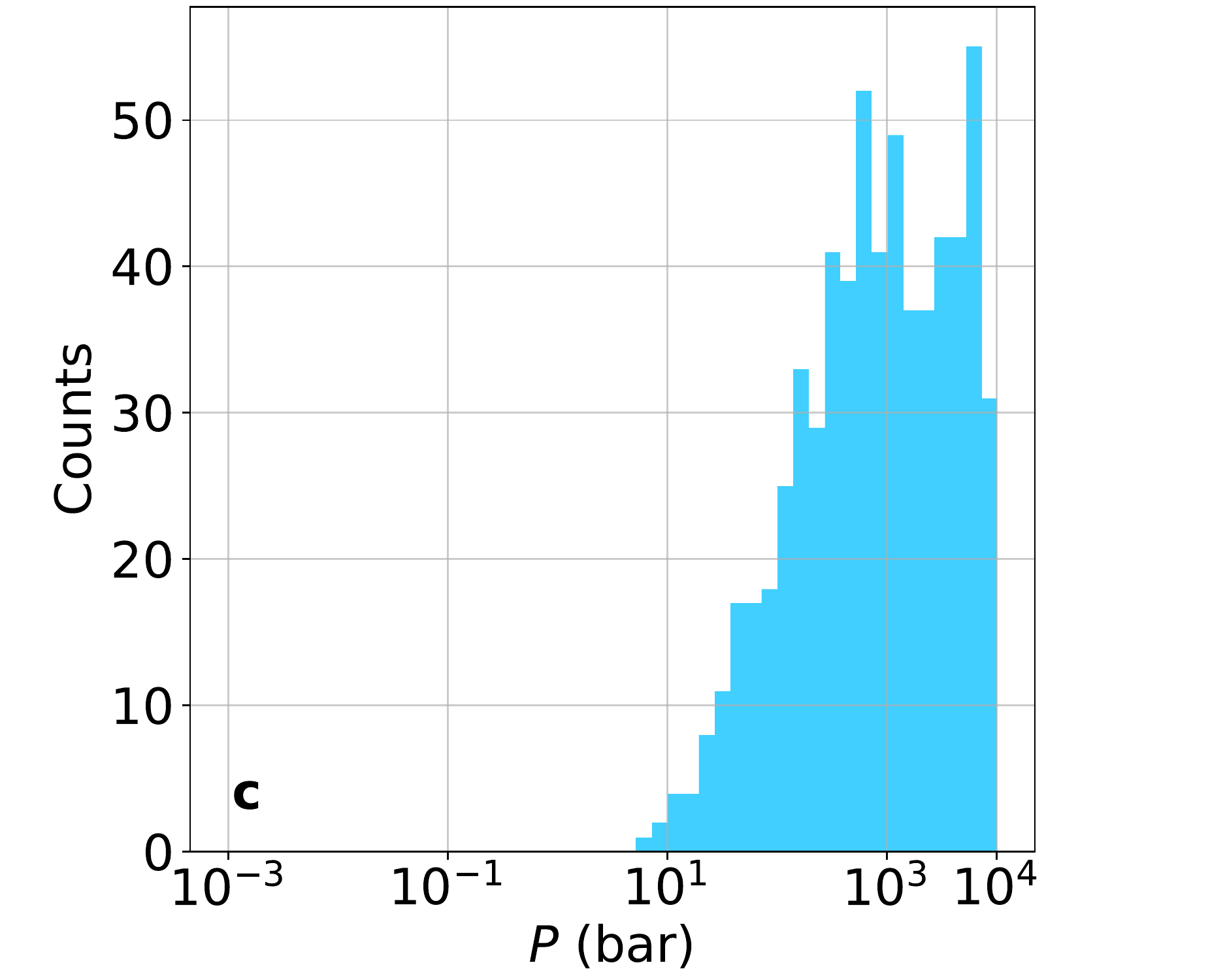}
\includegraphics[width=0.72\columnwidth]{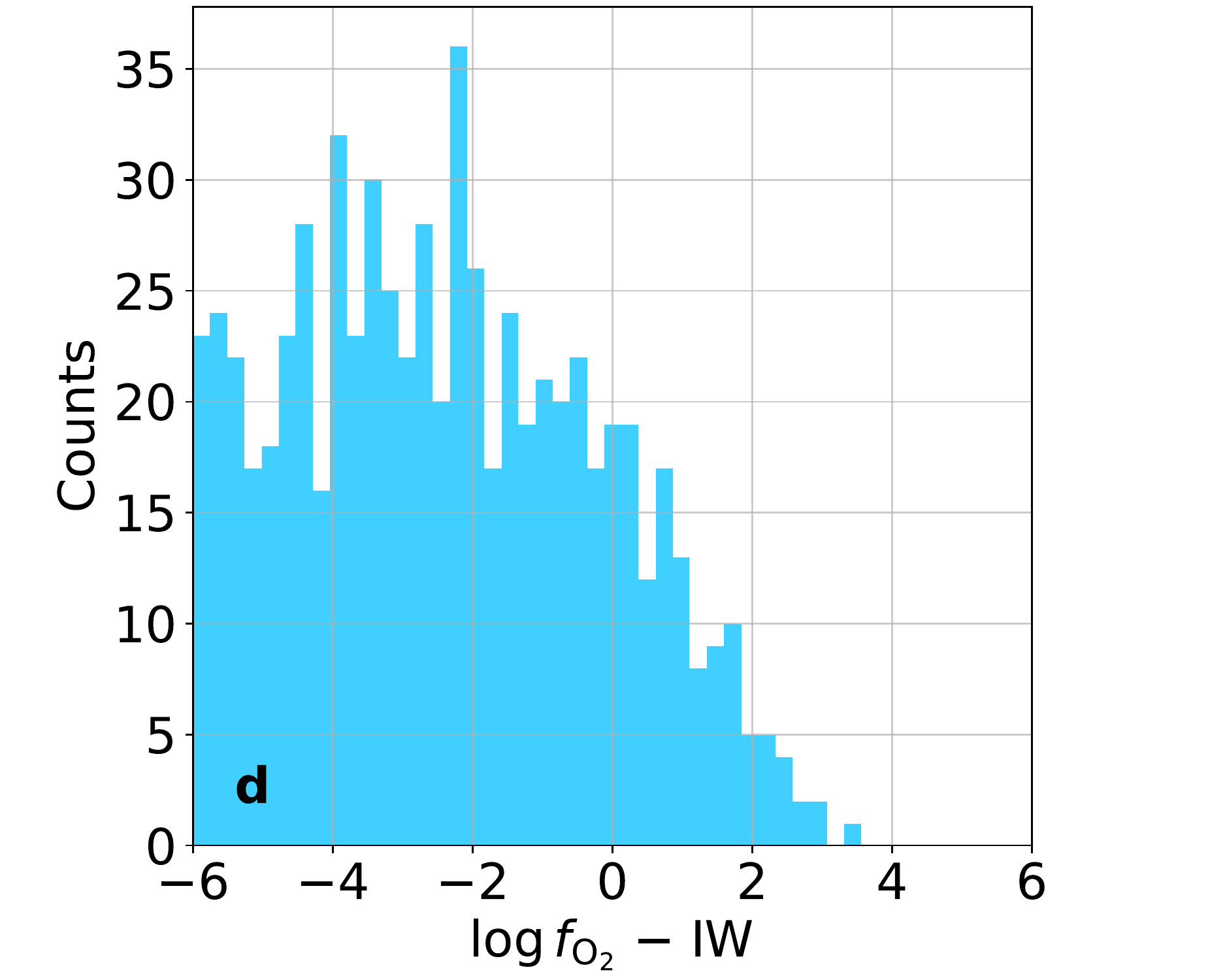}
\end{center}
% \vspace{-0.2in}
\caption{Distributions of Monte Carlo outcomes of the atmospheric partial pressure of molecular hydrogen (panel a), melt temperature (panel b), atmospheric surface pressure (panel c), and oxygen fugacity of the mantle (panel d) corresponding to methane-dominated, hybrid atmospheres computed using the C-H-O-N-S system.}
%\vspace{-0.15in}
\label{fig:methane-dominated_hybrid}
\end{figure*}

In our explorations of secondary and hybrid atmospheres, both within the C-H-O and C-H-O-N-S chemical systems, we noticed that methane-dominated atmospheres are somewhat rare.  From a theoretical standpoint, it seems difficult to make CH$_4$-dominated atmospheres.  The trends elucidated in Figures \ref{fig:CHO_secondary_vary_P}--\ref{fig:CHONS_secondary_vary_PN2} and \ref{fig:CHONS_hybrid_vary_PH2}--\ref{fig:CHONS_hybrid_vary_PN2} suggest that high surface pressures and reduced mantles are needed.  Motivated by these trends, we perform a set of Monte Carlo calculations, where we set $a_{\rm C}=1$ (i.e., graphite in presence) and sample the remaining parameters over the following ranges: $10^{-3} \le P/\mbox{ bar} \le 10^{4}$ (for secondary atmospheres only), $10^{-3} \le P_{\rm H_2}/\mbox{ bar} \le 10^{4}$ (for hybrid atmospheres only), $873 \le T/\mbox{ K} \le 1873$, $\mbox{IW}-6 \le \log f_{\rm O_2} \le \mbox{IW}+6$, $\mbox{PP}-10 \le \log f_{\rm S_2} \le \mbox{PP}$, $10^{-2} \le P_{\rm N_2}/P \le 1$ (for secondary atmospheres only) and $10^{-2} \le P_{\rm N_2}/P_{\rm H_2} \le 1$ (for hybrid atmospheres only).  A methane-dominated atmosphere is identified using the criterion,
\begin{equation}
X_{\rm CH_4} > \sum_i X_i,
\end{equation}
where the summation is performed over all of the molecular species besides methane.

Figures \ref{fig:methane-dominated_secondary} and \ref{fig:methane-dominated_hybrid} demonstrate that the production of methane-dominated atmospheres requires somewhat low melt temperatures ($T \lesssim 1000$ K), reduced mantles ($f_{\rm O_2} \lesssim \mbox{IW}$) and high atmospheric surface pressures exceeding $\sim 10$ bar.  The same calculations performed with C-H-O systems produced essentially the same outcomes (not shown). These rather specific conditions imply that, if a methane-dominated atmosphere is discovered by JWST, some of its atmospheric conditions will be strongly constrained if the methane is indeed sourced by geochemical outgassing \citep{thompson22}.

\section{Discussion}
\label{sect:discussion}

\subsection{Comparison to previous work}

In terms of its scientific intentions, the closest study to compare to is \cite{liggins20}, who used what they termed a ``magma degassing code" to study whether early Earth and Mars could have sustained an atmosphere with non-negligible amounts of molecular hydrogen.  As in the present study, \cite{liggins20} prescribed or parametrised the oxygen fugacity and total pressure in mixed-phase, chemical-equilibrium, carbon-hydrogen-oxygen-sulfur (C-H-O-S) calculations.  Unlike the present study, \cite{liggins20} explicitly considered outgassing fluxes to be balanced by atmospheric escape.  The former is generally difficult to calculate from first principles and \cite{liggins20} parametrised its value relative to that of modern Earth.  Upon specifying the total surface pressure as an input parameter, \cite{liggins20} estimated the atmospheric escape efficiency needed to satisfy the corresponding atmospheric abundance of H$_2$ and the assumed escape flux.  In that study, it remains the case that atmospheric escape involves complex radiative transfer processes and chemistry that are difficult to model from first principles and are instead parametrised by a ``fudge factor".  

\cite{liggins20} concluded that $\sim 1\%$ H$_2$ abundances (by number) is possible for early Earth, but $\sim 10\%$ H$_2$ abundances are unlikely given the plausible space of parameters explored.  For early Mars, about 2--8\% of H$_2$ (by number) is plausible but only if the magmas are water-rich.  \cite{liggins20} did not extend these explorations to exoplanets and their atmospheric chemical diversity.

In terms of methodology, previous studies closest to the current one are \cite{herbort20}, \cite{woitke21} and \cite{ortenzi20}. First, some form of volatile budget constraints has to be prescribed in all three studies, namely, various crustal or chondritic bulk compositions assumed in \cite{herbort20}, a large range of combinations of hydrogen, carbon, oxygen and nitrogen abundances in \cite{woitke21}, and initial H$_2$O and CO$_2$ budget in \cite{ortenzi20}. As reasoned earlier and suggested in Appendix \ref{append:solubility}, volatile inventories and/or bulk compositions on exoplanets are subject to large uncertainties, and assimilating these constraints into phase equilibrium modeling of exoplanetary atmospheres introduces more uncertainties; this is why \cite{woitke21} conducted an exhaustive exploration of C-O-H-N abundances. On the other hand, our approach here circumvents the necessity of bookkeeping elemental abundances, and thus enables transparency and simplicity amenable to analytical insights (e.g., graphite instability). Second, \cite{herbort20} surveyed the temperature range 600--3500 K, and one focus of the study is on the water lockup into hydrous minerals whose solid solution, however, remains unaccounted for. In addition to outgassing, \cite{ortenzi20} also modeled thermal evolution of rocky interiors, which enables simulation of temporal evolution of outgassing \emph{fluxes} for rocky planets with stagnant-lid convection, but their approach is parameterised and methane is completely excluded from the model. The lower limit of the temperature range investigated in \cite{woitke21} is below 600 K which, according to the melt temperature range 873--1873 K discussed in Section \ref{sec:ptrange}, corresponds to non-magmatic outgassing. Given the disparate model details and implementation, the three studies are comparable to the current one with respect to methodology rather than model results.

As for methane-rich atmospheres, \cite{kt18} proposed disequilibrium coexistence of abundant CH$_4$ and CO$_2$ as a potential biosignature for exoplanets, which could be further corroborated by absence of abundant CO. Our results can confirm that abundant CH$_4$ and CO$_2$ in coexistence represents chemical disequilibrium in the melt temperature range of 873--1873 K. However, as shown by \cite{woitke21}, at temperatures below 600 K, outgassing, likely metamorphic, can indeed produce abundant coexisting CH$_4$ and CO$_2$ in chemical equilibrium, rendering simultaneous detection of CH$_4$ and CO$_2$ a false positive biosignature for exoplanets. \cite{thompson22} explored the range of abiotic CH$_4$ \emph{fluxes} in the Solar System, and argues that the short photochemical lifetime of atmospheric CH$_4$ requires replenishing CH$_4$ fluxes higher than those from the abiotic outgassing, thus necessitating biological CH$_4$ fluxes. For an exoplanet, therefore, its planetary context and astrophysical environment needs to be carefully considered to gauge the magnitude of abiotic CH$_4$ fluxes and then the likelihood of CH$_4$ detection as a true biosignature. 
Since the current framework is incapable of computing outgassing fluxes, it is an opportunity for future work. We note that whether methane and its co-existence with abundant carbon dioxide may be regarded as a biosignature remains highly context-dependent.

\subsection{Will telescope data allow us to constrain the properties of secondary and hybrid atmospheres?}

\begin{figure*}
\vspace{-0.2in}
\begin{center}
\includegraphics[width=1.98\columnwidth]{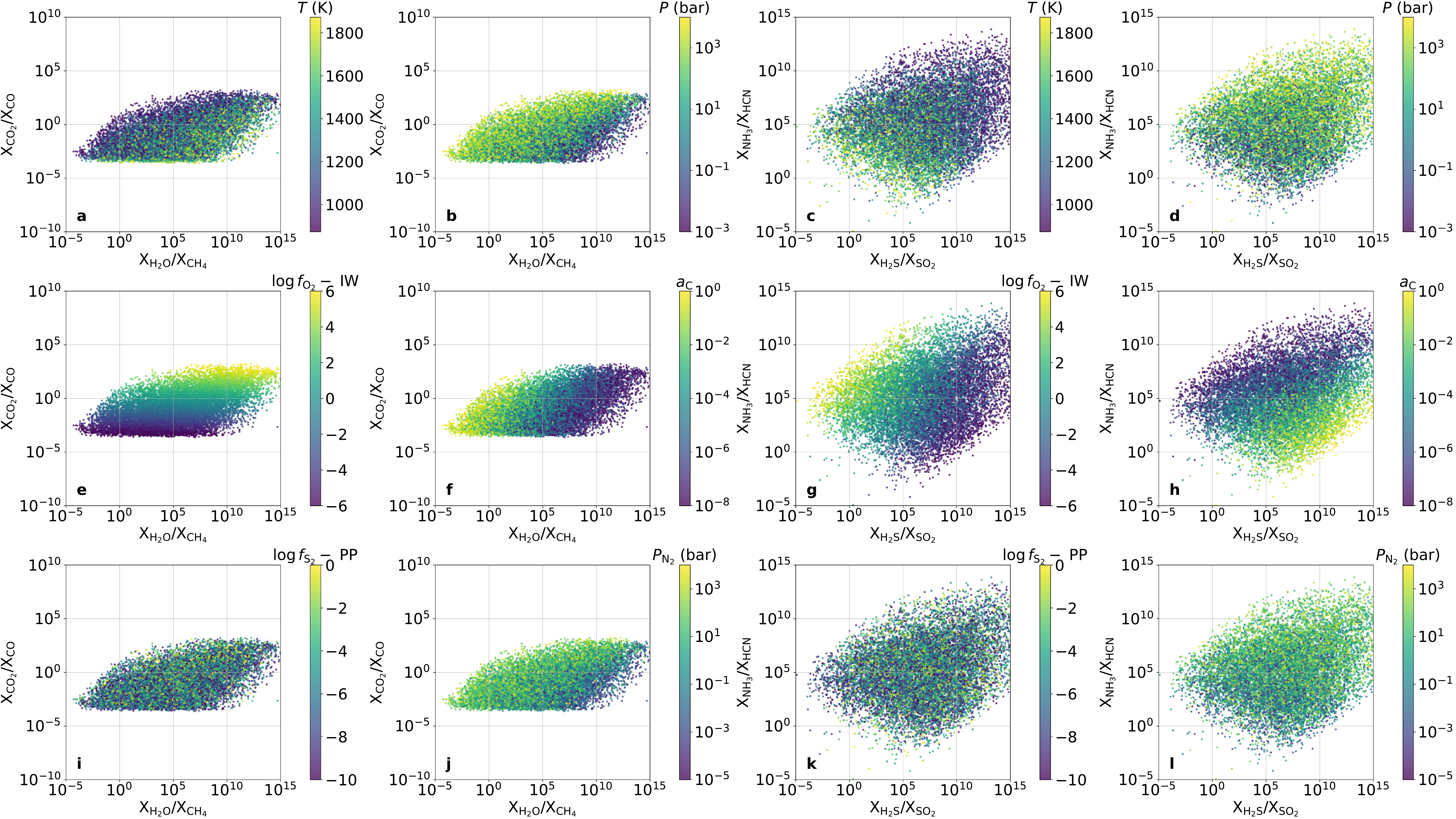}
\end{center}
\vspace{-0.1in}
\caption{Quantifying the relationships between observable ratios of chemical abundances and the various input parameters of the C-H-O-N-S system for secondary atmospheres.  These scatter plots are the outcomes of a suite of Monte Carlo calculations (see text for details on how the parameter values are sampled).}
\label{fig:Monte_Carlo_secondary}

\end{figure*}
\begin{figure*}
\vspace{-0.2in}
\begin{center}
\includegraphics[width=1.98\columnwidth]{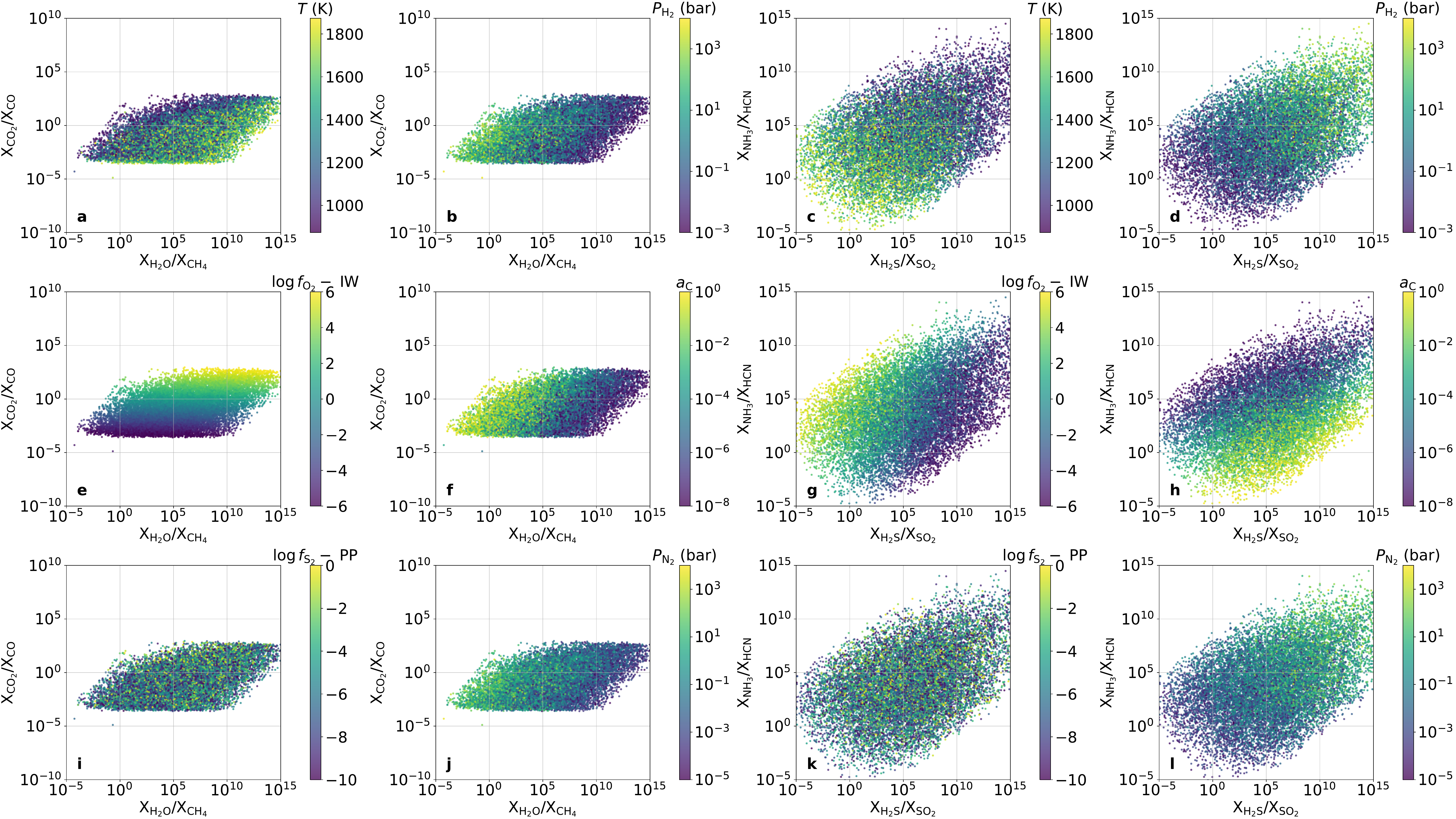}
\end{center}
\vspace{-0.1in}
\caption{Same as Figure \ref{fig:Monte_Carlo_secondary}, but for hybrid atmospheres.}
\label{fig:Monte_Carlo_hybrid}
\end{figure*}

The current study has demonstrated that secondary and hybrid atmospheres are expected to exhibit a rich diversity of chemistries.  Is it possible to constrain some of these input parameters, such as the oxygen fugacity of the mantle, from the measured spectra of these atmospheres?  Generally, the posterior distributions of chemical abundances may be extracted from measured spectra via the technique of Bayesian atmospheric retrieval (see \citealt{bh20} for a recent review).  It is expected that \textit{abundance ratios} may be extracted at a higher precision than absolute abundances.  Therefore, we are motivated to explore the theoretical relationships between the various input parameters and the ratio of abundances of simple molecules.

To accomplish this task, we perform Monte Carlo calculations that provide random realisations of ensembles of C-H-O-N-S chemical models.  We consider secondary and hybrid atmospheres separately.  In the absence of a robust theory for how these atmospheres formed and evolved, we sample each parameter uniformly either in a linear or logarithmic sense.  While it has been described elsewhere in the current study, we state the sampled ranges of values of the parameters here for the convenience of the reader (see Table \ref{tab:mc} for a summary):
\begin{itemize}

    \item Melt temperatures ($T$) are sampled uniformly from 873 to 1873 K.
    
    \item For the surface pressures ($P$) of secondary atmospheres, $\log{(P/{\rm bar})}$ is sampled uniformly from $-3$ to 4 and $\log\left({P_{\rm N_2}/P}\right)$ is sampled uniformly from $-2$ to 0.  For the hydrogen partial pressures ($P_{\rm H_2}$) of hybrid atmospheres, $\log{(P_{\rm H_2}/{\rm bar})}$ is sampled uniformly from -3 to 4 and $\log\left({P_{\rm N_2}/P_{\rm H_2}}\right)$ is sampled uniformly from $-2$ to 0. 
    
    \item For the oxygen fugacity ($f_{\rm O_2}$), $\log{f_{\rm O_2}}$ is sampled uniformly from $\mbox{IW}-6$ to $\mbox{IW}+6$.  In the absence of better knowledge, the sulfur fugacities are sampled in the same way as $\log{f_{\rm S_2}}$ from $\mbox{PP}-10$ to $\mbox{PP}$.
    
    \item The carbon activity ($a_{\rm C}$), which is a proxy for the carbon content of the mantle, is sampled uniformly as $\log{a_{\rm C}}$ from $-8$ to $0$.
    
\end{itemize}

Figures \ref{fig:Monte_Carlo_secondary} and \ref{fig:Monte_Carlo_hybrid} shows the outcomes of these Monte Carlo calculations for secondary and hybrid atmospheres, respectively.  As motivated by our findings reported earlier in the current study, the abundance ratio of carbon dioxide to carbon monoxide ($X_{\rm CO_2}/X_{\rm CO}$) is chosen as a diagnostic for the oxidation state of the mantle (oxygen fugacity); the abundance ratio of hydrogen sulfide to sulfur dioxide ($X_{\rm H_2S}/X_{\rm SO_2}$) is chosen as a supporting diagnostic of the oxidation state.  Since water and methane are expected to be readily detectable, their abundance ratio ($X_{\rm H_2O}/X_{\rm CH_4}$) is chosen as another observational diagnostic.  The abundance ratio of ammonia to hydrogen cyanide ($X_{\rm NH_3}/X_{\rm HCN}$), which are two important nitrogen carriers, complete the set of 4 observational diagnostics.

As expected, a combination of $X_{\rm CO_2}/X_{\rm CO}$ and $X_{\rm H_2S}/X_{\rm SO_2}$ potentially sets a strong constraint on the oxygen fugacity of the mantle of a rocky exoplanet.  Similarly, the carbon content (carbon activity) and melt temperature may be constrained using a combination of all 4 abundance ratios.  For secondary atmospheres, it is perhaps unsurprising that the atmospheric surface pressure cannot be constrained from measuring these abundance ratios.  However, the hydrogen content of hybrid atmospheres appears to be sensitive to these 4 abundance ratios.  For both types of atmospheres, it appears that the sulfur fugacity and nitrogen content are somewhat challenging to constrain from measuring these 4 abundance ratios.

It is worth noting that, in Cycle 1 of the JWST Guest Observer program\footnote{\scriptsize \texttt{https://www.stsci.edu/jwst/science-execution/} \\ \texttt{approved-programs/cycle-1-go}} alone, there are approved proposals targeting about 10 small exoplanets.  Therefore, it is conceivable that the mantle oxygen fugacities of some small exoplanets may be constrained in the near future.

\subsection{Are we thinking about the chemistry of rocky exoplanets in a physically realistic way?}

In the study of gas-giant exoplanets, astronomers often use the C/O ratio and the ``metallicity" as the control parameters.  The study of C/O ratio is motivated by its varying value with distance from the star within a protoplanetary disk, due to the differing condensation temperatures of simple molecules (so-called ``ice lines" or ``snow lines"; \citealt{oberg11}).  By measuring the C/O ratio of a gas-giant exoplanet, the hope is that one may then derive its original site of formation within a protoplanetary disk (see \citealt{oberg11} for caveats).   The ``metallicity" assumes that the entire set of elemental abundances, for both refractory and volatile elements, have ratios that are locked to their solar values.  Only in this way may a set of elemental abundances be described by a single number.  Motivated by an empirical trend in the Solar System, previous studies have claimed an empirical relationship between the metallicities and the masses of exoplanets (e.g., \citealt{wakeford17}), but this assumes that the derived abundances of water translate into oxygen abundances that scale with the abundances of refractory elements in a straightforward way \citep{heng18}.

Since secondary and hybrid atmospheres are at least partially sourced by geochemical outgassing, their chemistries do not straightforwardly trace their formation histories unlike in the case of gas-giant exoplanets.  This implies that the C/O ratio has less relevance for rocky exoplanets.  When the atmosphere is sourced by outgassing, the volatile and refractory elements are partitioned between their gaseous forms (atmosphere), the melt (which the gases may dissolve into) and rocks (which are composed of a mixture of minerals).  This implies that it will be complicated, if not impossible (from the viewpoint of interpreting astronomical data), to decipher the relative abundances of volatile and refractory elements, thus rendering the simplistic concept of metallicity (as defined in the preceding paragraph) suspect.  

Therefore, we suggest that a better way of thinking about rocky exoplanets is in terms of the oxygen fugacity and carbon content of their mantles.  In the Solar System, an empirical trend exists such that more massive bodies tend to have more oxidised mantles (e.g., \citealt{wadhwa08, cw19}).  Our current work suggests that there is a clear path towards deriving the oxygen fugacities ($f_{\rm O_2}$) of the rocky mantles of exoplanets by inferring accurate abundances of CO$_2$ and CO via atmospheric retrieval performed on high-quality spectra (by, for example, the JWST), but future work needs to correct for the effects of photochemistry.  An exciting prospect for exoplanet science will be to quantify the relationship between $f_{\rm O_2}$ and the masses of exoplanets for a sample of super Earths and sub-Neptunes.

\subsection{Opportunities for future work}

The current study is the first to consider secondary and hybrid atmospheres within the same theoretical framework.  But it will certainly not be the last, as there are many other processes to explore and other members of the model hierarchy of geochemical outgassing to construct and study.  The models constructed in the current study are fundamentally zero-dimensional (0D), meaning that the temperature-pressure profile of the mantle and atmosphere are not considered.  These thermal gradients could lead to chemical abundance gradients and processes such as cold traps.  Extending these calculations beyond 0D requires coupling the chemical models to radiative transfer calculations that consider relevant dynamical processes such as convection and large-scale atmospheric circulation.  Another process we have neglected in the current study is the general solubility of gases in melts.

The full or ideal model would quantify the joint evolution of the interior, mantle and atmosphere of the exoplanet, as well as the radiative influence of its host star.  Its chemical geodynamics would be elucidated such that one could calculate the oxygen fugacity and carbon content of the mantle from first principles, rather than parametrise them by single numbers.  In the current study, we have parametrised the carbon content via the carbon activity, from which it is difficult to directly compute the carbon content by mass.  The atmospheric surface pressure and H$_2$ partial pressure should be outcomes of a calculation balancing the outgassing flux and atmospheric escape (e.g., \citealt{liggins20}).  The latter process involves performing radiative transfer calculations alongside ionic chemistry in parts of the atmosphere where the fluid approximation no longer holds.  In essence, we have strived for simplicity over sophistication in the current study by choosing to parametrise these formidable processes using two numbers ($f_{\rm O_2}$ and $P$ or $P_{\rm H_2}$).

%It is possible to convince ourselves that the hydrogen-dominated atmospheres (both secondary and hybrid) of rocky exoplanets are not only chemically plausible, but likely to survive on geological timescales.  At the order-of-magnitude level, the retention of a H$_2$-dominated atmosphere is possible if the thermal velocity of the gas is less than its escape velocity.  Ignoring factors of order unity, we obtain
%\begin{equation}
%T < 1.8 \times 10^4 \mbox{ K} ~\left( \frac{M}{M_\oplus} \right)^{0.72},
%\end{equation}
%if we use the empirically-calibrated mass-radius relationship for rocky exoplanets derived by \cite{ck17}.  The temperature should be interpreted as the temperature at the exobase, where hydrogen transitions from being collisional to collisionless.  The temperature threshold of $\sim 10^4$ K is an order of magnitude higher than the estimated exobase temperatures for Earth, Venus and Mars \citep{sa96}, implying that diffusive Jeans escape will not be the dominant mechanism on mature exoplanets.  While the exobase temperatures of exoplanets are generally unknown, it gives us some confidence on the long-term survival of hydrogen-dominated atmospheres on some rocky exoplanets.  Future work should focus, on a case-by-case basis, on the details of the radiative transfer that will determine if long-term survival is indeed plausible.

%\subsection{Open-source outgassing code for the exoplanet community?}

\acknowledgments

%\vspace{0.3in}
\textit{\scriptsize We acknowledge partial financial support from the Chair of Theoretical Astrophysics of Extrasolar Planets at the LMU-Munich and the European Research Council (ERC) via a 2018 Consolidator Grant awarded to KH (number: 771620; acronym: EXOKLEIN).  KH is especially grateful to Ralf Bender, Gudrun Niggl and Daniel Gr\"{u}n for their collegial support during his transition from the University of Bern to the LMU.}

\vspace{0.1in}
\textit{\scriptsize MT and KH jointly designed the current study over dozens of discussions.  MT clarified the concepts and derivations associated with thermodynamics, introduced the relevant geochemical literature to KH, curated the thermodynamic data, wrote the computer code necessary to perform the calculations, produced all of the figures and co-wrote the paper.  KH led the writing and structuring of the paper based on these conversations, introduced the relevant exoplanet literature to MT, derived the generalised outgassing model equations and provided general scientific direction for the paper.}

\clearpage
\appendix
\section{Illustrative model of including gas solubility in the H-O subsystem}
\label{append:solubility}
The purpose of this appendix is twofold: to illustrate that our current modeling framework is easily extensible to include gas dissolution into melts, and to discuss the effect of gas dissolution on atmosphere chemistry using a simple model. The necessity of taking the hierarchical modeling approach will also be revealed toward the end of this appendix.

\subsection{Dissolution-free case} \label{sec:b-dissol}
In order to find analytical solutions that help build our intuitive understanding, we consider the simplest possible system comprising only two elements: hydrogen (H) and oxygen (O), and only three molecular species, H$_2$, O$_2$, and H$_2$O. The linear algebra analysis method in \cite{powell98} tells us that there exists only one \emph{independent} chemical reaction for this subsystem:
\begin{equation} \label{eq:h2orxn}
    \mbox{H}_2 + 0.5 \mbox{O}_2 \leftrightarrows \mbox{H}_2 \mbox{O}.
\end{equation}

Aiming for analytical solutions, we first ignore non-ideality and thus all fugacity quantities become equivalent to pressure quantities, i.e., $f_i = P_i$. Writing out the equilibrium constant for reaction \eqref{eq:h2orxn}:
\begin{equation} \label{eq:K}
    K = \frac{P_{\rm H_2O}}{P_{\rm H_2} f_{\rm O_2}^{\frac{1}{2}}},
\end{equation}
where the equilibrium constant can be calculated in the way in main text:
\begin{equation} \label{eq:KCalc}
    K = \exp{\left( \frac{-\Delta_{r, A1} G}{RT} \right)}.
\end{equation}

One nice property of this simple H-O system is that, in equation~\eqref{eq:K}, $P_{\rm H_2O}$ and $P_{\rm H_2}$ have the same exponent of $1$, which makes it convenient to convert equation~\eqref{eq:K} to an expression of volume mixing ratios:
\begin{equation} \label{eq:KX}
    K = \frac{P_{\rm H_2O}}{P_{\rm H_2} f_{\rm O_2}^{\frac{1}{2}}} = \frac{\left( P_{\rm H_2O}/P \right)}{\left( P_{\rm H_2}/P \right) f_{\rm O_2}^{\frac{1}{2}}} = \frac{X_{\rm H_2O}}{X_{\rm H_2} f_{\rm O_2}^{\frac{1}{2}}},
\end{equation}
where $P$ is the total gas pressure of the subsystem.

Another nice property of reaction \eqref{eq:h2orxn} is that it involves only gaseous species, and thus $\Delta_{r, A1} G$ and $K$ are only temperature dependent, i.e., they don't depend on total pressure $P$ (see the discussion in Section \ref{subsect:french1966}). We can thus write $K = K(T)$. Rearranging equation~\eqref{eq:KX} and considering unity sum lead to:
\begin{equation} \label{eq:sys}
  \begin{cases}
    \frac{X_{\rm H_2O}}{X_{\rm H_2}} = K(T) f_{\rm O_2}^{\frac{1}{2}}, & \\
    X_{\rm H_2O} + X_{\rm H_2} = 1, & \text{unity sum} .
  \end{cases}
\end{equation}
Equation~\eqref{eq:sys} suggests that as along as temperature and oxygen fugacity ($T$ and $f_{\rm O_2}$) are given, mixing ratios of H$_2$ and H$_2$O are fully determined for the simple H-O system. In other words, the solution of mixing ratios is pressure-independent; the simplicity of this subsystem allows us to peel off all pressure variables in the system and thus reduce its degree of freedom by one. In spite of this, to compute partial pressures $P_{\rm H_2O}$ and $P_{\rm H_2}$, we still need a total pressure $P$ such that $P_{\rm H_2O}=X_{\rm H_2O} P$ and $P_{\rm H_2}=X_{\rm H_2} P$. This total pressure could be derived from volatile budget constraints (see below).

\subsection{Dissolution} \label{sec:dissol}
Now we start considering gas solubilities in the H-O system. At the outset, when accounting for volatile dissolution into molten rocks, solubility laws are generally needed that dictate how volatiles are partitioned between the atmosphere reservoir and the molten silicate reservoir. Naturally, total volatile budgets need also to be prescribed that are to be thermodynamically partitioned between the two reservoirs. With these two more constraints, one can \emph{solve for} how much mass of each volatile are distributed among the two reservoirs. For the atmosphere reservoir, if the surface area ($A$) and gravity ($g$) of a planet are further known or assumed, one can convert volatile mass $M_{i{\rm,atm}}$ in the atmosphere to its partial pressure $P_i$ by $P_i = M_{i{\rm,atm}} g/A$ (e.g., \citealt{ortenzi20, bower22}).

\subsubsection{Hydrogen budget}

For the simple H-O system, no oxygen budget needs to be provided because the silicate melt can be regarded as an infinite oxygen reservoir (e.g., \citealt{gs14, bower22, gaillard22}), hence only a hydrogen budget is needed, which can be expressed in terms of total moles of H$_2$. An Earth ocean contains about $1.4 \times 10^{21}$ kg of water \citep{hamano13}, which is equivalent to $7.8\times 10^{22}$ mol H$_2$. For illustration, we will assume an Earth's ocean of hydrogen as unit of hydrogen budget in this model (e.g., \citealt{bower22})

\subsubsection{Solubility laws}
With silicate melt as infinite oxygen reservoir and the negligible dissolution of molecular hydrogen \citep{bower22}, only the dissolution of water into melt needs to be considered through its solubility law:
\begin{equation}
\label{eq:sol-law}
    X_{\rm H_2O} = \alpha P_{\rm H_2O}^{0.5}. 
\end{equation}
where $X_{\rm H_2O}$ is the mass fraction of H$_2$O in the melts, and $\alpha = 534$ ppmw/bar$^{0.5}$ is experimentally determined for peridotitic melts \citep{sossi23}.

\subsubsection{Mass balance}
Accompanying the two constraints from hydrogen budget and water solubility comes one more equation of hydrogen mass balance:
\begin{equation} \label{eq:mass-mol}
    \frac{M_{\rm H_2O, melt}}{m_{\rm H_2O}} + \frac{M_{\rm H_2O, atm}}{m_{\rm H_2O}} + \frac{M_{\rm H_2, atm}}{m_{\rm H_2}}= \mathcal{M}_{\rm H_2, total\; in\; mol},
\end{equation}
where $m_i$ is the molecular weight of species $i$ and $\mathcal{M}_{\rm H_2, total\; in\; mol}$ is the prescribed hydrogen budget. The first term is the moles of H$_2$ contained in H$_2$O dissolved in the melt, the second is the amount of H$_2$ contained in H$_2$O in the atmosphere, and the third term is the H$_2$ amount in the atmosphere; the H$_2$ dissolved in the melt is ignored due to its low solubility.

First, according to the simple relation $P_i = M_{i{\rm,atm}} g/A$, we can replace $M_{i{\rm,atm}}$ with $P_i$-bearing terms:
\begin{equation} \label{eq:mass-mol}
    \frac{M_{\rm H_2O, melt}}{m_{\rm H_2O}} + 10^5 \times \frac{P_{\rm H_2O}A}{g m_{\rm H_2O}} + 10^5 \times \frac{P_{\rm H_2} A}{g m_{\rm H_2}}= \mathcal{M}_{\rm H_2, total\; in\; mol},
\end{equation}
where the pre-factor $10^5$ arises from unit conversion of partial pressures from bar to Pa.

Second, the solubility law for H$_2$O can be substituted:
\begin{equation} \label{eq:mass-pressure}
    \frac{M_m \alpha P_{\rm H_2O}^{0.5} }{m_{\rm H_2O}} + 10^5 \times \frac{P_{\rm H_2O}A}{g m_{\rm H_2O}} + 10^5 \times \frac{P_{\rm H_2} A}{g m_{\rm H_2}}= \mathcal{M}_{\rm H_2, total\; in\; mol},
\end{equation}
where $M_m$ is the total mass of melt (molten rocks).

Third, chemical equilibrium relation $P_{\rm H_2O} = K(T) P_{\rm H_2} f_{\rm O_2}^{\frac{1}{2}}$ from equation~\eqref{eq:K} can be inserted:
\begin{equation} \label{eq:mass-ph2}
    \frac{M_m \alpha K^{\frac{1}{2}} f_{\rm O_2}^{\frac{1}{4}} }{m_{\rm H_2O}} P_{\rm H_2}^{\frac{1}{2}} + 10^5 \times \frac{K f_{\rm O_2}^{\frac{1}{2}} A}{g m_{\rm H_2O}} P_{\rm H_2} + 10^5 \times \frac{A}{g m_{\rm H_2}} P_{\rm H_2} = \mathcal{M}_{\rm H_2, total\; in\; mol},
\end{equation}
which, if letting $x = P_{\rm H_2}^{\frac{1}{2}}$, becomes a quadratic equation in $x$.

\subsubsection{Solution method}
\begin{figure}[h!]
\begin{center}
%\vspace{-0.2in} 
\includegraphics[width=0.45\columnwidth]{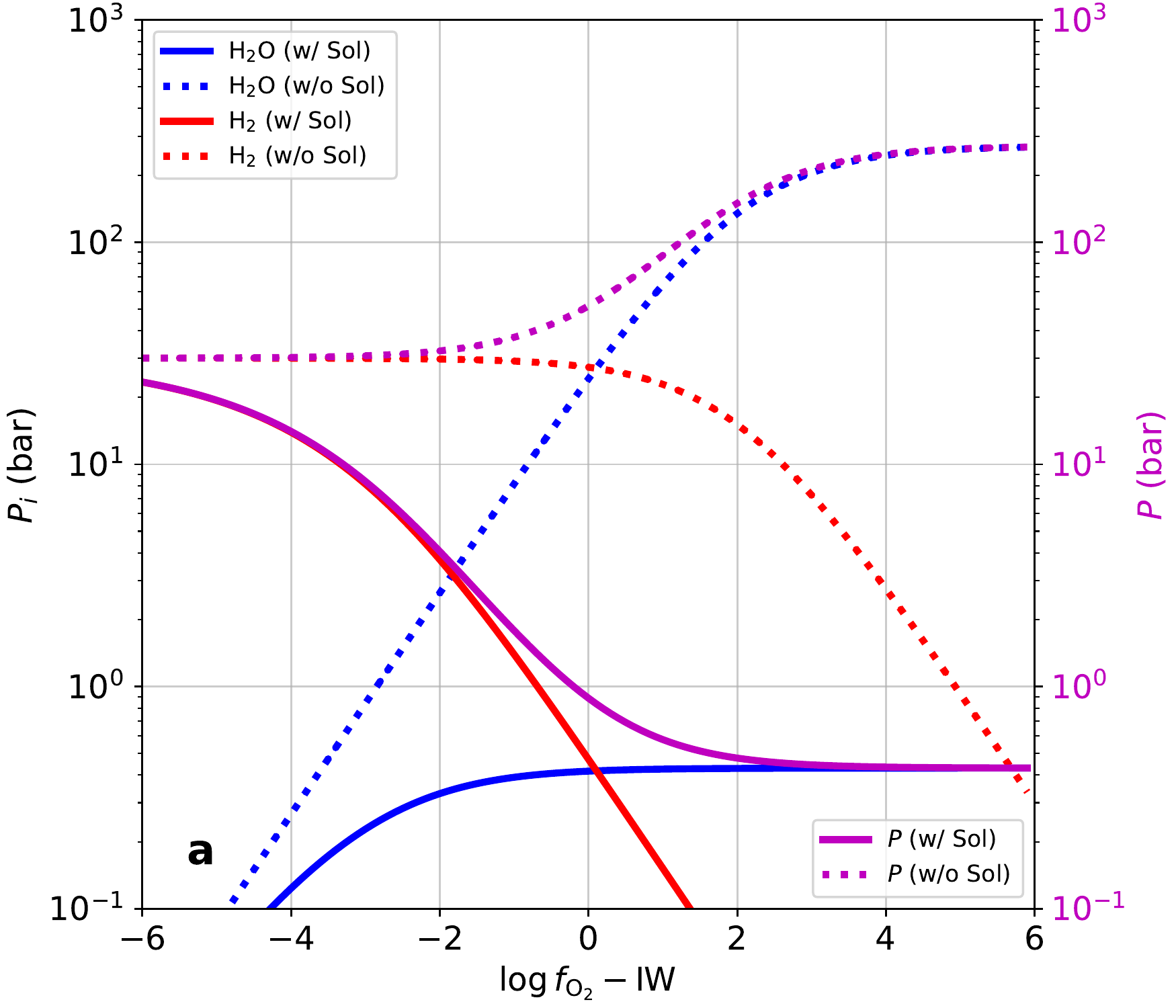}
\includegraphics[width=0.45\columnwidth]{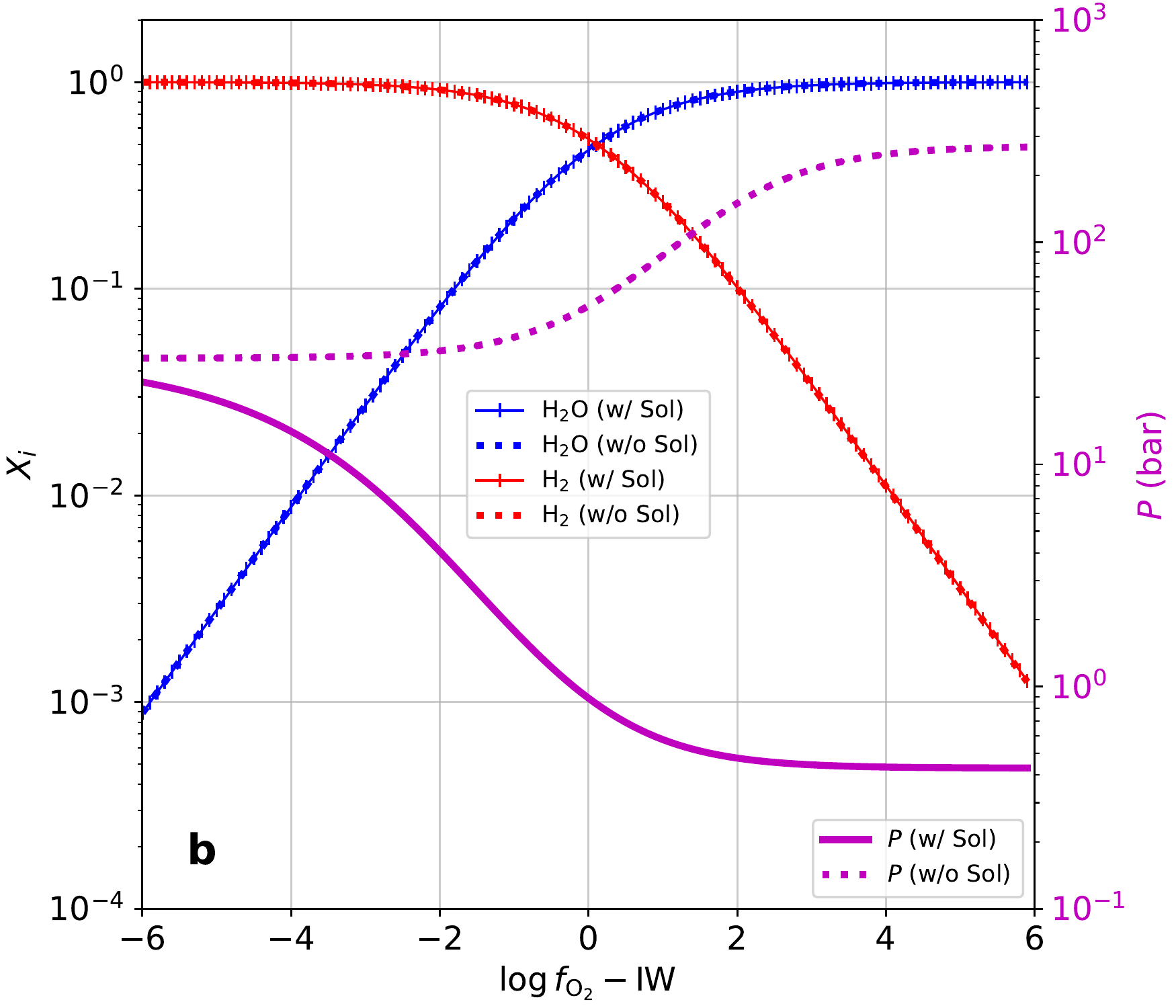}
\end{center}
%\vspace{-0.2in}
\caption{Outgassed atmosphere composition, partial and total pressures for the simple H-O system with $T=1600$ K at the atmosphere-melt interface. (a) Variation of H$_2$O and H$_2$ partial pressures with oxygen fugacity. Left y-axis is for partial pressures $P_i$ and right y-axis is for total pressure $P$. Dotted curves correspond to dissolution-free ($\alpha = 0$) cases, whereas solid curves correspond to cases that consider H$_2$O dissolution ($\alpha \neq 0$). (b) Same results as in (a), but replotted to show the variation of mixing ratios ($X_i$) with oxygen fugacity. Note that cases with and without gas dissolution share the same trend of volume mixing ratios, despite disparate trends of total pressure (magenta curves).}
%\vspace{-0.15in}
\label{fig:app-vary-fo2-fix-T}
\end{figure}

The preceding derivation suggests that equation~\eqref{eq:mass-ph2} accounts altogether for chemical speciation \eqref{eq:h2orxn}, gas dissolution \eqref{eq:sol-law} and total hydrogen budget of the simple H-O system. With a planet radius $R_p$ and density $\rho$, the surface area $A = 4 \pi R_p^2$, planet mass $M_p = \rho (4/3) \pi R_p^3$ and surface gravity $g=GM_p/R_p^2$ can be obtained. The total melt mass is $M_m = X_m M_p$ if the mass fraction of molten rocks is denoted by $X_m$. For simplicity and as proof of concept, we take Earth values for $R_p$ and $\rho$, and assume a whole-mantle magma ocean leading to $X_m \approx 0.67$. Once a total H$_2$ budget is further prescribed, $P_{\rm H_2}$ is solvable from equation~\eqref{eq:mass-ph2} at given temperature and oxygen fugacity values. A representative temperature $T=1600$ K, a hydrogen budget of one Earth ocean ($7.8\times 10^{22}$ mol H$_2$), and a range of oxygen fugacity values are selected to observe model trend (Figure~\ref{fig:app-vary-fo2-fix-T}). With solved $P_{\rm H_2}$, water partial pressure $P_{\rm H_2O}$ and total pressure $P$ are further computed through $P_{\rm H_2O} = K(T) P_{\rm H_2} f_{\rm O_2}^{\frac{1}{2}}$ and $P = P_{\rm H_2}+P_{\rm H_2O}$, respectively. To calculate a case without gas dissolution, simply set the solubility coefficient $\alpha$ to zero in equation~\eqref{eq:mass-ph2}. Note that in dissolution-free cases, the constraint from H$_2$ budget is fundamentally to only determine various pressures ($P_{\rm H_2O}$, $P_{\rm H_2}$, and $P$) because volume mixing ratios ($X_{\rm H_2O}$ and $X_{\rm H_2}$) are already determined by temperature and oxygen fugacity only (see Section \ref{sec:b-dissol}).

\subsection{Results and comparison with dissolution-free model}

Figure \ref{fig:app-vary-fo2-fix-T} shows the results with (solid curves) and without (dotted curves) considering gas dissolution into melts. Under low oxygen fugacities ($\sim$IW$-6$), both the dissolution and dissolution-free cases feature almost pure H$_2$ atmosphere (red curve in Figure~\ref{fig:app-vary-fo2-fix-T}b). Moreover, due to the negligible H$_2$ solubility, the entire hydrogen budget (one Earth ocean) resides exclusively in the atmosphere reservoir in the form of H$_2$ gas. This enables us to estimate the corresponding $P_{\rm H_2}$ under low oxygen fugacities, i.e., $P_{\rm H_2} = M_{{\rm H_2}{\rm,atm}} g/A \approx 30$ bar, where $M_{{\rm H_2}{\rm,atm}}$ is equivalent to the prescribed hydrogen budget in mass unit. The estimated $P_{\rm H_2} \approx 30$ bar is verified by the numerical solutions for both dissolution and dissolution-free cases in Figure~\ref{fig:app-vary-fo2-fix-T}a (red curves). In contrast, under high oxygen fugacity conditions ($\sim$IW$+6$), the dissolution and dissolution-free cases both feature almost pure H$_2$O atmosphere (blue curve in Figure~\ref{fig:app-vary-fo2-fix-T}b). In this case, ignoring H$_2$O dissolution means nearly all the hydrogen budget is oxidised to H$_2$O and resides exclusively in the atmosphere. This enables estimating $P_{\rm H_2O} = M_{{\rm H_2O}{\rm,atm}} g/A \approx 30 \times 9 = 270$ bar, where the multiplier 9 is the ratio of molecular weight between H$_2$O and H$_2$ and it stems from the total hydrogen budget being completely oxidised to H$_2$O. Such an estimate of $P_{\rm H_2O} \approx 270$ bar at high $f_{\rm O_2}$ is verified by the numerical solution of the dissolution-free case in Figure~\ref{fig:app-vary-fo2-fix-T}a (dotted blue curve). When H$_2$O dissolution is enabled at high $f_{\rm O_2}$, its sequestration into coexisting melts greatly reduces its atmospheric partial pressure, as suggested by Figure~\ref{fig:app-vary-fo2-fix-T}a (compare solid and dotted blue curves). Furthermore, under oxidising conditions, dissolution-induced drop of $P_{\rm H_2O}$ drives $P_{\rm H_2}$ to even lower levels, as revealed by comparison of solid and dotted red curves in Figure~\ref{fig:app-vary-fo2-fix-T}a. Overall, because the H$_2$O converted from H$_2$ oxidation tends to dissolve in melts and thus decreases total pressure, the trend of total pressure variation with oxygen fugacity is flipped from dissolution-free to dissolution cases (dotted and solid magenta curves in Figure~\ref{fig:app-vary-fo2-fix-T}a).

Despite the considerable effect of gas solubility on atmosphere pressures, it is worth emphasising that for the H-O system investigated here, the variational trend of atmospheric volume mixing ratios with oxygen fugacity remains immune to the effect of gas (H$_2$O in this study) solubility, as attested to by the fully overlapped curves in Figure~\ref{fig:app-vary-fo2-fix-T}b. This is unsurprising because it is analytically elucidated in Section \ref{sec:b-dissol} that this trend depends on temperature only; gas solubility and the associated hydrogen budget play the mere role of further constraining partial and total pressure quantities. This immunity of H$_2$O and H$_2$ mixing ratios to gas solubilities and total hydrogen budget can also be robustly confirmed by our dissolution-free results and previous studies. For example, the analysis in Figure~\ref{fig:app-vary-fo2-fix-T}b shows that the crossover of $X_{\rm H_2O}$ and $X_{\rm H_2}$ occurs at $f_{\rm O_2}\sim$ IW. This crossover at $\sim$IW is first immune to expanding the system from H-O to C-H-O and C-H-O-N-S, as confirmed by closer inspection of Figures~\ref{fig:CHO_secondary_vary_fO2}, \ref{fig:CHO_hybrid_vary_fO2}, \ref{fig:CHONS_secondary_vary_fO2}, and \ref{fig:CHONS_hybrid_vary_fO2}. Second, as shown by Figure 1 in \cite{ortenzi20} and Figures 2 \& 8 in \cite{sossi23}, considering or not gas solubilities does not alter the crossover position either.

It is worth clarifying that, despite the above invariability for the H-O system, mixing ratios of other gas species, e.g., CO, CO$_2$, are indeed affected by dissolution behaviours \citep{ortenzi20}. Nevertheless, we reiterate that the invariability would not be discovered had we not followed the hierarchical modeling approach to begin with dissolution-free models.

\section{Supplementary figures} \label{append:fig}
In this appendix, we present for completeness the results for hybrid atmosphere simulations for the C-H-O-N-S system, which reveal no new trends beyond what we already learned from examining C-H-O systems and C-H-O-N-S secondary atmospheres.

\begin{figure*}[h!]
\begin{center}
% \vspace{-0.2in} 
% \includegraphics[width=0.6\columnwidth]{f15a.pdf}
% \includegraphics[width=0.6\columnwidth]{f15b.pdf}
% \includegraphics[width=0.6\columnwidth]{f15c.pdf}
% \includegraphics[width=0.6\columnwidth]{f15d.pdf}
% \includegraphics[width=0.6\columnwidth]{f15e.pdf}
% \includegraphics[width=0.6\columnwidth]{f15f.pdf}
\includegraphics[width=0.95\columnwidth]{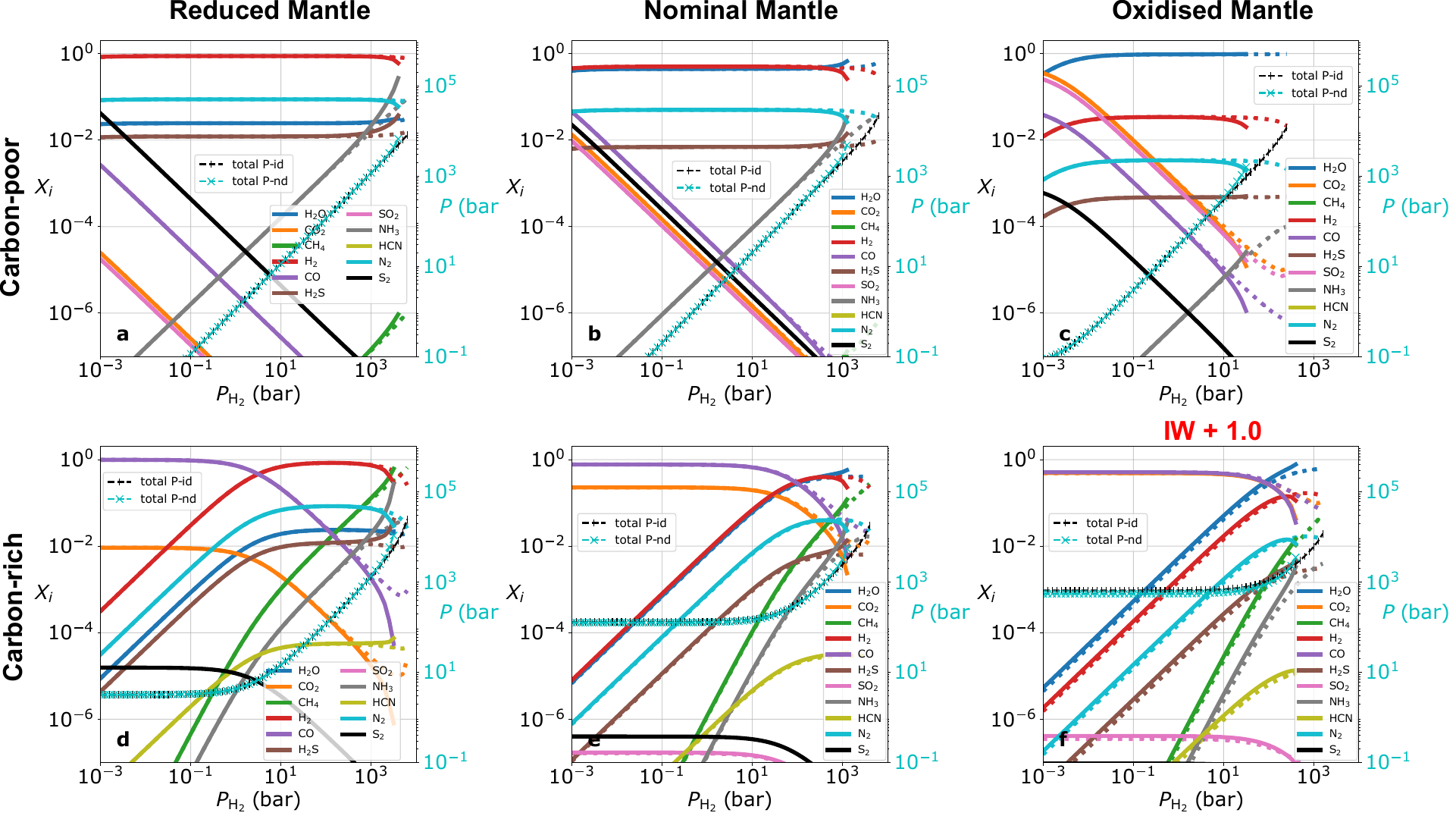}
\end{center}
% \vspace{-0.2in}
\caption{Examples of hybrid atmospheres in the C-H-O-N-S chemical system, where the volume mixing ratios (relative abundances by number) of gases are shown as a function of the prescribed atmospheric partial pressure of molecular hydrogen.  The sulfur fugacity is arbitrarily chosen to be $\log f_{\rm S_2} = {\rm PP} - 10$.  The top and bottom rows are for low- and high-carbon content in the mantle, respectively.  The first, second and third columns are for reduced, nominal and oxidised mantles, respectively.  See text for specific parameter values.  Regions of the plots where no curves exist are because the computed partial pressures of various gases exceed the computed total pressure, implying that no mathematical solutions exist.  Solid and dotted curves correspond to calculations with partially non-ideal effects (see text for details) and the assumption of an ideal gas with ideally-mixed constituents, respectively. The solved total pressures for ideal (P-id) and non-ideal (P-nd) cases overlap with each other until $P_{\rm H_2} \gtrsim 10^3$ bar. }
%\vspace{-0.15in}
\label{fig:CHONS_hybrid_vary_PH2}
\end{figure*}

\begin{figure*}[h!]
\begin{center}
\vspace{-0.1in} 
\includegraphics[width=0.95\columnwidth]{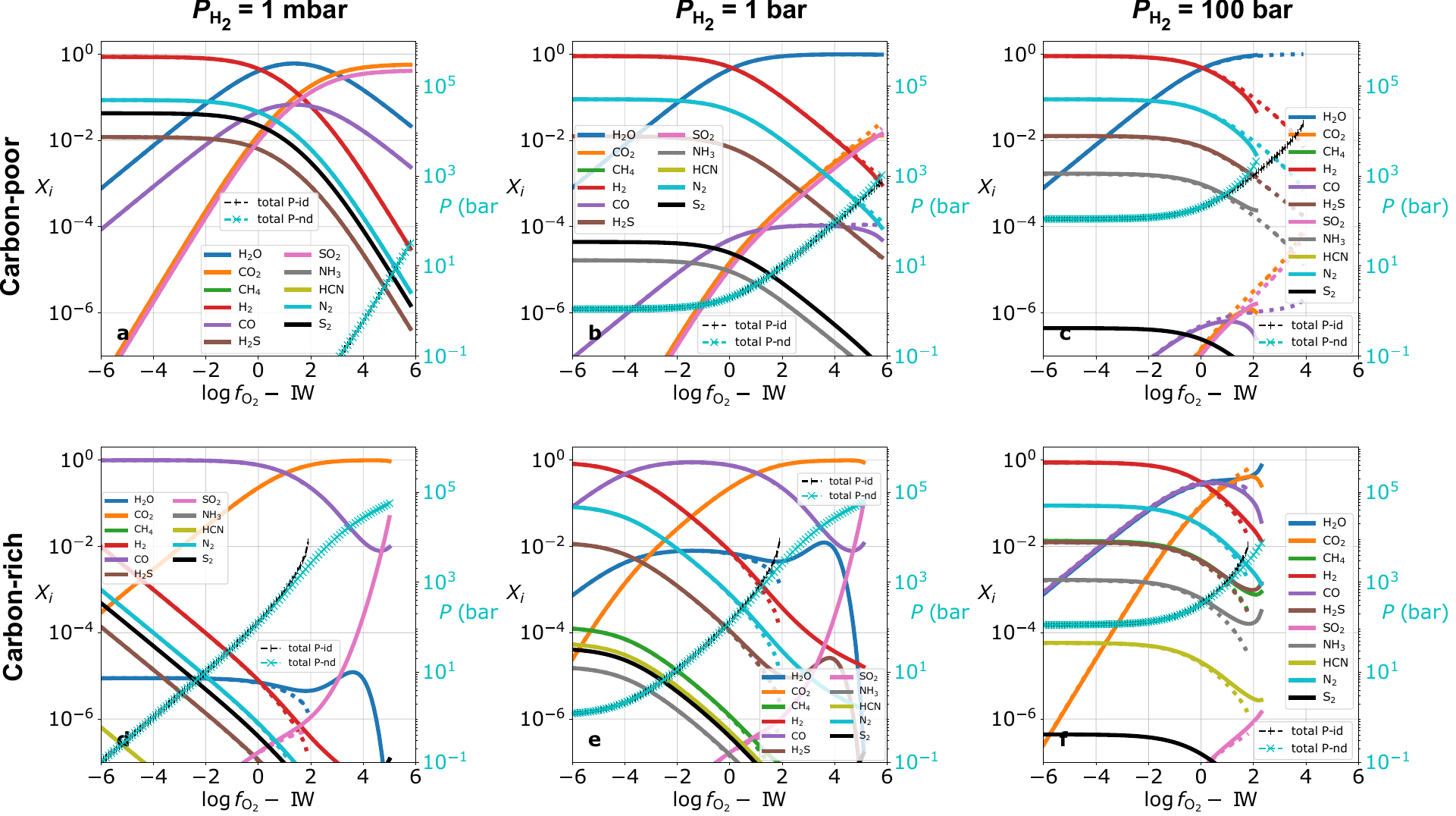}
\end{center}
% \vspace{-0.2in}
\caption{Same as Figure \ref{fig:CHONS_hybrid_vary_PH2} for hybrid atmospheres in the C-H-O-N-S chemical system, but with volume mixing ratios as a function of the oxygen fugacity of the mantle.  The sulfur fugacity is arbitrarily chosen to be $\log f_{\rm S_2} = {\rm PP} - 10$.  The first, second and third columns are for atmospheric partial pressures of molecular hydrogen of 1 mbar, 1 bar and 100 bar, respectively. The solved total pressures for ideal (P-id) and non-ideal (P-nd) cases overlap with each other until $\log f_{\rm O_2} \gtrsim {\rm IW}+2$. }
%\vspace{-0.15in}
\label{fig:CHONS_hybrid_vary_fO2}
\end{figure*}

\begin{figure*}[h!]
\begin{center}
% \vspace{-0.2in} 
% \includegraphics[width=0.6\columnwidth]{f17a.pdf}
% \includegraphics[width=0.6\columnwidth]{f17b.pdf}
% \includegraphics[width=0.6\columnwidth]{f17c.pdf}
% \includegraphics[width=0.6\columnwidth]{f17d.pdf}
% \includegraphics[width=0.6\columnwidth]{f17e.pdf}
% \includegraphics[width=0.6\columnwidth]{f17f.pdf}
\includegraphics[width=0.95\columnwidth]{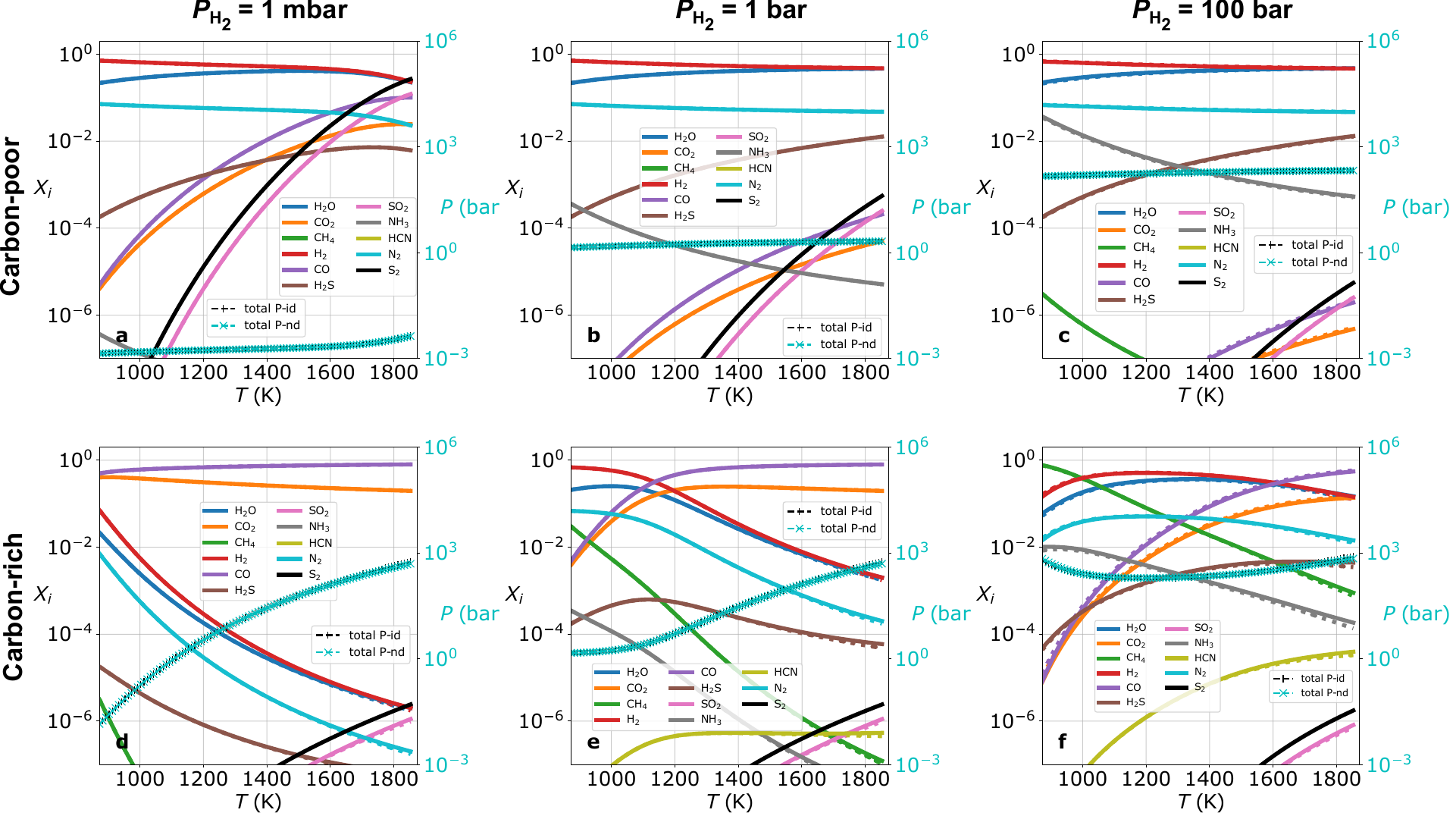}
\end{center}
% \vspace{-0.2in}
\caption{Same as Figure \ref{fig:CHONS_hybrid_vary_fO2} for hybrid atmospheres in the C-H-O-N-S chemical system, but with volume mixing ratios as a function of the melt temperature.  The oxygen and sulfur fugacities of the mantle are arbitrarily chosen to be $\log f_{\rm O_2} = \mbox{IW}$ and $\log f_{\rm S_2} = {\rm PP} - 10$, respectively. The solved total pressures for ideal (P-id) and non-ideal (P-nd) cases overlap for the entire $T$ range. }
%\vspace{-0.15in}
\label{fig:CHONS_hybrid_vary_T}
\end{figure*}

\begin{figure*}[h!]
\begin{center}
\vspace{-0.1in} 
\includegraphics[width=0.95\columnwidth]{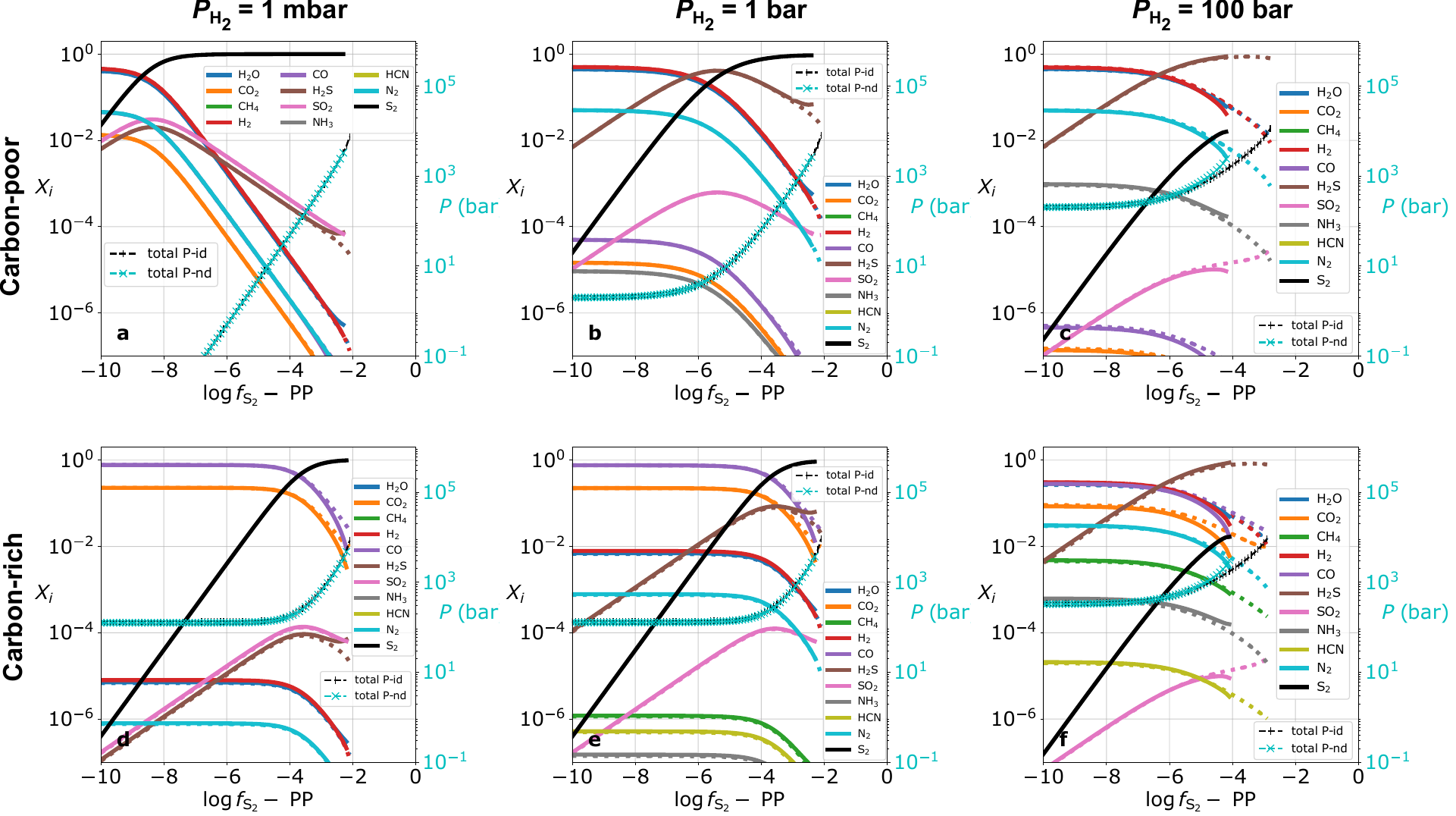}
\end{center}
% \vspace{-0.2in}
\caption{Same as Figure \ref{fig:CHONS_hybrid_vary_fO2} for hybrid atmospheres in the C-H-O-N-S chemical system, but with volume mixing ratios as a function of the sulfur fugacity of the mantle.  The oxygen fugacity of the mantle is arbitrarily chosen to be $\log f_{\rm O_2} = \mbox{IW}$. At $P_{\rm H_2}=100$ bar, the solved total pressures for ideal (P-id) and non-ideal (P-nd) cases overlap with each other until $\log f_{\rm S_2} \gtrsim {\rm PP} -4$. }
%\vspace{-0.15in}
\label{fig:CHONS_hybrid_vary_fS2}
\end{figure*}

\begin{figure*}[h!]
\begin{center}
% \vspace{-0.2in} 
% \includegraphics[width=0.6\columnwidth]{f19a.pdf}
% \includegraphics[width=0.6\columnwidth]{f19b.pdf}
% \includegraphics[width=0.6\columnwidth]{f19c.pdf}
% \includegraphics[width=0.6\columnwidth]{f19d.pdf}
% \includegraphics[width=0.6\columnwidth]{f19e.pdf}
% \includegraphics[width=0.6\columnwidth]{f19f.pdf}
\includegraphics[width=0.95\columnwidth]{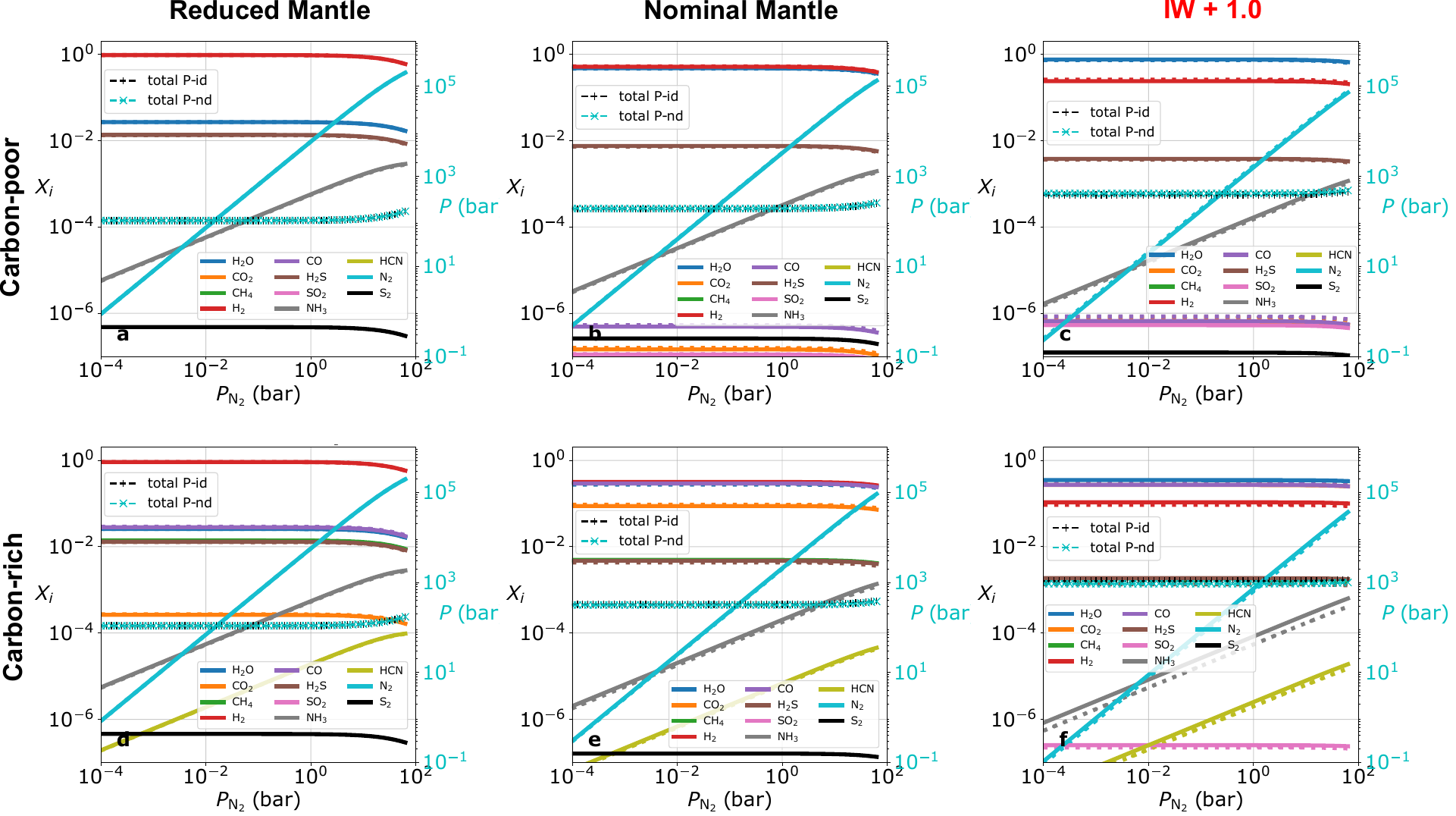}
\end{center}
% \vspace{-0.2in}
\caption{Same as Figure \ref{fig:CHONS_hybrid_vary_fO2} for hybrid atmospheres in the C-H-O-N-S chemical system, but with volume mixing ratios as a function of the atmospheric partial pressure of molecular nitrogen.  The oxygen and sulfur fugacities of the mantle are arbitrarily chosen to be $\log f_{\rm O_2} = \mbox{IW}$ and $\log f_{\rm S_2} = {\rm PP} - 10$, respectively. The solved total pressures for ideal (P-id) and non-ideal (P-nd) cases overlap for the entire $P_{\rm N_2}$ range. }
%\vspace{-0.15in}
\label{fig:CHONS_hybrid_vary_PN2}
\end{figure*}

\label{lastpage}


\clearpage
\begin{thebibliography}{99}

\bibitem[Ague et al.(2022)]{ague22} Ague, J.J., Tassara, S., Holycross, M.E., et al. \ 2022, Nature Geoscience, 15, 320

\bibitem[Ballhaus et al.(1991)]{ballhaus91} Ballhaus, C., Berry, R.F., \& Green, DH. \ 1991, Contributions to Mineralogy and Petrology, 107, 27

\bibitem[Barstow \& Heng(2020)]{bh20} Barstow, J.K., \& Heng, K. \ 2020, Space Science Reviews, 216, 82

\bibitem[Bean et al.(2021)]{bean21} Bean, J.L., Raymond, S.N., \& Owen, J.E. \ 2021, Journal of Geophysical Research, 126, e06639

\bibitem[Benneke et al.(2019)]{benneke19} Benneke, B., Knutson, H.A., Lothringer, J., et al. \ 2019, Nature Astronomy, 3, 813

\bibitem[Bockrath et al.(2004)]{bockrath04} Bockrath, C., Ballhaus, C., \& Holzheid, A. \ 2004, Chemical Geology 208, 265

\bibitem[Bower et al.(2022)]{bower22} Bower, D.J., Hakim, K., Sossi, P.A., \& Sanan, P. \ 2022, PJS, 3, 93

\bibitem[Burrows \& Sharp(1999)]{bs99} Burrows, A., \& Sharp, C.M. \ 1999, ApJ, 512, 843

%\bibitem[Chen \& Kipping(2017)]{ck17} Chen, J., \& Kipping, D. \ 2017, ApJ, 834, 17

\bibitem[Cartier \& Wood(2019)]{cw19} Cartier, C., \& Wood, B. J.  \ 2019, Elements, 15, 39

\bibitem[Connolly \& Cesare(1993)]{cc93} Connolly, J.A.D, \& Cesare, B. \ 1993, Journal of Metamorphic Geology, 11, 379

\bibitem[Delano(2001)]{delano01} Delano, J.W. \ 2001, Origins of Life and Evolution of Biospheres, 31, 311

\bibitem[Denbigh(1981)]{denbigh81} Denbigh, K.G. \ 1981, The Principles of Chemical Equilibrium, 4th edition (Cambridge University Press)

\bibitem[DeVoe(2020)]{devoe20} DeVoe, H. \ 2020, Thermodynamics and Chemistry, 2nd edition (University of Maryland)

\bibitem[Evans et al.(2010)]{evans10} Evans, K.A., Powell, R., \& Holland, T.J.B. \ 2010, Journal of Metamorphic Geology, 28, 667

\bibitem[Fegley et al.(2016)]{fegley16} Fegley, B., Jr., Jacobson, N.S., Williams, K.B., et al. \ 2016, ApJ, 824, 103

\bibitem[Ferry \& Baumgartner(1987)]{fb87} Ferry, J.M.\& Baumgartner, L. 1987, Reviews in Mineralogy and Geochemistry, 17, 323

\bibitem[French(1966)]{french66} French, B.M. \ 1966, Reviews of Geophysics, 4, 223

\bibitem[Fressin et al.(2013)]{fressin13} Fressin, F., Torres, G., Charbonneau, D., et al. \ 2013, ApJ, 766, 81

\bibitem[Fisher \& Heng(2018)]{fh18} Fisher, C., \& Heng, K. \ 2018, MNRAS, 481, 4698

\bibitem[Froese \& Gunter(1976)]{fr76} Froese, E. \& Gunter, A.E. \ 1976, Economic Geology, 8, 1589

\bibitem[Frost(1991)]{frost91} Frost, B.R.\ 1991, Reviews in Mineralogy and Geochemistry, 25, 1

\bibitem[Frost \& McCammon(2008)]{fm08} Frost, D.J., \& McCammon, C.A. \ 2008, Annual Review of Earth and Planetary Sciences, 36, 389

\bibitem[Fulton et al.(2017)]{fulton17} Fulton, B.J., Petigura, E.A., Howard, A.W., et al. \ 2017, AJ, 154, 109

\bibitem[Gaillard \& Scaillet(2014)]{gs14} Gaillard, F., \& Scaillet, B. \ 2014, Earth \& Planetary Science Letters, 403, 307

\bibitem[Gaillard et al.(2022)]{gaillard22} Gaillard, F., Bernadou, F., Roskosz, M., et al. \ 2022, Earth \& Planetary Science Letters, 577, 117255

\bibitem[Hamano et al.(2013)]{hamano13} Hamano, K., Abe, Y., \& Genda, H. \ 2013, Nature, 497, 607

\bibitem[Held(2005)]{held05} Held, I.M. \ 2005, Bulletin of the American Meteorological Society, 86, 1609

\bibitem[Heng et al.(2016)]{heng16} Heng, K., Lyons, J.R., \& Tsai, S.-M. \ 2016, ApJ, 816, 96

\bibitem[Heng \& Lyons(2016)]{henglyons} Heng, K., \& Lyons, J.R. \ 2016, ApJ, 817, 149

\bibitem[Heng(2018)]{heng18} Heng, K. \ 2018, Research Notes of the American Astronomical Society, 2, 128 

\bibitem[Herbort et al.(2020)]{herbort20} Herbort, O., Woitke, P., Helling, Ch. \& Zerkle, A. \ 2020, A\&A, A71, 18

\bibitem[Hirchmann(2012)]{hirschmann12} Hirschmann., M.M. \ 2012, Earth \& Planetary Science Letters, 341-344, 48

\bibitem[Holland \& Powell(1991)]{hp91} Holland, T.J.B, \& Powell, R. \ 1991, Contributions to Mineralogy and Petrology, 109, 265

\bibitem[Holland \& Powell(1998)]{hp98} Holland, T.J.B, \& Powell, R. \ 1998, Journal of Metamorphic Geology, 16, 209

\bibitem[Holland \& Powell(2003)]{hp03} Holland, T.J.B, \& Powell, R. \ 2003, Contributions to Mineralogy and Petrology, 145, 492

\bibitem[Holland \& Powell(2011)]{hp11} Holland, T.J.B, \& Powell, R. \ 2011, Journal of Metamorphic Geology, 29, 333

\bibitem[Holloway(1977)]{holloway77} Holloway, J.R. \ 1977, Thermodynamics in Geology, 161

\bibitem[Holloway(1981)]{holloway81} Holloway, J.R. \ 1981, Thermodynamics of Minerals and Melts, 273

\bibitem[Holloway(1987)]{holloway87} Holloway, J.R. \ 1987, Reviews in Mineralogy, 17, 211

\bibitem[Hort(1998)]{hort98} Hort, M. \ 1998, Journal of Petrology, 39, 1063

\bibitem[Howard et al.(2012)]{howard12} Howard, A.W., Marcy, G.W., Bryson, S.T., et al. \ 2012, ApJS, 201, 15

\bibitem[H{\"o}ning et al.(2021)]{honing21} H{\"o}ning., D., Baumeister, P., Grenfell, J.L., Tosi, N., \& Way, M.J. \ 2021, JGR-Planets, 126, e2021JE006895

\bibitem[Hu et al.(2015)]{hu15} Hu, R., Seager, S., \& Yung, Y.L. \ 2015, ApJ, 807, 8

\bibitem[Huang \& Wyllie(1973)]{hw73} Huang, W.L. \& Wyllie, P.J. \ 1973, Contributions to Mineralogy and Petrology, 42, 1

\bibitem[Huizenga(2011)]{huizenga11} Huizenga, J. \ 2011, Mineralium Deposita, 46, 23

\bibitem[Iacono-Marziano et al.(2012)]{Iacono12} Iacono-Marziano, G., Gaillard, F., Scaillet, B., et al. \ 2012, Earth \& Planetary Science Letters, 357-358, 319

\bibitem[J{\'e}go \& Dasgupta(2014)]{sd14} J{\'e}go S. \& Dasgupta, R. \ 2014, Journal of Petrology, 55, 1019

\bibitem[Keith et al.(2014)]{keith14} Keith, M., Hasse, K.M., Schwarz-Schampera, U., et al. \ 2014, Geology, 42, 699

\bibitem[Keppler \& Golabek(2019)]{kg19} Keppler, H. \& Golabek, G. \ 2019, Geochem. Persp. Let., II, 12

\bibitem[Kite et al.(2019)]{kite19} Kite, E.S., Fegley, B., Schaefer, L., \& Ford, E.B. \ 2019, ApJL, 887, L33

\bibitem[Kite et al.(2020)]{kite20} Kite, E.S., Fegley, B., Schaefer, L., \& Ford, E.B. \ 2020, ApJ, 891, 111

\bibitem[Krissansen-Totton et al.(2018)]{kt18} Krissansen-Totton, J., Olson, S., \& Catling, D.C. \ 2018, Sci. Adv., 4, eaao5747

\bibitem[Lewis(1901)]{lewis1901} Lewis, G.N. \ 1901, Proceedings of the American Academy of Arts and Sciences, 31, 49

\bibitem[Liggins et al.(2020)]{liggins20} Liggins, P., Shorttle, O., \& Rimmer, P.B. \ 2020, Earth \& Planetary Science Letters, 550, 116546

\bibitem[Lodders \& Fegley(2002)]{lodders02} Lodders, K., \& Fegley, B. \ 2002, Icarus, 155, 393

\bibitem[Mikal-Evans et al.(2021)]{mevans21} Mikal-Evans, T., Crossfield, I.J.M., Benneke, B., et al. \ 2021, AJ, 161, 18

\bibitem[Moses et al.(2011)]{moses11} Moses, J.I., et al. \ 2011, ApJ, 737, 15

\bibitem[\"{O}berg et al.(2011)]{oberg11} \"{O}berg, K.I., Murray-Clay, R., \& Bergin, E.A. \ 2011, ApJL, 743, L16

\bibitem[Ohmoto \& Kerrick(1977)]{ok77} Ohmoto, H., \& Kerrick, D. \ 1977, American Journal of Science, 277, 1013

\bibitem[O'Neill(1987)]{oneill87} O'Neill, H.S.C. \ 1987, American Mineralogist, 72, 67

\bibitem[Ortenzi et al. (2020)]{ortenzi20} Ortenzi, G., Noack, L., Sohl, F., Guimond, C. M., Grenfell, J. L., Dorn, C., Schmidt, J. M., Vulpius, S., Katyal, N., Kitzmann, D., \& Rauer, H. \ 1993, Philosophical Transactions of the Royal Society of London, Series A, 342, 105

\bibitem[Owen(2019)]{owen19} Owen, J.E. \ 2019, Annual Review of Earth and Planetary Sciences, 47, 67

\bibitem[Petigura et al.(2013)]{petigura13} Petigura, E.A., Marcy, G.W., \& Howard, A.W. \ 2013, ApJ, 770, 69

\bibitem[Philpotts \& Ague(2009)]{pa09} Philpotts, A.R., \& Ague, J.J. \ 2009 Principles of Igneous and Metamorphic Petrology, 2nd edition (Cambridge University Press)

\bibitem[Poulson \& Ohmoto(1977)]{po89} Poulson, S.R, \& Ohmoto, H. \ 1989, Contributions to Mineralogy and Petrology, 101, 418

\bibitem[Powell(1978)]{powell78} Powell, R. \ 1978, Equilibrium thermodynamics in Petrology: An introduction (Cambridge University Press)

\bibitem[Powell et al.(1998)]{powell98} Powell, R., Holland, T., \& Worley, B. \ 1998, Journal of Metamorphic Geology, 16, 577

\bibitem[Rogers(2015)]{rogers15} Rogers, L.A. \ 2015, ApJ, 801, 41

\bibitem[Rogers et al.(2021)]{rogers21} Rogers, J.G., Gupta, A., Owen, J.E., \& Schlichting, H.E. \ 2021, MNRAS, 508, 5886

\bibitem[Schaefer \& Fegley(2003)]{sf03} Schaefer, L. \& Fegley, B., Jr. \ 2003, Icarus, 169, 216

\bibitem[Schaefer et al.(2012)]{schaefer12} Schaefer, L., Lodders, K., \& Fegley, B., Jr. \ 2012, ApJ, 755, 41

\bibitem[Schaefer \& Fegley(2017)]{schaefer17} Schaefer, L., \& Fegley, B., Jr. \ 2017, ApJ, 843, 120

\bibitem[Seager et al.(2013)]{seager13} Seager, S., Bains, W., \& Hu, R. \ 2013, ApJ, 777, 95

%\bibitem[Shizgal \& Arkos(1996)]{sa96} Shizgal, B.D., \& Arkos, G.G. \ 1996, Reviews of Geophysics, 34, 483

\bibitem[Sossi et al.(2020)]{sossi20} Sossi, P.A., Burham, A.D, Badro, J., Lanzirotti, A, Newville M., \& O'Neill, H.St.C. \ 2020, Science Advances, 6, eabd1387

\bibitem[Sossi et al.(2023)]{sossi23} Sossi, P.A., Tollan, P. M. E., Badro, J., \& Bower, D. J. \ 2023, Earth \& Planetary Science Letters, 601, 117894

\bibitem[Sun \& Lee(2022)]{sl22} Sun, C. \& Lee, C.T.A. \ 2022, Geochimica et Cosmochimica Acta, 338, 302

\bibitem[Symonds \& Reed(1993)]{sr93} Symonds, R.B. \& Reed, M.H. \ 1993, American Journal of Science, 293, 758

\bibitem[Takahashi et al.(1993)]{takahashi93} Takahashi, E., Shimazaki, T., Tsuzaki, Y., Yoshida, H., Cox, K.G., McKenzie, D.P., \& White, R.S. \ 1993, Philosophical Transactions of the Royal Society of London, Series A, 342, 105

\bibitem[Thompson et al. (2022)]{thompson22} Thompson, M. A., Krissansen-Totton, J., Wogan, N., Telus, M., \& Fortney, J. J. \ 2022, PNAS, 119, e2117933119

\bibitem[Toulmin \& Barton(1964)]{tb64} Toulmin, P., \& Barton, P.B. \ 1964, Geochimica et Cosmochimica Acta, 28, 641

\bibitem[Wade \& Wood(2005)]{ww05} Wade, J. \& Wood, B.J. \ 2005, Earth and Planetary Science Letters, 236, 78

\bibitem[Wadhwa(2008)]{wadhwa08} Wadhwa, M. \ 2008, Reviews in Mineralogy \& Geochemistry, 68, 493

\bibitem[Wakeford et al.(2017)]{wakeford17} Wakeford, H.R., Sing, D.K., Kataria, T., et al. \ 2017, Science, 356, 628

\bibitem[Woitke et al. (2021)]{woitke21} Woitke, P., Herbort, O., Helling, Ch., St{\"u}eken, E., Dominik, M., Barth, P., \&Samra, D. \ 2021, A \& A, 646, A43

\end{thebibliography}
\end{document}